\documentclass[a4paper,12pt]{article}
\usepackage{jheppub}
\usepackage[dvipsnames]{xcolor}
\usepackage[T1]{fontenc}
\usepackage{amsthm,amsmath,latexsym,amssymb,amsfonts,amscd}
\usepackage{mathtools,commath}
\usepackage{graphics,lscape,fancyhdr,array,euscript}

\usepackage[compat=1.1.0]{tikz-feynman}
\usepackage{tikz}
\usetikzlibrary{calc,decorations.markings}

\usepackage{empheq}
\numberwithin{equation}{section} 
\usepackage{hyperref}
\hypersetup{colorlinks=true,linkcolor=magenta,anchorcolor=green,citecolor=cyan,filecolor=black,menucolor=black,urlcolor=brown}
\makeatletter

\pgfdeclarelayer{bg}    % declare background layer
\pgfsetlayers{bg,main}  % set the order of the layers (main is the standard layer)

\usepackage{float}
\allowdisplaybreaks
\definecolor{linkblue}{rgb}{0.1,0.3,.7}
\definecolor{forestgreen(web)}{rgb}{0.13, 0.55, 0.13}
\definecolor{lava}{rgb}{0.81, 0.06, 0.13}
\hypersetup{
	breaklinks,
	colorlinks,
	citecolor=forestgreen(web),
	filecolor=linkblue,
	linkcolor=lava,
	urlcolor=linkblue
}

\newcommand{\pl}{\partial}
\newcommand{\besubeqs}{\begin{subequations}}%
	\newcommand{\esubeqs}{\end{subequations}}

\def\Li#1(#2){\textrm{Li}_{#1}\left(#2\right)}
\def\cLi_#1(#2){\mathcal{L}_{#1}\left(#2\right)}
\def\bLi_#1(#2){\mathbf{L}_{#1}\left(#2\right)}

\newcommand{\dd}{\mathrm{d}}
\newcommand{\eul}{\mathrm{e}}

\newcommand{\Op}{\mathcal{O}}
\newcommand{\zetab}{\bar{\zeta}}
\newcommand{\cH}{\mathcal{H}}
\newcommand{\cW}{\mathcal{W}}

\def\u1{u_1}
\def\uZ{u_{\zeta}}
\def\|#1|{\norm{#1}}
\def\Mpl#1#2{\textrm{Li}_{#1}\left(#2\right)}

\preprint{\vbox{\hbox{\hphantom{XXXXX}IPhT-t22/001}\hbox{\hphantom{XXX}LMU-ASC 03/22}}}

\title{Analytical evaluation of AdS${}_4$ Witten diagrams as flat space multi-loop Feynman integrals
}

\author[a]{Till Heckelbacher,}
\author[a]{Ivo Sachs,}
\author[b,c]{Evgeny Skvortsov}
\author[d,e]{and Pierre Vanhove}

\affiliation[a]{Arnold-Sommerfeld-Center for Theoretical Physics, Ludwig-Maximilians-Universit\"at M\"unchen,
	Theresienstr. 37, D-80333 Munich, Germany}
\affiliation[b]{Service de Physique de l’Univers, Champs et Gravitation, Universit\'e de Mons, 20 place du Parc, 7000 Mons, Belgium}
\affiliation[c]{Lebedev Institute of Physics, Leninsky ave. 53, 119991 Moscow, Russia}
\affiliation[d]{Institut de Physique Theorique, Universit\'e Paris-Saclay, CEA, CNRS, F-91191 Gif-sur-Yvette Cedex, France}
\affiliation[e]{National Research University Higher School of
	Economics, Russian Federation}
	

\abstract{ 
We describe a systematic approach for the evaluation of Witten diagrams for multi-loop scattering amplitudes of a conformally coupled scalar $\phi^4$-theory in Euclidean AdS$_4$, by recasting the Witten diagrams as flat space Feynman integrals. We derive closed form expressions for the anomalous dimensions for all double-trace operators up to the second order in the coupling constant. We explain the relation between the flat space unitarity methods and the discontinuities of the short        distance expansion on the boundary of Witten diagrams. 
}
\begin{document}
	\maketitle
	\flushbottom
	
	\newpage
	%%%%%%%%%%%%%%%%%%%%%%%%%%%%%%%%%%%%%%%%%%%%%%%%%%%%%%%%%%
	\section{Introduction}
	%%%%%%%%%%%%%%%%%%%%%%%%%%%%%%%%%%%%%%%%%%%%%%%%%%%%%%%%%%
	
	Although quantum field theory in curved space time has a long tradition, progress beyond classical, or tree-level calculations has been slow, compared to flat space calculations because of technical complications due to the absence of translation invariance, in particular. 
	
	The simplest space-times that are not flat are arguably de Sitter and anti-de Sitter space (AdS), since they have the same number of isometries as Minkowski space. Furthermore, there is a natural analog of $S$-matrix elements for these spaces in the form of conformal boundary correlators for anti-de Sitter space~\cite{Witten:1998qj,Maldacena:1997re,Gubser:1998bc,Freedman:1998tz} and the wave function of the universe for de Sitter~\cite{Hartle:1983ai,Hertog:2011ky,Harlow:2011ke,Strominger:2001pn,Maldacena:2002vr,Anninos:2014lwa}. However, calculations in these spaces are still technically challenging since no Fourier transformation into four-momentum space exists and one has to evaluate amplitudes in position space. Recently there have been several attempts at loop calculations in AdS, introducing new techniques like Mellin space, differential representation and using the AdS/CFT correspondence in combination with conformal bootstrap methods (see e.g.~\cite{Penedones:2010ue,Aprile:2017bgs,Aprile:2017qoy,Herderschee:2021jbi,Aharony:2016dwx,Giombi:2017hpr,Liu:2018jhs,Alday:2017xua,Alday:2017vkk,Alday:2018pdi,Alday:2018kkw,Alday:2019nin,Alday:2021ajh,Drummond:2019hel,Carmi:2020ekr,Yuan:2017vgp,Yuan:2018qva,Albayrak:2020bso,Ghosh:2019lsx,Meltzer:2019nbs,Aprile:2019rep,Ponomarev:2019ofr,Carmi:2018qzm,Carmi:2019ocp,Carmi:2021dsn,Sleight:2019hfp,Meltzer:2020qbr,Meltzer:2019pyl,Meltzer:2021bmb,Costantino:2020vdu,Fichet:2021xfn,Eberhardt:2020ewh}).
	
	Explicit loop calculations in these backgrounds in position space have been successfully carried out for de Sitter~\cite{Heckelbacher:2020nue} and for anti-de Sitter space~\cite{Bertan:2018khc,Bertan:2018afl}. This has lead to new results for loop corrections to the  operator product expansion (OPE) coefficients and dimensions of ``double-trace'' operators of the three-dimensional conformal field theory that governs the wave-function or boundary correlation functions---the analog of branching ratios and the mass spectrum in flat space QFT. Nevertheless, it did not provide a systematic formalism for pushing these calculations to higher-loop orders. 
	
	One of the aims of the present paper is  to fill this gap. We do so by mapping the AdS calculations to flat space calculations where an impressive machinery for an analytic evaluation of Feynman integrals has been developed in recent years~\cite{Smirnov:2019qkx,Panzer:2015ida,Vanhove:2018mto,Bogner:2018bvz,Bonisch:2021yfw,Mastrolia:2018uzb,Weinzierl:2022eaz}. Here, we bring the four-point function of a conformally coupled scalar $\phi^4$-theory in euclidean AdS$_4$ into a form we can interpret as a flat space Feynman diagram and evaluate it analytically. In doing so, we are able to identify some well known structures of flat space momentum integrals in position space calculations in AdS, in particular, various types of special functions and the corresponding  complex geometry manifested in the integrands. 
	
For a conformally coupled scalar field, the dimension $\Delta$ of the ``single-trace'' operator whose two- and four-point function we compute in this paper equals $\Delta=1$ or $\Delta=2$, depending on the choice of boundary conditions. We explain that the results for $\Delta=2$ can be obtained from the ones  for $\Delta=1$ by using a descent procedure obtained by the action of  differential operators on the Feynman integrals. To the loop order that is considered in this work,  the Witten diagrams for $\Delta=1$ and $\Delta=2$ are expressed in terms of single-valued multiple polylogarithms~\cite{Brown:2013gia,Schnetz:2013hqa,Charlton:2021uhu} and elliptic polylogarithms~\cite{Bloch:2013tra,Bloch:2014qca,Bloch:2016izu,Broedel:2019hyg,Adams:2014vja,Adams:2015gva,Broedel:2017siw,Broedel:2017kkb}.

For a field of dimension $\Delta\geq3$, the large distance behaviour of the bulk AdS propagator has a logarithmic behaviour, which can be treated by introducing an auxiliary analytic regulator. This leads to non-integer powers of propagators, familiar from the analytic regularisation~\cite{Smirnov:2004ym}.

	As far as concrete results are concerned, the framework developed in this paper allows us to derive a closed form of the anomalous dimensions for all ``double-trace'' operators that arise in the deformation of the generalized free field in three euclidean dimensions. In the absence of bulk interactions the three-dimensional conformal field theory data is that of a generalized free field $\Op_\Delta$ of dimension $\Delta=1$ and $\Delta=2$ for which the double-trace operators are schematically of the form $:\Op_\Delta\square^n \pl^l \Op_\Delta:$ with dimension $\Delta_{(n,l)}=2\Delta+2n+l$ and spin $l$. A $\phi^4$-interaction in the bulk amounts to a shift in the dimensions of double-trace primaries as well as a deformation of the OPE-coefficients in accordance with the conformal bootstrap~\cite{Heemskerk:2009pn}. As an application of our formalism we derive a closed form of the anomalous dimensions for the complete set of double-trace primaries, up to second order (or one-loop) in the deformation parameter which can be taken to be the renormalised $\phi^4$ coupling $\lambda_R$. To give a flavor of our results, we list some of the anomalous dimensions derived in this paper. Writing  $\Delta_{(n,l)}=2\Delta+2n+l+\gamma^{(1)}_{n,l}(\Delta)+\gamma^{(2)}_{n,l}(\Delta)$, we determine 
	\begin{equation}
    \gamma_{n,0}^{(1)}(\Delta)=\frac{\lambda_R}{16\pi^2}(1+\delta_{\Delta,1}\delta_{n,0}) \,;\qquad\gamma^{(2)}_{n>0,l>0}(\Delta)=\frac{\lambda_R^2}{(16\pi^2)^2}T^\Delta_{n,l}\,,
\end{equation}
with
{\footnotesize
\begin{equation}
T^\Delta_{n,l}=-\frac{2(l^2+(2\Delta+2n-1)(\Delta+n+l-1))}{l(l+1)(2\Delta+2n+l-1)(2\Delta+2n+l-2)}
-\frac{2(-1)^\Delta(H^{(1)}_l-H^{(1)}_{2\Delta+2n+l-2})}{(2\Delta+2n+2l-1)(\Delta+n-1)}\,,
\end{equation}}%
\noindent where $H^{(1)}_{i}=\sum_{n=1}^i n^{-1}$ is the harmonic sum. We derive similar results for all values of $n$ and $l$ in section~\ref{sec:conformal_block_expansion}. 

To apply the flat space formalism in this work, we use dimensional regularisation. While preserving flat space Poincar\'e invariance, this regularisation does not preserve the AdS-invariance. This is in contrast to the regularisation scheme introduced in~\cite{Bertan:2018khc} which, however, does not easily combine with mapping to flat space techniques used in this work. We find a way to implement dimensional regularisation, which restores the AdS-invariance of the renormalised four-point function

A natural question is whether the results for the anomalous dimensions are scheme independent. Given that they correspond to what would be mass relations in a flat space QFT one would expect this. On the other hand, since dimensional regularisation is not AdS-invariant, one may question its validity. There are two ways to test scheme invariance. One is to compare with the AdS-invariant scheme in~\cite{Bertan:2018khc,Bertan:2018afl,Heckelbacher:2020nue}. For this purpose, we perform the calculations in both schemes and then show how to relate them. Another test follows from the conformal bootstrap, or conformal block expansion. It implies that the square of the first order anomalous dimension $\gamma^{(1)}(\Delta)$, that multiplies a logarithm of the cross-ratio in the short distance expansion of the tree-level cross diagram, enters in the coefficient of the logarithm square term of the one-loop diagram~\cite{Bertan:2018afl,Heckelbacher:2020nue}. This can also be seen to follow from unitarity by relating sequential discontinuities of amplitudes at different loop order.\footnote{In AdS one can equally consider the double discontinuity~\cite{Fitzpatrick:2011dm,Liu:2018jhs,Meltzer:2019nbs,Meltzer:2020qbr,Meltzer:2021bmb} to relate the logarithm square terms at one-loop to $\gamma^{(1)}(\Delta)$ at tree-level.} While not strictly necessary here, since we compute the one-loop amplitude explicitly, we propose a simple method, based on the Cutkosky rules~\cite{Cutkosky:1960sp,Abreu:2014cla,Bourjaily:2020wvq}, to extract the sequential discontinuities of the Witten diagrams. We illustrate the success of this method by comparing two examples to our exact results and emphasize that this method could be useful for calculating higher-loop corrections to anomalous dimensions.

	The paper is organized as follows: In section~\ref{sec:cop} we present the conventions as well as the definitions and normalizations of the propagators used in the sequel. In particular the relation to boundary conditions in AdS is discussed in detail. We show how to rewrite the AdS propagators as a combination of flat space propagators which will be important in mapping AdS position space loop calculations to momentum space calculations in flat space. In section~\ref{sec:QFT_in_AdS} we describe the quantization of an interacting scalar field on the Poincar\'e patch of AdS$_4$, in particular, the choice of vacuum as well as the relation of bulk- and boundary $n$-point functions to conformal correlators. We also specify the Witten diagrams which we will compute in the sequel as well as two regularisation prescriptions, natural for AdS- and flat space calculations, respectively and the relation between them. Finally, we introduce differential operator relations which will be important to interpolate between different boundary conditions in loop calculations. In section~\ref{sec:calculation_witten_diagram} we evaluate the four-point correlation function at tree-level and one-loop, both in the AdS-invariant regularisation and dimensional regularisation relevant for the flat space approach to AdS-correlators. In section~\ref{sec:unitarity_methods} we
        evaluate the flat space unitarity cuts of the cross diagram and a one-loop diagram,
        and show their relation to the discontinuities of the short
        distance expansion on the boundary. We discuss renormalisation of UV-divergences in both schemes and, in combination with appendix~\ref{sec:special functions}, provided closed expressions for the finite parts that will provide the data to determine the anomalous dimension of ``double-trace'' operators in section~\ref{sec:conformal_block_expansion}.   In section~\ref{sec:conclusions} we discuss some further application of the methods presented in this work. We have collected in the appendix~\ref{sec:MPLs} various expressions in terms of single-valued multiple polylogarithms that enter our analytic evaluations. In appendix~\ref{sec:exapnsion_cross} we collect various results about  the tree-level (cross) Witten diagram. In section~\ref{sec:cross_mpl} we give the evaluation for the cross diagram for all $\Delta$ in terms of single-valued polylogarithms of weight at most two, in appendix~\ref{subsec:Cross_exact} we evaluate the Witten cross diagram in dimensional regularisation, and in appendix~\ref{subsec:Cross_expand} we give the expansion of the cross for general conformal dimension $\Delta$. In appendix~\ref{sec:special functions} we collect our results for the evaluation of the one-loop Witten diagram, and in appendix~\ref{sec:OPE} we recall the expressions for 
	the OPE coefficients for a generalized free field in $d=3$ dimensions with external conformal dimension $\Delta$ and the series representation of the conformal blocks in three dimensions.

%-------------------------------------------------------------------------%

	%%%%%%%%%%%%%%%%%%%%%%%%%%%%%%%%%%%%%%%%%%%%%%%%%%%%%%%%%%
	\section{Coordinate systems and propagators in AdS}\label{sec:cop}
	%%%%%%%%%%%%%%%%%%%%%%%%%%%%%%%%%%%%%%%%%%%%%%%%%%%%%%%%%%
	In this section we review the basic geometric properties of anti-de Sitter space as well as the propagators of a scalar field theory in AdS. This defines the framework that will be used for mapping
	the computation of Witten diagrams in the bulk of AdS space to expressions that are familiar from momentum space Feynman integrals in flat space. In the rest of this paper, except for section \ref{sec:unitarity_methods}, we will exclusively work with the Wick rotated geometry also known as EAdS. 
	%%%%%%%%%%%%%%%%%%%%%%%%%%%%%%%%%%%%%%%%%%%%%%%%%%%%%%%%%%%%%%%%%%%%%%%%%%%%%%%%%%%%%%%%%%%%%%%%
	\subsection{Coordinate systems}
	\label{sec:coordinates}
	%%%%%%%%%%%%%%%%%%%%%%%%%%%%%%%%%%%%%%%%%%%%%%%%%%%%%%%%%%%%%%%%%%%%%%%%%%%%%%%%%%%%%%%%%%%%%%%%
	Euclidean anti-de Sitter space or Lobachevsky space in four
        dimensions is a maximally symmetric space which can be
        embedded as a disconnected hyperboloid in the five-dimensional
        ambient space equipped with the mostly plus metric
        $(\eta_{AB})=(+,\cdots ,+,-)$
	\begin{equation}
		\label{eq:quadric}
		\mathbf X^2: = \eta_{AB} \mathbf X^A \mathbf X^B = (\mathbf X^0)^2+\cdots +(\mathbf X^3)^2-(\mathbf X^4)^2 = -\frac{1}{a^2},
	\end{equation}
	where $a$ is the inverse of the anti-de Sitter radius. There is a map from this space to the upper-half space,
	\begin{equation}
		\label{eq:Hplusdef}
		\mathcal H^+_4:=\left\{X:=(\vec x,z), \, \vec x\in\mathbb R^3,
		z>0\right\}, 
	\end{equation}
	equipped with the Poincar\'e metric
	\begin{equation}
		\label{eq:poincmetricc}
		\mathrm{d}s^2 =\frac{1}{a^2 z^2} (\mathrm{d}z^2 + \mathrm{d}{\vec x}^2).
	\end{equation}
	The Poincar\'e coordinates are related to the embedding coordinates by ($1\leq i\leq 3$)
	\begin{align}
		\mathbf X^0 = \frac{1}{\sqrt{2} a z} (1 - \frac{\vec x^2}{2} -\frac{z^2}{2}),&&
		\mathbf X^i = \frac{x^i}{a z}, &&
		\mathbf X^4 = \frac{1}{\sqrt{2}a z} (1 + \frac{\vec x^2}{2} +\frac{z^2}{2})\,.\label{eq:AdScoord}
	\end{align}
			The $SO(4,1)$ invariant geodesic distance 
	\begin{align}
		d(\mathbf X,\mathbf Y)&=\frac{1}{a}\mathrm{arccosh}\left(-a^2\mathbf{X}\cdot\mathbf{Y}\right),
	\end{align}
	involving hyperbolic functions complicates calculations unnecessarily. We use instead the hyperbolic ``angle'',
	\begin{equation}
		\label{eq:Kdef}
		K(\mathbf X,\mathbf Y):=-\frac{1}{a^2 \mathbf X
			\cdot \mathbf Y} = \frac{2zw}{(\vec x - \vec y)^2 + z^2 + w^2},
                    \end{equation}
                    with
                    	\begin{align}
		\mathbf Y^0 = \frac{1}{\sqrt{2} a w} (1 - \frac{\vec y^2}{2} -\frac{w^2}{2}),&&
		\mathbf Y^i = \frac{y^i}{a w}, &&
		\mathbf Y^4 = \frac{1}{\sqrt{2}a w} (1 + \frac{\vec y^2}{2} +\frac{w^2}{2})\,.\label{eq:AdScoordY}
	\end{align}
	We introduce, the anti-podal map
	\begin{equation}
		\label{eq:sigmadef}
		\sigma(\vec x,z):= (\vec x,-z)\,,
	\end{equation}
	whose fixed locus is given by the conformal
	boundary of anti-de Sitter space, located at $z=0$.
	This operation
	exchanges the upper-half space in~\eqref{eq:Hplusdef}, with $z>0$, and
	the lower-half space $\mathcal H^-_4:=\left\{(\vec x,z), \, \vec x\in\mathbb R^3, z<0\right\}$.

It is easy to see that for coincident points in the bulk   $\mathbf
Y=\mathbf X$, i.e. $(\vec y,w)=(\vec x,z)$, 
        we have $K(\mathbf X,\mathbf X)=1$ and that for coincident
        antipodal points in bulk $\mathbf
        Y=\sigma(\mathbf X)$, i.e. $(\vec y,w)=(\vec x,-z)$, one finds
        $K(\mathbf X,\sigma(\mathbf X))=-1$.

	%%%%%%%%%%%%%%%%%%%%%%%%%%%%%%%%%%%%%%%%%%%%%%%%%%%%%%%%%%%%%%%%%%%%%%%%%%%%%%%%%%%%%%%%%%%%%%%%
	\subsection{Propagators in \texorpdfstring{$EAdS_4$}{Lg}}
	\label{sec:propagators}
	%%%%%%%%%%%%%%%%%%%%%%%%%%%%%%%%%%%%%%%%%%%%%%%%%%%%%%%%%%%%%%%%%%%%%%%%%%%%%%%%%%%%%%%%%%%%%%%%
	
	The bulk-to-bulk propagator $\Lambda(\mathbf X,\mathbf Y;\Delta)$ for a
	scalar field of mass $m$ between two bulk points $\mathbf X$ and
	$\mathbf Y$ in EAdS is a solution of 
	\begin{equation}
		\label{inhel}
		\left( - \Box + m^2\right) \Lambda(\mathbf X,\mathbf Y) =
		\frac{1}{\sqrt{|g|}} \delta^{4}(\mathbf X-\mathbf Y)\,,
	\end{equation}
	subject to boundary conditions at $z=0$.
By expressing the d'Alembertian in terms of the hyperbolic ``angle''
$K=K(\mathbf X, \mathbf Y)$ gives
	\begin{align}
		\left[K^2(1-K^2)\frac{\dd^2}{\dd K^2}-2K\left(1+K^2\right)\frac{\dd}{\dd K}+\frac{m^2}{a^2}\right]\Lambda(K)=\frac{1}{\sqrt{g}}\delta^{4}(\mathbf{X}-\mathbf{Y})\,.
	\end{align}
It is clear that the solution only depends on the hyperbolic ``angle''. We therefore find the general solution
	\begin{align}
	\Lambda(\mathbf X,\mathbf Y;\Delta_+) =&C_+(\Delta_+) K^{\Delta_+}{}_2F_1\left(\frac{\Delta_+}{2},\frac{\Delta_+ +1}{2};\Delta_+ -\frac{1}{2};K^2\right)\nonumber\\
		&+C_-(\Delta_+) K^{\Delta_-}{}_2F_1\left(\frac{\Delta_-}{2},\frac{\Delta_- +1}{2};\Delta_- -\frac{1}{2};K^2\right),\label{eq:LambdaGen}
	\end{align}
expressed in terms of the Gauss' hypergeometric function
${}_2F_1(a,b;c;z)$. The coefficients $C_{\pm}$ are fixed by the boundary conditions and 
	\begin{align}
		\label{DeltaDef}
		\Delta_{+} = \frac{3}{2} + \sqrt{\frac{9}{4} +
          \frac{m^2}{a^2}}, \qquad \Delta_-=3-\Delta_+\,,
	\end{align}
	determines the conformal weight of the dual operator on the boundary. The leading contribution of the Green function for $z\to0$ is simply
	\begin{align}\notag
	\lim_{z\to0}	\Lambda(\mathbf  X,\mathbf
          Y;\Delta_+)= C_+(\Delta_+)\left(\frac{(2zw)^{\Delta_+}}{((\vec{x}-\vec{y})^2+w^2)^{\Delta_+}}+\cdots\right)+\nonumber\\*
          +C_-(\Delta_+)\left(\frac{(2zw)^{\Delta_-}}{((\vec{x}-\vec{y})^2+w^2)^{\Delta_-}}+\cdots\right)\,.
	\end{align}

	In what follows we will consider two boundary conditions\footnote{	
	Choosing a Green function corresponds to choosing a
        vacuum. We will argue in section~\ref{sec:QFT_in_AdS} why the
        Dirichlet and Neumann boundary conditions correspond to the
        correct choice of the vacuum.} by either
      setting $C_+(\Delta_+)$ or $C_-(\Delta_+)$ to zero.
     The choice $C_+(\Delta_+)=0$ corresponds to a Neumann boundary condition, while $C_-(\Delta_+)=0$ corresponds to a Dirichlet boundary condition. For given boundary conditions, the behaviour
          of the propagator under the antipodal map in~\eqref{eq:sigmadef} is determined by the factor $K^{\Delta}$ appearing~\eqref{eq:LambdaGen} and is therefore either even or odd depending whether $\Delta$ is even or odd.

	To fix the normalization of the Dirichlet and Neumann Green
        function we demand that in the flat space limit, $a\to 0$, their singularity agrees with the flat space Green function. 
The properly normalized propagator is (see
	e.g.~\cite{BURGES1986285})
	\begin{equation}\label{eq:Lambdadef}
		\Lambda(\mathbf X,\mathbf Y;\Delta_{\pm}):=
		\mathcal{N}_{\Delta_\pm}K(\mathbf X,\mathbf Y)^{\Delta_{\pm} } \,
			_2F_1\left(\frac{\Delta_{\pm} }{2},\frac{\Delta_{\pm} +1}{2};\Delta_{\pm}
			-\frac12;K(\mathbf X,\mathbf Y)^2\right),
	\end{equation}
	with
	\begin{equation}\label{e:Ndef}
		\mathcal{N}_{\Delta}=\left(\frac{a}{2\pi}\right)^2\frac{  \Gamma \left(\frac{\Delta}{2}\right) \Gamma
			\left(\frac{\Delta +1}{2}\right) }{ \Gamma \left(\Delta-\frac{1}{2}\right)}\,.
	\end{equation}
The short distance singularity of the propagator for coincident points
is given by 
\begin{equation}
  \lim_{\mathbf Y\to\mathbf X}  \Lambda(\mathbf X,\mathbf Y;\Delta_{\pm})\simeq
    \left(\frac{a}{2\pi}\right)^2 \frac{zw}{(\vec x-\vec y)^2+(z-w)^2},
  \end{equation}
  and for 
 coincident antipodal points is given by 
\begin{equation}
  \lim_{\mathbf Y\to\sigma(\mathbf X)}  \Lambda(\mathbf X,\mathbf Y;\Delta_{\pm})\simeq
 (-1)^{\Delta}   \left(\frac{a}{2\pi}\right)^2 \frac{-zw}{(\vec x-\vec y)^2+(z+w)^2}.
  \end{equation}
Using the functional equation for the Gauss hypergeometric function
for $|k|\leq 1$
\begin{multline}
  k^{\Delta } \, _2F_1\left(\frac{\Delta }{2},\frac{\Delta +1}{2};\Delta
   -\frac{1}{2};k^2\right)= \frac{\Gamma (2-\Delta ) \Gamma (3-\Delta )}{2^{\Delta } \left(1+(-1)^{-2
       \Delta }\right) \Gamma (3-2 \Delta )}\cr
 \times\left({}_2F_1\left(3-\Delta ,\Delta ;2;\frac{k-1}{2 k}\right)+(-1)^{-\Delta } \,
   {}_2F_1\left(3-\Delta ,\Delta ;2;\frac{k+1}{2 k}\right)\right) , 
\end{multline}
we  obtain an equivalent expression for the normalized propagator~\eqref{eq:Lambdadef} 
	\begin{multline}
		\Lambda(\mathbf X,\mathbf Y;\Delta)=-\left(a\over2\pi\right)^2\frac{\pi(\Delta-1)(\Delta-2)
                              }{2\tan(\pi\Delta) (1+(-1)^{-2\Delta})}\cr\times\left((-1)^{-\Delta } \, _2F_1\left(3-\Delta ,\Delta ;2;\frac{K(\mathbf X,\mathbf Y)+1}{2 K(\mathbf X,\mathbf Y)}\right)+
   {}_2F_1\left(3-\Delta ,\Delta ;2;\frac{K(\mathbf X,\mathbf Y)-1}{2 K(\mathbf X,\mathbf Y)}\right)\right) \,.\label{eq:euclidean_propagator}
	\end{multline}
        The first term is singular for coincident points $\mathbf
        X=\mathbf Y$, i.e. $K(\mathbf X,\mathbf Y)=1$, and
        the second term is singular for coincident antipodal points $\mathbf
        X=\sigma(\mathbf Y)$, i.e. $K(\mathbf X,\mathbf Y)=-1$.

        Since $K(\mathbf X,\sigma(\mathbf Y))=-K(\mathbf X,\mathbf
        Y)$, we have the alternative representation for the bulk-to-bulk propagator
        	\begin{multline}
		\Lambda(\mathbf X,\mathbf Y;\Delta)=-\frac{\pi(\Delta-1)(\Delta-2)
                              2^{\Delta}}{\tan(\pi\Delta)}\cr\times\frac12\left((-1)^{-\Delta } \, _2F_1\left(3-\Delta ,\Delta ;2;\frac{K(\mathbf X,\mathbf Y)+1}{2 K(\mathbf X,\mathbf Y)}\right)+
  {}_2F_1\left(3-\Delta ,\Delta ;2;\frac{K(\mathbf X,\sigma(\mathbf Y))+1}{2 K(\mathbf X,\sigma(\mathbf Y))}\right)\right).
	\end{multline}
      Since the action of the antipodal map on a field of dimension
        $\Delta$ is $(-1)^\Delta$, we conclude  that the Dirichlet and
        Neumann propagators are obtained by the method of images under
        the action of the antipodal map.

The bulk-to-boundary propagator, which determines the evolution of a field on the boundary into the bulk, is therefore given by taking the Dirichlet or Neumann Green function and pulling one point to the boundary:
	\begin{align}
	\lim\limits_{z\to0}z^{-\Delta}\Lambda(\mathbf X,\mathbf Y;\Delta)=
		\frac{ a^2 \Gamma \left(\frac{\Delta}{2}\right) \Gamma
			\left(\frac{\Delta+1}{2}\right) }{(2\pi)^2 \Gamma \left(\Delta-\frac{1}{2}\right)}
		\frac{(2w)^{\Delta}}{((\vec{x}-\vec{y})^2+w^2)^{\Delta}}\,.
	\end{align}

	%%%%%%%%%%%%%%%%%%%%%%%%%%%%%%%%%%%%%%%%%%%%%%%%%%%%%%%%%%%%%%%%%%%%%%%%%%%%%%%%%%%%%%%%%%%%%%%%
	\subsection{Mapping to flat space propagators}
	\label{subsec:mappropagators}
	%%%%%%%%%%%%%%%%%%%%%%%%%%%%%%%%%%%%%%%%%%%%%%%%%%%%%%%%%%%%%%%%%%%%%%%%%%%%%%%%%%%%%%%%%%%%%%%%
	In this section, we make the 
        relationship between EAdS bulk-to-bulk propagators and flat space
        propagators explicit. We will use this relationship for the analytic evaluation of the
        Witten diagrams at loop orders.

        \subsubsection{The cases \texorpdfstring{$\Delta=1,2$}{Lg}}
        
   For $\Delta=1,2$, the propagator~\eqref{eq:Lambdadef} reads
	\begin{equation}
		\label{eq:LambdaSpecialDef}
		\Lambda(\mathbf X,\mathbf Y;\Delta)=\left(\frac{a}{2\pi}\right)^2
		\frac{K(\mathbf X,\mathbf Y)^\Delta}{1-K(\mathbf
                  X,\mathbf Y)^2}\,.
	\end{equation}
For this we first introduce the Euclidean norm 
	\begin{equation}
		\label{eq:normE}
		\norm{X}^2:= \vec x^2+z^2\,,
	\end{equation}
	as well as 
	\begin{equation}
		\label{eq:Gdef}
		G(X,Y):=\frac{zw}{\norm{X-Y}^2}=-\frac14\,
        {}_2F_1\left(1,2;2;\frac{K(\mathbf X,\mathbf Y)+1}{2K(\mathbf X,\mathbf Y)}\right)\,.
	\end{equation}
        We call $G(X,Y)$ \emph{the conformal flat space
          propagator}\footnote{A Feynman
          $i\varepsilon$ prescription will be introduced in the
          unitarity cuts section~\ref{sec:unitarity-cuts}.} due to the following transformation properties:
\begin{itemize}
	\item Invariance under translation of boundary points, $X_0=(\vec x,0)$:
	\begin{equation}
		\label{e:Gtranslation}
		G(X+X_0,Y+X_0)=G(X,Y);   \qquad G(X+X_0,Y)=G(X,Y-X_0).
	\end{equation}
	\item Scale invariance:
	\begin{equation}
		\label{e:Gscaling}
		G(\lambda X,\lambda Y)=G(X,Y)\;, \quad\lambda\in \mathbb{R}-\{0\}\,.
	\end{equation}
	\item Invariance under the inversion: 
	\begin{equation}
		\label{e:Ginversion}
		G\left(\frac{X'}{\norm{X'}^2},\frac{Y'}{\norm{Y'}^2}\right)=G(X,Y).
	\end{equation}
	\item 
	The antipodal map in~\eqref{eq:sigmadef} acts as
	\begin{equation}
		G(\sigma(X),Y)=G(X,\sigma(Y))= -\frac{zw}{\norm{X-\sigma(Y)}^2}=\frac14\,
        {}_2F_1\left(1,2;2;\frac{K(\mathbf X,\mathbf Y)-1}{2K(\mathbf X,\mathbf Y)}\right)\,.
	\end{equation}
	\item 
	An identity will be useful when simplifying the expressions for the
	multi-loop Witten diagrams
	\begin{equation}\label{eq:Gidentity}
		G(X,Y) G(X,\sigma(Y))= \frac14\left(G(X,Y)+G(X,\sigma(Y))\right)  .
	\end{equation}
\end{itemize}
To continue we note that the hyperbolic ``angle''
in~\eqref{eq:Kdef} can be expressed in terms of the conformal flat
space propagator
	\begin{equation}\label{e:KtoG}
		\frac{1}{K(\mathbf X,\mathbf Y)}=\frac14\left(\frac{1}{G(
			X, Y)}-\frac{1}{G( X,\sigma(Y))}\right).
	\end{equation}
For  $\Delta=1$ and $\Delta=2$ the bulk-to-bulk propagator ~\eqref{eq:LambdaSpecialDef} is then  expressed in terms of the conformal flat space propagator as 
	\begin{align}
		\label{eq:LambdaSpecial}
		\Lambda(\mathbf X,\mathbf Y;1)&=-\left(\frac{a}{2\pi}\right)^2
		\left(G(X,Y)-G(X,\sigma(Y))\right),\cr
		\Lambda(\mathbf X,\mathbf Y;2)&=-\left(\frac{a}{2\pi}\right)^2
		\left(G(X,Y)+G(X,\sigma(Y)\right)\,.
	\end{align}
	Moreover, by sending $X=(\vec x,z)$ to  the boundary point $(\vec x,0)$ we have 
	\begin{equation}
		\label{eq:Kbdry}
		\bar K(\vec{x},Y):= \lim_{z\to 0} \frac{K(\mathbf X,\mathbf Y)}{z}=\frac{2w}{\|\vec{x}-Y|^2}=\lim_{z\to0} \frac{2G(X,Y)}{z},
	\end{equation}
	 in terms of the conformal flat space propagator which is again odd under the action of the antipodal map $\sigma$. 

         % -----------------------------------------------------------------------------------------
	\subsubsection{The cases  \texorpdfstring{$\Delta\geq3$}{Lg}}
	\label{sec:bulkpropagator}
	%%%%%%%%%%%%%%%%%%%%%%%%%%%%%%%%%%%%%%%%%%%%%%%%%%%%%%%%%%
	
	For $\Delta\geq3$ integers the bulk-to-bulk propagators take a
        more complicated form. We have  for  $\Delta=2n+1\geq3$
	\begin{multline}
		\Lambda(\mathbf X,\mathbf Y;2n+1)=  \Lambda(\mathbf X,\mathbf Y;1) +\left(\frac{a}{2\pi}\right)^2
		\frac{1}{K(\mathbf X,\mathbf Y)} P_1^{(n-2)}\left(\frac{1}{K(\mathbf X,\mathbf Y)^2}\right)
		\cr+\left(\frac{a}{2\pi}\right)^2 Q_1^{(n-1)}\left(\frac{1}{K(\mathbf X,\mathbf Y)^2}\right) \log\left(
		\frac{-G(X,\sigma(Y))}{G(X,Y)}\right)  ,
            \end{multline}
            and for $\Delta=2n\geq4$
            	\begin{multline}
		\Lambda(\mathbf X,\mathbf Y;2n)= \Lambda(\mathbf X,\mathbf Y;2) +\left(\frac{a}{2\pi}\right)^2
		P_2^{(n-2)}\left(\frac{1}{K(\mathbf X,\mathbf Y)^2}\right)
		+\cr
		+\left(\frac{a}{2\pi}\right)^2 \frac{1}{K(\mathbf X,\mathbf Y)} Q_2^{(n-2)}\left(\frac{1}{K(\mathbf X,\mathbf Y)^2}\right)\log\left(
		\frac{-G(X,\sigma(Y))}{G(X,Y)}\right)  ,
	\end{multline}
	where $ P_i^{(r)}(x)$ and $Q_i^{(r)}(x)$   are polynomial of
        degree $r$ in $x$. Using the relation in~\eqref{e:KtoG} these propagators can be written as a
        combination of the conformal flat space propagators $G(X,Y)$
        and $G(X,\sigma(Y))$.

       The short distance singularities for coincident bulk points or
       antipodal points is the same as for $\Delta=1,2$ but the
       general structure differs due to the presence of logarithms of the
       conformal flat space propagator.
Using that $x^\eta= 1+\eta \log(x)+O(\eta^2)$ we can consider the
$\eta$-deformed propagators by making the replacement 
\begin{equation}
  \log  \left(
		\frac{-G(X,\sigma(Y))}{G(X,Y)}\right)\to \left(
		\frac{-G(X,\sigma(Y))}{G(X,Y)}\right)^\eta  ,
            \end{equation}
   in the above expressions for $	\Lambda(\mathbf X,\mathbf Y;2n+1)$ and $	\Lambda(\mathbf X,\mathbf Y;2n)$.

       In this representation we end up with expressions for the
       Witten diagrams in terms of flat space like QFT Feynman integrals
       with generalised powers of the propagators 
       \begin{equation}
            G(X,Y)^\eta=\left(\frac{zw}{\|X-Y|^2}\right)^\eta, \qquad    G(X,\sigma(Y))^\eta=\left(\frac{-zw}{\|X-\sigma(Y)|^2}\right)^\eta\,,
       \end{equation}
       which can be treated, using familiar analytic regularisation methods~\cite{Smirnov:2004ym}. The parameter $\eta$ will introduce some generalized powers of the
       propagators in addition to the one generated by the breaking of the conformal invariance due to the dimensional
       regularisation, as shown in section~\ref{subsubsec:conformal_mappings}.

	%%%%%%%%%%%%%%%%%%%%%%%%%%%%%%%%%%%%%%%%%%%%%%%%%%%%%%%%%%%%%%%%%%%%%%%%%%%%%%%%%%%%%%%%%%%%%%%%
	\section{Perturbative QFT in AdS}
	\label{sec:QFT_in_AdS}
	%%%%%%%%%%%%%%%%%%%%%%%%%%%%%%%%%%%%%%%%%%%%%%%%%%%%%%%%%%%%%%%%%%%%%%%%%%%%%%%%%%%%%%%%%%%%%%%%
	Let us begin by reviewing some points about the perturbative quantization of a conformally coupled scalar field on the Poincar\'e patch of AdS. Since the Poincar\'e patch is conformal  to the upper-half space, which in turn is obtained from $\mathbb{R}^4$ by the antipodal map described in the last section, we may start with the propagator  
		\begin{equation}
		\label{eq:conformal_vacuum}
		\Lambda_F(X,X')=\Omega(X)^{-1}\Omega(X')^{-1}G_F(X,X'),
	\end{equation}
	where $G_F(X,X')$ is the Feynman Green function in flat space and $\Omega(X)$ is the scale factor relating AdS to flat space such that $g_{\mu\nu}^{AdS}=\Omega^2\eta_{\mu\nu}$. Equation \eqref{eq:conformal_vacuum} defines the conformal vacuum \cite{Birrell:1982ix}. The Wick rotated version of ~\eqref{eq:conformal_vacuum} then reproduces the euclidean propagator ~\eqref{eq:Gdef} and the restriction to the upper-half space with the help of the antipodal map returns ~\eqref{eq:LambdaSpecial} for Neumann and Dirichlet boundary conditions respectively.  
	
	Now that we specified the vacuum we can calculate correlation functions in the same way as in flat space by performing an analytic continuation $t\to i x_4$ such that we are in EAdS and differentiate the perturbative expansion of the partition function with respect to some external current
	\begin{equation}
		\langle\phi(X_1)\phi(X_2)\cdots\phi(X_n)\rangle=\frac{\delta^n}{\delta j(X_1)\delta j(X_2)\cdots\delta j(X_n)}Z[j]\vert_{j=0}.
	\end{equation}
In this paper we will compute two and four-point functions on the Poincar\'e patch of EAdS in a loop expansion and, moreover, map this calculation to an equivalent calculation in flat space. The bulk amplitudes on AdS, evaluated on the conformal boundary, define correlation functions of primary fields $\mathcal{O}$ in some conformal field theory~\cite{Bertan:2018afl,Bertan:2018khc} whose operator content and OPE coefficients can be extracted with the help of the conformal block expansion~\cite{Dolan:2000ut}. The dimension of $\mathcal{O}$ is determined by the propagator with a fixed boundary condition $\Delta$ and the boundary correlation function is obtained by taking the limit that moves the external bulk points to the boundary while rescaling by a factor of $z_i^{-\Delta}$ for each boundary point. This is the basis of the correspondence~\cite{Maldacena:1997re, Gubser:1998bc,Witten:1998qj} between conformal field theory and field theory in AdS.

	The perturbative expansion of the correlation function is
        given by the well-known Witten diagrams \cite{Witten:1998qj}. They have the
        following graphical
        representation. Each boundary point
        lies on the outer circle and the bulk points are located
        inside the circle.
        The lines connecting bulk and boundary points represent bulk-to-boundary propagators and lines connecting two bulk points represent a bulk-to-bulk propagator. The vertices can be read off the Lagrangian and the symmetry factors can be obtained in the same way as for the corresponding Feynman diagrams. We will elaborate on this in section~\ref{subsec:definition_witten_diagrams}. Concretely, in this paper we consider a scalar field theory defined by the action
	\begin{equation}
		S=\int_{\mathrm{AdS}_4} \sqrt{g} \left( \frac12 (\partial\phi)^2 +\frac{m^2}{2} \phi^2+\frac{\lambda}{4!} \phi^4\right)\,,
	\end{equation}
	which for $m^2=-2a^2$ describes a conformally coupled scalar
        in AdS. The boundary two point function for an operator
        $\Op_{\Delta}$ has the perturbative expansion  
	\begin{align}
		\label{eq:conformal_2point_funct}
		&\langle\Op_{\Delta}(x_1)\Op_{\Delta}(x_2)\rangle=:\begin{tikzpicture}[baseline=(x1)]
			\begin{feynman}
				\tikzfeynmanset{every vertex=dot}
				\vertex [label=180:$x_1$] (x1);
				\vertex [right=1.5cm of x1, label=0:$x_2$] (x2);
				\diagram* {
					(x1) -- (x2),
					(x1) --[out=85, in=95, min distance=0.9cm,color=blue] (x2) --[out=265, in=275, min distance=0.9cm,color=blue] (x1),
				};
			\end{feynman}
                      \end{tikzpicture}\cr
           &=
		\begin{tikzpicture}[baseline=(x1)]
			\begin{feynman}
				\tikzfeynmanset{every vertex=dot}
				\vertex [label=180:$x_1$] (x1);
				\vertex [right=1.5cm of x1, label=0:$x_2$] (x2);
				\diagram* {
					(x1) --[scalar] (x2),
					(x1) --[out=85, in=95, min distance=0.9cm,color=blue] (x2) --[out=265, in=275, min distance=0.9cm,color=blue] (x1),
				};
			\end{feynman}
		\end{tikzpicture}
		-\frac{\lambda}{2}	\begin{tikzpicture}[baseline=(x)]
			\begin{feynman}
				\tikzfeynmanset{every vertex=dot}
				\vertex [label=180:$x_1$] (x1);
				\vertex [right=0.75cm of x1] (x) ;
				\vertex [right=0.75cm of x, label=0:$x_2$] (x2);
				\tikzfeynmanset{every vertex={empty dot,minimum size=0mm}}
				\vertex [above=0.4cm of x] (y);
				\diagram* {
					(x1) --[scalar] (x) --[scalar] (x2),
					(x) --[scalar,out=135, in=180, min distance=0.1cm] 
					(y) --[scalar,out=0, in=45, min distance=0.1cm](x),
					(x1) --[out=85, in=95, min distance=0.9cm,color=blue] (x2) --[out=265, in=275, min distance=0.9cm,color=blue] (x1),
				};
			\end{feynman}
		\end{tikzpicture}\cr
		&+\frac{\lambda^2}{4}\begin{tikzpicture}[baseline=(x)]
			\begin{feynman}
				\tikzfeynmanset{every vertex=dot}
				\vertex [label=180:$x_1$] (x1);
				\vertex [right=1cm of x1] (x) ;
				\vertex [right=1cm of x, label=0:$x_2$] (x2);
				\vertex [above=0.4cm of x] (y);
				\tikzfeynmanset{every vertex={empty dot,minimum size=0mm}}
				\vertex [above=0.4cm of y] (z);
				\diagram* {
					(x1) --[scalar] (x) --[scalar] (x2),
					(x) --[scalar,out=135, in=225, min distance=0.1cm] (y)  --[scalar,out=315, in=45, min distance=0.1cm] (x),
					(y) --[scalar,out=135, in=180, min distance=0.1cm] (z) --[scalar,out=0,in=45, min distance=0.1cm] (y),
					(x1) --[out=85, in=95, min distance=1.2cm,color=blue] (x2) --[out=265, in=275, min distance=1.2cm,color=blue] (x1),
				};
			\end{feynman}
		\end{tikzpicture}+\frac{\lambda^2}{4}\begin{tikzpicture}[baseline=(x)]
			\begin{feynman}
				\tikzfeynmanset{every vertex=dot}
				\vertex [label=180:$x_1$] (x1);
				\vertex [right=0.75cm of x1] (x) ;
				\vertex [right=0.5cm of x] (y);
				\vertex [right=0.75cm of y, label=0:$x_2$] (x2);
				\tikzfeynmanset{every vertex={empty dot,minimum size=0mm}}
				\vertex [above=0.4cm of x] (z);
				\vertex [above=0.4cm of y] (w);
				\diagram* {
					(x1) --[scalar] (x) -- [scalar] (y) -- [scalar] (x2),
					(x) --[scalar, out=135, in=180, min distance=0.1cm]
					(z) --[scalar, out=0, in=45, min distance=0.1cm](x),
					(y) --[scalar,out=135, in=180, min distance=0.1cm] (w) --[scalar,out=0, in=45, min distance=0.1cm](y),
					(x1) --[out=85, in=95, min distance=1.2cm,color=blue] (x2) --[out=265, in=275, min distance=1.2cm,color=blue] (x1),
				};
			\end{feynman}
		\end{tikzpicture}+\frac{\lambda^2}{6}\begin{tikzpicture}[baseline=(x)]
			\begin{feynman}
				\tikzfeynmanset{every vertex=dot}
				\vertex [label=180:$x_1$] (x1);
				\vertex [right=0.75cm of x1] (x);
				\vertex [right=0.5cm of x] (y);
				\vertex [right=0.75cm of y, label=0:$x_2$] (x2);
				\diagram* {
					(x1) --[scalar]  (x) -- [scalar]  (y) -- [scalar] (x2),
					(x) -- [scalar,out=85, in=95, min distance=0.3cm] (y) -- [scalar,out=265, in=275, min distance=0.3cm] (x),
					(x1) --[out=85, in=95, min distance=1.2cm,color=blue] (x2) --[out=265, in=275, min distance=1.2cm,color=blue] (x1),
				};
			\end{feynman}
		\end{tikzpicture}\cr
		&	-\frac{\lambda^3}{8}\begin{tikzpicture}[baseline=(x)]
			\begin{feynman}
				\tikzfeynmanset{every vertex=dot}
				\vertex [label=180:$x_1$] (x1);
				\vertex [right=0.5cm of x1] (x) ;
				\vertex [right=0.5cm of x] (y);
				\vertex [right=0.5cm of y] (y2);
				\vertex [right=0.5cm of y2, label=0:$x_2$] (x2);
				\tikzfeynmanset{every vertex={empty dot,minimum size=0mm}}
				\vertex [above=0.4cm of x] (z);
				\vertex [above=0.4cm of y] (w);
				\vertex [above=0.4cm of y2] (w2);
				\diagram* {
					(x1) --[scalar] (x) -- [scalar] (y) -- [scalar] (x2),
					(x) --[scalar,out=135, in=180, min distance=0.1cm] (z) --[scalar,out=0, in=45, min distance=0.1cm](x),
					(y) --[scalar,out=135, in=180, min distance=0.1cm] (w) --[scalar,out=0, in=45, min distance=0.1cm](y),
					(y2) --[scalar,out=135, in=180, min distance=0.1cm] (w2) --[scalar,out=0, in=45, min distance=0.1cm](y2),
					(x1) --[out=85, in=95, min distance=1.2cm,color=blue] (x2) --[out=265, in=275, min distance=1.2cm,color=blue] (x1),
				};
			\end{feynman}
		\end{tikzpicture}-\frac{\lambda^3}{8}\begin{tikzpicture}[baseline=(x)]
			\begin{feynman}
				\tikzfeynmanset{every vertex=dot}
				\vertex [label=180:$x_1$] (x1);
				\vertex [right=0.75cm of x1] (x) ;
				\vertex [right=0.5cm of x] (y);
				\vertex [right=0.75cm of y, label=0:$x_2$] (x2);
				\vertex [above=0.4cm of x] (z);
				\tikzfeynmanset{every vertex={empty dot,minimum size=0mm}}
				\vertex [above=0.4cm of y] (w);
				\vertex [above=0.4cm of z] (z2);
				\diagram* {
					(x1) -- [scalar] (x) --  [scalar]  (y) --[scalar] (x2),
					(x) --[scalar,out=135, in=225, min distance=0.1cm] (z)  --[scalar,out=315, in=45, min distance=0.1cm] (x),
					(z) --[scalar,out=135, in=180, min distance=0.1cm] (z2) --[scalar,out=0, in=45, min distance=0.1cm](z),
					(y) --[scalar,out=135, in=180, min distance=0.1cm] (w) --[scalar,out=0, in=45, min distance=0.1cm](y),
					(x1) --[out=85, in=95, min distance=1.2cm,color=blue] (x2) --[out=265, in=275, min distance=1.2cm,color=blue] (x1),
				};
			\end{feynman}
		\end{tikzpicture}-\frac{\lambda^3}{12}\begin{tikzpicture}[baseline=(x)]
			\begin{feynman}
				\tikzfeynmanset{every vertex=dot}
				\vertex [label=180:$x_1$] (x1);
				\vertex [right=0.5cm of x1] (x);
				\vertex [right=0.5cm of x] (y);
				\vertex [right=0.5cm of y] (y2);
				\vertex [right=0.5cm of y2, label=0:$x_2$] (x2);
				\tikzfeynmanset{every vertex={empty dot,minimum size=0mm}}
				\vertex [above=0.4cm of y2] (w2);
				\diagram* {
					(x1) --[scalar] (x) --[scalar] (y) --[scalar] (y2)--[scalar] (x2),
					(x) -- [scalar,out=85, in=95, min distance=0.3cm] (y) -- [scalar,out=265, in=275, min distance=0.3cm] (x),
					(x1) --[out=85, in=95, min distance=1.2cm,color=blue] (x2) --[out=265, in=275, min distance=1.2cm,color=blue] (x1),
					(y2) --[scalar,out=135, in=180, min distance=0.1cm] (w2) --[scalar,out=0, in=45, min distance=0.1cm](y2),
				};
			\end{feynman}
		\end{tikzpicture}\cr
		&-\frac{\lambda^3}{12}\begin{tikzpicture}[baseline=(x)]
			\begin{feynman}
				\tikzfeynmanset{every vertex=dot}
				\vertex (x);
				\vertex[right=1cm of x, label=0:$x_2$](x2);
				\vertex[left=1cm of x, label=180:$x_1$](x1);
				\vertex[above left=0.5cm and 0.2cm of x] (y1);
				\vertex[above right=0.5cm and 0.2cm of x] (y2);
				\tikzfeynmanset{every vertex={empty dot,minimum size=0mm}}
				\vertex[above=0.5cm of  x] (z);
				\diagram* {
					(x1)--[scalar]  (x)--[scalar] (x2),
					(x)--[scalar,out=120, in=270](y1),
					(x)--[scalar,out=60, in=270](y2),
					(y1)--[scalar](y2),
				};
			\end{feynman}
			\draw[blue] (x) circle (1cm);
			\draw [dash pattern=on 2pt off 2pt] (z) circle [radius=0.2cm];
		\end{tikzpicture}+\mathcal O(\lambda^4).
	\end{align}
	
The renormalised propagator is represented by a solid line and the
bare propagator by a dash line.

	It is obvious that the loop corrections produce short distance divergences at
	colliding bulk points and colliding antipodal bulk points~\cite{Akhmedov:2020jsi,Akhmedov:2018lkp} which will have to be regulated. We will show two
	different regularisation schemes and compare them in sections~\ref{subsubsec:covaraint_reg} and~\ref{subsubsec:dimreg}.
	
	In~\cite{Bertan:2018afl,Bertan:2018khc} it was shown that the loop corrections to the two point function considered in~\eqref{eq:conformal_2point_funct} are all proportional to the mass shift. The mass is usually fixed to the physically relevant mass measured in an experiment. In our case the only physically meaningful quantity related to the mass of the field is the scaling dimension of the operator on the boundary. Therefore, by fixing the scaling dimension on the boundary to be $\Delta$, we automatically renormalise the mass and can ignore loop corrections to the two point function in the following calculations

	The contributions to the four-point function to order $\lambda^3$ without tadpoles are:
	\begin{align}
	\label{eq:four_point_Wittendiagrams}
		&\langle\Op_{\Delta}(x_1)\Op_{\Delta}(x_2)\Op_{\Delta}(x_3)\Op_{\Delta}(x_4)\rangle=\left(\begin{tikzpicture}[baseline=(z)]
			\begin{feynman}[inline=(z)]
				\vertex (z);
				\tikzfeynmanset{every vertex=dot}	
				\vertex [above left=0.63cm and 0.71cm of z, label=180:$x_2$] (x2);
				\vertex [below left=0.63cm and 0.71cm of z, label=180:$x_1$] (x1);
				\vertex [above right=0.63cm and 0.71cm of z, label=0:$x_3$] (x3);
				\vertex [below right=0.63cm and 0.71cm of z, label=0:$x_4$] (x4);
				\tikzfeynmanset{every vertex={empty dot,minimum size=0mm}}
				\diagram* {
					(x2)--(x3),
					(x1)--(x4),
				};
			\end{feynman}
			\begin{pgfonlayer}{bg}
				\draw[blue] (z) circle (1cm);
			\end{pgfonlayer}
		\end{tikzpicture}+(x_2\leftrightarrow x_3)+(x_2\leftrightarrow
		x_4)\right)\cr
			&\qquad\quad\;\;-\lambda\begin{tikzpicture}[baseline=(z)]
			\begin{feynman}[inline=(z)]
				\tikzfeynmanset{every vertex=dot}
				\vertex (z);	
				\vertex [above left=0.71cm and 0.71cm of z, label=180:$x_2$] (x2);
				\vertex [below left=0.71cm and 0.71cm of z, label=180:$x_1$] (x1);
				\vertex [above right=0.71cm and 0.71cm of z, label=0:$x_3$] (x3);
				\vertex [below right=0.71cm and 0.71cm of z, label=0:$x_4$] (x4);
				\tikzfeynmanset{every vertex={empty dot,minimum size=0mm}}
				\diagram* {
					(x2)--(z),
					(x1)--(z),
					(x3)--(z),
					(x4)--(z),
				};
			\end{feynman}
			\begin{pgfonlayer}{bg}
				\draw[blue] (z) circle (1cm);
			\end{pgfonlayer}
		\end{tikzpicture}+\frac{\lambda^2}{2}\left(\begin{tikzpicture}[baseline=(z)]
			\begin{feynman}[inline=(z)]
				\vertex (z);	
				\tikzfeynmanset{every vertex=dot}
				\vertex [above left=0.71cm and 0.71cm of z, label=180:$x_2$] (x2);
				\vertex [below left=0.71cm and 0.71cm of z, label=180:$x_1$] (x1);
				\vertex [above right=0.71cm and 0.71cm of z, label=0:$x_3$] (x3);
				\vertex [below right=0.71cm and 0.71cm of z, label=0:$x_4$] (x4);
				\vertex [left=0.3cm of z] (y1);
				\vertex [right=0.3cm of z] (y2);
				\tikzfeynmanset{every vertex={empty dot,minimum size=0mm}}
				\diagram* {
					(x2)--(y1),
					(x1)--(y1),
					(x3)--(y2),
					(x4)--(y2),
				};
			\end{feynman}
			\begin{pgfonlayer}{bg}
				\draw[blue] (z) circle (1.05cm);
				\draw (z) circle (0.34cm);
			\end{pgfonlayer}
		\end{tikzpicture}+(x_2\leftrightarrow x_3)+(x_2\leftrightarrow
		x_4)\right)\cr
		&-\frac{\lambda^3}{4}\left(\begin{tikzpicture}[baseline=(z)]
			\begin{feynman}[inline=(z)]	
				\tikzfeynmanset{every vertex=dot}
				\vertex (z);
				\vertex [above left=0.71cm and 0.71cm of z, label=180:$x_2$] (x2);
				\vertex [below left=0.71cm and 0.71cm of z, label=180:$x_1$] (x1);
				\vertex [above right=0.71cm and 0.71cm of z, label=0:$x_3$] (x3);
				\vertex [below right=0.71cm and 0.71cm of z, label=0:$x_4$] (x4);
				\vertex [left=0.4cm of z] (y1);
				\vertex [right=0.4cm of z] (y2);
				\tikzfeynmanset{every vertex={empty dot,minimum size=0mm}}
				\vertex [left=0.2cm of z] (z1);
				\vertex [right=0.2cm of z] (z2);	
				\diagram* {
					(x2)--(y1),
					(x1)--(y1),
					(x3)--(y2),
					(x4)--(y2),
				};
			\end{feynman}
			\begin{pgfonlayer}{bg}
				\draw[blue] (z) circle (1cm);
				\draw (z1) circle (0.2cm);
				\draw (z2) circle (0.2cm);
			\end{pgfonlayer}
		\end{tikzpicture}+(x_2\leftrightarrow x_3)+(x_2\leftrightarrow
		x_4) \;+2\!\!\begin{tikzpicture}[baseline=(z)]
			\begin{feynman}[inline=(z)]
				\vertex (z);	
				\tikzfeynmanset{every vertex=dot}
				\vertex [above left=0.71cm and 0.71cm of z, label=180:$x_2$] (x2);
				\vertex [below left=0.71cm and 0.71cm of z, label=180:$x_1$] (x1);
				\vertex [above right=0.71cm and 0.71cm of z, label=0:$x_3$] (x3);
				\vertex [below right=0.71cm and 0.71cm of z, label=0:$x_4$] (x4);
				\vertex [left=0.4cm of z] (y1);
				\vertex [above right=0.17cm and 0.17cm of z] (y2);
				\vertex [below right=0.17cm and 0.17cm of z] (y3);
				\tikzfeynmanset{every vertex={empty dot,minimum size=0mm}}
				\vertex [left=0.2cm of z] (z1);
				\vertex [right=0.2cm of z] (z2);	
				\diagram* {
					(x2)--(y1),
					(x1)--(y1),
					(y1)--(y2),
					(y1)--(y3),
					(x3)--(y2),
					(x4)--(y3),
				};
			\end{feynman}
			\begin{pgfonlayer}{bg}
				\draw[blue] (z) circle (1.05cm);
				\draw (z2) circle (0.2cm);
			\end{pgfonlayer}
		\end{tikzpicture}  +\text{perm.}   \right)
                   ,
	\end{align}
	where the explicit forms of the permutations are given in section~\ref{sec:calculation_witten_diagram}.
	
%%%%%%%%%%%%%%%%%%%%%%%%%%%%%%%%%%%%%%%%%%%%%%%%%%%%%%%%%%%%%%%%%%%%%%%%%%%%%%%%%%%%%%%%%%%%%%%
	\subsection{Definition of Witten diagrams}
	\label{subsec:definition_witten_diagrams}
%%%%%%%%%%%%%%%%%%%%%%%%%%%%%%%%%%%%%%%%%%%%%%%%%%%%%%%%%%%%%%%%%%%%%%%%%%%%%%%%%%%%%%%%%%%%%%%

		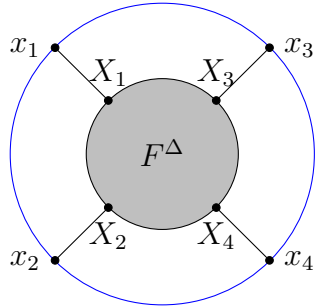
\begin{figure}[h]
		\begin{center}
			\begin{tikzpicture}[baseline=(x)]
				\begin{feynman}[inline=(x)]
					\tikzfeynmanset{every vertex=dot}
					\vertex [label=180:$x_1$] (x1);
					\tikzfeynmanset{every vertex={empty dot,minimum size=0mm}}
					\vertex [below right=1.41cm and 1.41cm of x1] (z);
					\vertex [below =0.3cm of z, label=90:$F^{\Delta}$] (z2);
					\tikzfeynmanset{every vertex=dot}
					\vertex [below left=1.41cm and 1.41cm of z, label=180:$x_2$] (x2);
					\vertex [above right=1.41cm and 1.41cm of z, label=0:$x_3$] (x3);
					\vertex [below right=1.41cm and 1.41cm of z, label=0:$x_4$] (x4);
					\vertex [above left=0.71cm and 0.71cm of z, label=90:$X_1$] (y1);
					\vertex [above right=0.71cm and 0.71cm of z, label=90:$X_3$] (y3);
					\vertex [below left=0.71cm and 0.71cm of z, label=270:$X_2$] (y2);
					\vertex [below right=0.71cm and 0.71cm of z, label=270:$X_4$] (y4);
					\tikzfeynmanset{every vertex={empty dot,minimum size=0mm}}
					\vertex [below right=0.6cm and 0.5cm of x1] (x);
					\diagram* {
						(x1)--(y1),
						(x2)--(y2),
						(x3)--(y3),
						(x4)--(y4),
					};
				\end{feynman}
				\begin{pgfonlayer}{bg}
					\draw[blue] (z) circle (2cm);
					\draw[fill=lightgray] (z) circle (1cm);
				\end{pgfonlayer}
			\end{tikzpicture}
			\caption{General four-point Witten diagram}
			\label{fig:Witten_general}
		\end{center}
	\end{figure}

A generic four-point Witten diagram $\Gamma_W$ depicted in
figure~\ref{fig:Witten_general} with $L+1$ bulk vertices and $L$
loops is associated to the following integral:
\begin{multline}
	\label{eq:Witten_diagram_definition}
	W^{\Delta}_L(\vec{x}_1,\vec{x}_2,\vec{x}_3,\vec{x}_4)=2^{4\Delta}\left(\mathcal{N}_{\Delta}\right)^{2L+4}\int\limits_{(\mathcal{H}_4^+)^{L+1}}\prod\limits_{i=1}^{L+1}
	\frac{\dd^4X_i}{(az_i)^4}
	F^{\Delta}(X_1,\dots,X_{L+1})\cr\times\sum\limits_{\rho\in\mathfrak
		S_4} \frac{\delta(\Gamma_W)}{|\Gamma_W|}f^{\Delta}(X_{\rho(1)},\dots,X_{\rho(4)};\vec{x}_{1},\dots,\vec{x}_{4})\,,
\end{multline}
	where the normalization $\mathcal N_\Delta$ of the propagators in~\eqref{e:Ndef} has been pulled out of the integral. The delta-function $\delta(\Gamma_W)$ denotes the identification of the bulk points according the topology of the graph and $|\Gamma_W|$ is the symmetry factor of the graph.
	
	The term $F^{\Delta}(X_1,...,X_{L+1})$ involves only bulk-to-bulk propagators. Its explicit form is determined by the loop order and topology of the concrete graph. Together with the integration measure it is invariant under AdS isometries.
		The term $f^{\Delta}(X_1,..,X_4;\vec{x}_1,...,\vec{x}_4)$ consists of bulk-to-boundary propagators and, depending on the loop order and the topology of the graph, some of the bulk points $X_i$ may be identical. The sum is performed over different scattering channels corresponding to permutations of the bulk points $X_1,...,X_4$. In its most general form this term is given by
	\begin{equation}
		\label{eq:Witten_diagram_bulk_to_boundary}
		f^{\Delta}(X_1,\dots,X_4;\vec{x}_1,\dots,\vec{x}_4)=\prod\limits_{i=1}^4\left(\frac{z_i}{\norm{X_i-\vec{x}_i}^2}\right)^{\Delta}\,.
	\end{equation}
	The integral in~\eqref{eq:Witten_diagram_definition} is divergent in general and thus needs to be regulated before it can be manipulated. In what follows we will consider two regularisations. The first, considered in~\cite{Bertan:2018afl,Bertan:2018khc}, preserves the AdS symmetry. The dimensional regularisation discussed next, while being natural from the flat space perspective, breaks AdS invariance.

%%%%%%%%%%%%%%%%%%%%%%%%%%%%%%%%%%%%%%%%%%%%%%%%%%%%%%%%%%%%%%%%%%%%%%%%%%%%%%%%%%%%%%%%%%%%%%%%%%%%%%
	\subsubsection{AdS invariant regularisation}
	\label{subsubsec:covaraint_reg}
%%%%%%%%%%%%%%%%%%%%%%%%%%%%%%%%%%%%%%%%%%%%%%%%%%%%%%%%%%%%%%%%%%%%%%%%%%%%%%%%%%%%%%%%%%%%%%%%%%%%%%

	An AdS invariant regularisation method, given by the deformation
	\begin{equation}\label{e:Kdelta}
		K^\delta(\mathbf X,\mathbf Y):=\frac{K(\mathbf X,\mathbf
			Y)}{1+\delta}, \qquad \textrm{with}~ \delta>0,
	\end{equation}
	was developed and used for regulating loops in AdS
	space in~\cite{Bertan:2018afl,Bertan:2018khc} and  applied to loops in de Sitter
	space in~\cite{Heckelbacher:2020nue}. This preserves the
        conformal symmetry on the boundary and we will use
        it in section~\ref{subsubsec:AdSinvariant_oneloop}.

        For $\Delta=1$ the regularised propagator reads
	\begin{equation}
		\label{eq:LambdaDelta1Reg}
		\Lambda(\mathbf X,\mathbf Y;1,\delta)=\left(\frac{a}{2\pi}\right)^2\frac12
		\Bigg(\frac{K(\mathbf X,\mathbf Y)}{1+\delta-K(\mathbf X,\mathbf Y)}
		+\frac{K(\mathbf X,\mathbf Y)}{1+\delta+K(\mathbf X,\mathbf Y)}\Bigg),
	\end{equation}
and for $\Delta=2$
	\begin{equation}
		\Lambda(\mathbf X,\mathbf Y;2,\delta)=\left(\frac{a}{2\pi}\right)^2\frac12
		\Bigg(\frac{K(\mathbf X,\mathbf Y)}{1+\delta-K(\mathbf X,\mathbf Y)}
		-\frac{K(\mathbf X,\mathbf Y)}{1+\delta+K(\mathbf X,\mathbf Y)}\Bigg),
              \end{equation}
with similar expressions for propagators with $\Delta\geq3$.
              
              We will denote the regularised Witten
	diagrams~\eqref{eq:Witten_diagram_definition} by 
	\begin{multline}\label{eq:Witten_diagram_delta}
		W^{\Delta,\delta}_L(\vec{x}_1,\vec{x}_2,\vec{x}_3,\vec{x}_4)=2^{4\Delta}\frac{\left(\mathcal{N}_{\Delta}\right)^{2L+4}}{a^{4L+4}}
		\int\limits_{(\mathcal{H}_4^+)^{L+1}}\prod\limits_{i=1}^{L+1}
		\frac{\dd^{4} X_i}{z_i^4}
		F^{\Delta,\delta}(X_1, \dots
		,X_{L+1})\cr
		\times\sum\limits_{\rho\in\mathfrak
			S_4} \frac{\delta(\Gamma_W)}{|\Gamma_W|}f^{\Delta}(X_{\rho(1)},\dots,X_{\rho(4)};\vec{x}_1,\vec{x}_2,\vec{x}_3,\vec{x}_4)\,,
                    \end{multline}
with  normalization $\mathcal N_\Delta$ given in~\eqref{e:Ndef}.

%%%%%%%%%%%%%%%%%%%%%%%%%%%%%%%%%%%%%%%%%%%%%%%%%%%
	\subsubsection{Dimensional regularisation}
	\label{subsubsec:dimreg}
%%%%%%%%%%%%%%%%%%%%%%%%%%%%%%%%%%%%%%%%%%%%%%%%%%%

	For $\Delta=1$ and $\Delta=2$ we have shown in
	section~\ref{subsec:mappropagators}, that the propagators
	in~\eqref{eq:LambdaSpecial} can be expressed as a sum of two euclidean propagators.
	Therefore the bulk-to-bulk part $F^{\Delta}(X_1,\dots,X_{L+1})$ ~\eqref{eq:Witten_diagram_definition} can always be expressed as a sum over products of flat space propagators.
	
	Let us now discuss the domain of integration, which for ~\eqref{eq:Witten_diagram_definition} is the
	upper-half space $\mathcal{H}_4^+$. In flat space momentum space on the other hand, one integrates over the  entire
	space $\mathbb{R}^4$. It is clear from
	the previous discussion in section~\ref{sec:propagators} that the
	propagator is in general not invariant under the antipodal map due to
	the $z^{\Delta}$ term in the numerator which changes the sign for odd
	$\Delta$. However, since we focus on the  $\lambda\phi^4$ theory, each vertex
	joins four propagators, meaning that each radial coordinate in the numerator
	only appears as $z^{4\Delta-4}$, where the $-4$ is due to the
	integration measure. For $\Delta\in\mathbb{N}$ this is always an
        even number and
	therefore invariant under the antipodal map. We thus  conclude
	that the entire Witten diagram~\eqref{eq:Witten_diagram_definition} is
	invariant under mapping every bulk point to its antipodal point and
	the domain of integration can be extended to $\mathbb{R}^4$. Note that this can also be done for the AdS-invariant regularisation method in equation~\eqref{eq:Witten_diagram_delta}.
	
	To continue we note that in~\eqref{eq:Witten_diagram_definition} powers of the radial coordinates, $z_i$ appear in the denominator, originating from the AdS-invariant measure as well as in the numerator of the propagators. It is convenient to ``covariantize'' these 
	contributions by writing them as  linear propagators  $z=u\cdot X_i$ with the help of 
	the auxiliary unit vector $u=(\vec{0},1)$, where the dot product is understood with respect to the euclidean metric. This auxiliary vector is orthogonal to the boundary and is therefore
	perpendicular to any vector $X_i=(\vec{x}_i,0)$ parametrizing points on
	the boundary. In particular, for $\Delta=1,2$ the propagators
	in~\eqref{eq:LambdaSpecial} take a tensorial from
	\begin{align}
		\label{eq:LambdaSpecial_iep2}
		\Lambda(\mathbf X,\mathbf Y;1)&=-\left(\frac{a}{2\pi}\right)^2
		\left(\frac{u\cdot X\, u\cdot Y}{\norm{X-Y}^2}+\frac{u\cdot X\, u\cdot \sigma(Y)}{\norm{X-\sigma(Y)}^2}\right),\cr
		\Lambda(\mathbf X,\mathbf Y;2)&=-\left(\frac{a}{2\pi}\right)^2
		\left(\frac{u\cdot X\, u\cdot Y}{\norm{X-Y}^2}-\frac{u\cdot X\, u\cdot \sigma(Y)}{\norm{X-\sigma(Y)}^2}\right),
	\end{align}
        with similar expression for the propagator with $\Delta\geq3$.
        
	We then define the dimensionally regulated Witten diagrams~\eqref{eq:Witten_diagram_definition} by evaluating the integration measure in $D$ dimensions,
	\begin{multline}\label{eq:Witten_diagram_dim_reg}
		W^{\Delta,D}_L(\vec{x}_1,\vec{x}_2,\vec{x}_3,\vec{x}_4)=2^{4\Delta}\frac{\left(\mathcal{N}_{\Delta}\right)^{2L+4}}{(2a^D)^{L+1}}
		\int\limits_{(\mathbb{R}^D)^{L+1}}\prod\limits_{i=1}^{L+1}
		\frac{\dd^{D} X_i}{(u\cdot X_i)^4}
		F^{\Delta}(X_1, \dots
		,X_{L+1})\cr
		\times\sum\limits_{\rho\in\mathfrak
			S_4} \frac{\delta(\Gamma_W)}{|\Gamma_W|}f^{\Delta}(X_{\rho(1)},\dots,X_{\rho(4)};\vec{x}_1,\vec{x}_2,\vec{x}_3,\vec{x}_4),
	\end{multline}
	where we have pulled out a factor of $(a^{-D})^{L+1}$ and
        rescaled every point with $a$ such that the only dimensional
        dependence is in the prefactor. Upon substitution of~\eqref{eq:Witten_diagram_bulk_to_boundary} and~\eqref{eq:LambdaSpecial_iep2} the Witten diagram~\eqref{eq:Witten_diagram_dim_reg} takes the form of  standard
        flat space tensorial integrals with linear propagators.

        \paragraph{Remark 1:}
        Note  that we have used a dimensional regularisation scheme by
        changing the dimension of integration without changing the
        measure factor from the AdS metric, which breaks the conformal
        invariance.
        An AdS preserving integration measure
        $
          \prod_{i=1}^{L+1} \frac{\dd^DX_i}{(u\cdot X_i)^D}
       $
        in~\eqref{eq:Witten_diagram_dim_reg} will not regulate the integral as a consequence of conformal symmetry.
        
        \paragraph{Remark 2:} When $D$ approaches 4 the Witten diagrams develop divergences with leading behaviour  $\frac{1}{(D-4)^L}$ at $L$-loop order.  In order to preserve the conformal symmetry, which is broken by the dimensional regularisation, we need to parametrize $D=4-\frac{4\epsilon}{L+1}$ at each loop order.  With $\epsilon<0$  since the only divergences come from coinciding bulk points.

%%%%%%%%%%%%%%%%%%%%%%%%%%%%%%%%%%%%%%%%%%%%%%%%%	 
	\subsubsection{Conformal mapping of the regularised integrals}
	\label{subsubsec:conformal_mappings}
%%%%%%%%%%%%%%%%%%%%%%%%%%%%%%%%%%%%%%%%%%%%%%%%%	 

	We will now use invariance of the diagram under translation of the boundary points and inversion to write the four-point diagram in terms of three-dimensional conformal cross-ratios. First we apply these transformations to the integrand. The non-invariance of the regularised measure will be taken into account in a second step.   
	
To begin with, we shift every boundary point by $\vec{x}_3$ and then invert every point. The latter leaves the bulk-to-bulk propagators invariant while the bulk-to-boundary propagators transform as 
\begin{align}\label{eq:TI1}
	\frac{z}{\norm{X-\vec{x}}^2}=\frac{1}{x^2}\frac{z'}{\norm{X'-\vec{y}}^2}\qquad\text{with }X'=\frac{X}{\norm{X}^2}\;\text{ and }\;
	\vec{y}=\frac{\vec{x}}{x^2}\,,
\end{align}
where we have set $\|\vec x|^2\equiv x^2$. After these transformations~\eqref{eq:Witten_diagram_bulk_to_boundary} becomes
\begin{multline}
	f^{\Delta}(X_1,\dots,X_4;\vec{x}_1, \dots ,\vec{x}_4)=\frac{z_3^{\Delta}}{(x^2_{13}x^2_{23}x^2_{34})^{\Delta}}\left(\frac{z_1}{\norm{X_1-y_{13}}^2}\right)^{\Delta}\cr\times\left(\frac{z_2}{\norm{X_2-y_{23}}^2}\right)^{\Delta}\left(\frac{z_4}{\norm{X_4-y_{43}}^2}\right)^{\Delta},
\end{multline}
where we have set $x_{ij}:=\vec{x}_i-\vec{x}_j$ and $y_{ij}:=x_{ij}/x_{ij}^2$. To continue we shift every bulk point as $X_i\to X_i+y_{13}$ and use scale invariance to rescale every bulk point by 
$X_i\to \norm{y_{43}-y_{13}}X_i$. This gives

\begin{multline}
	f^{\Delta}(X_1,\dots,X_4;\vec{x}_1, \dots
	,\vec{x}_4)=\frac{1}{(x^2_{14}x^2_{23})^{\Delta}}\cr\times\left(\frac{z_1 z_2 z_3 z_4}{\norm{X_1}^2\norm{X_2-\frac{y_{23}-y_{13}}{\norm{y_{43}-y_{13}}}}^2
	\norm{X_4-\frac{y_{43}-y_{13}}{\norm{y_{43}-y_{13}}}}^2}\right)^{\Delta}.
\end{multline}
Finally, we may use the fact that the AdS group acts on points of the conformal boundary
as the conformal group to implement the familiar conformal operations on the
boundary points that map $\vec{x}_4$ to infinity, $\vec{x}_3$ to the
origin $(0,0,0,0)$ and $\vec{x}_1\to(-1,0,0,0)$. The remaining point $\vec{x}_2$ can be
chosen to lie in the 1-4 plane, parametrized by the complex coordinate
$\zeta$, that is
\begin{align}
    \vec{x}_2=\left(\frac{\zeta+\bar\zeta -2}{2
	(1-\zeta)(1-\bar\zeta)}, \frac{\zeta-\bar\zeta}{2i
	(1-\zeta)(1-\bar\zeta)}, 0,0\right)\,.
\end{align}
This takes equation~\eqref{eq:Witten_diagram_bulk_to_boundary} to the final form
\begin{equation}
	\label{eq:Witten_bulk_to_boundary_tranformed}
	f^{\Delta}(X_1,\dots,X_4;\vec{x}_1, \dots
	,\vec{x}_4)=\frac{v^\Delta}{(x^2_{12}x^2_{34})^{\Delta}}\left(\frac{z_1 z_2 z_3 z_4}{\norm{X_1}^2\norm{X_2-\uZ}^2
		\norm{X_4-u_1}^2}\right)^{\Delta}\,,
\end{equation}
with
\begin{equation}
  \label{e:udef}
  u_1=(1,0,0,0), \qquad u_{\zeta}=\left(\frac{\zeta+\bar\zeta}{2},\frac{\zeta-\bar\zeta}{2i},0,0\right),\qquad v= \zeta\bar\zeta=\frac{x_{12}^2x_{34}^2}{x_{14}^2x_{23}^2}.
\end{equation}

Let us now turn to the measure. The dimensional regularisation implemented in~\eqref{eq:Witten_diagram_dim_reg} breaks the AdS invariance of the integration measure. We therefore have to take into account the Jacobian of the  transformations implemented above. Since the regularised measure is still invariant under shifts in the $z=$ const hyperplane, the first tranformation leaves the latter invariant. The second tranformation in ~\eqref{eq:TI1} is an inversion ($X_i\to\frac{X_i}{\norm{X_i}^2}$) which induces a Jacobian 
\begin{align}
	\prod\limits_{i=1}^{L+1}\frac{\dd^{D} X_i}{(u\cdot X_i)^4}\to\prod\limits_{i=1}^{L+1}\frac{\dd^{D} X_i}{(u\cdot X_i)^4}\frac{1}{\norm{X_i}^{2(D-4)}}\,.
\end{align}
This is followed by a shift of all bulk points by $y_{13}$, under which 
\begin{align}
	\prod\limits_{i=1}^{L+1}\frac{\dd^{D} X_i}{(u\cdot X_i)^4}\frac{1}{\norm{X_i}^{2(D-4)}}\to
	\prod\limits_{i=1}^{L+1}\frac{\dd^{D} X_i}{(u\cdot X_i)^4}\frac{1}{\norm{X_i+y_{13}}^{2(D-4)}}\,.
\end{align}
Finally, the rescaling by $\norm{y_{43}-y_{13}}$ gives 
\begin{align}
	\prod\limits_{i=1}^{L+1}\frac{\dd^{D} X_i}{(u\cdot X_i)^4}\frac{1}{\norm{X_i+y_{13}}^{2(D-4)}}\to
	\prod\limits_{i=1}^{L+1}\frac{\dd^{D} X_i}{(u\cdot X_i)^4}\frac{\norm{y_{43}-y_{13}}^{4-D}}{\norm{X_i+\frac{y_{13}}{\norm{y_{43}- y_{13}}}}^{2(D-4)}}\,.
\end{align}
Rewriting the inverted boundary points in terms of the original coordinates and choosing $\vec{x}_1, \vec{x}_2,\vec{x}_3$ and $\vec{x}_4$ as described above we get
\begin{align}
	&\norm{y_{43}-y_{13}}=\frac{\norm{x_{41}}}{\norm{x_{43}}\norm{x_{13}}}\to 1\quad\text{and}\quad \frac{y_{13}}{\norm{y_{43}-y_{13}}}\to\frac{x_{13}}{\norm{x_{13}}^2}=
	-u_1\,,
\end{align}
and therefore the complete Jacobian is given by
\begin{align}
	\label{eq:transformation_measure}
	\prod\limits_{i=1}^{L+1}\frac{\dd^{D} X_i}{(u\cdot X_i)^4}\frac{1}{\norm{X_i-u_1}^{2(D-4)}}\,.
\end{align}
From~\eqref{eq:Witten_bulk_to_boundary_tranformed}  and~\eqref{eq:transformation_measure} it is then clear that the Witten
  diagrams will depend only on $\zeta$ and $\bar\zeta$ or,
  equivalently, the conformal cross-ratios introduced in~\cite{Bertan:2018khc,Bertan:2018afl}
\begin{equation}
	\label{eq:crossratio}
	v=\frac{x_{12}^2x_{34}^2}{x_{14}^2x_{23}^2}= \zeta\bar
	\zeta; \qquad
	1-Y=\frac{x_{13}^2x_{24}^2}{x_{14}^2 x_{23}^2}=(1-\zeta)(1-\bar \zeta)\,.
\end{equation}
  To summarize
we have the $\delta$-regularised
Witten diagram (removing the prefactor $2^{4\Delta}\left(\mathcal{N}_{\Delta}\right)^{2L+4}/(a^4)^{L+1}$ with   $\mathcal N_\Delta$ given in~\eqref{e:Ndef})
	\begin{multline}
\label{eq:Witten_diagram_delta_trans}
	\cW_L^{\Delta,\delta}(\zeta,\zetab):=\frac{1}{2^{L+1}}\frac{v^\Delta}{(x^2_{12}x^2_{34})^\Delta}
	\int\limits_{(\mathbb{R}^4)^{L+1}}\prod\limits_{i=1}^{L+1}
	\frac{\dd^4 X_i}{z_i^4}
	F^{\Delta,\delta}(X_1, \dots
	,X_{L+1})\cr
	\times\sum\limits_{\rho\in\mathfrak
		S_4} \frac{\delta(\Gamma_W)}{|\Gamma_W|}\left(\frac{z_{\rho(1)}}{\|X_{\rho(1)}|^2}\right)^\Delta\left(\frac{z_{\rho(2)}}{\|X_{\rho(2)}-u_\zeta|^2}\right)^\Delta
	z^\Delta_{\rho(3)}\left(\frac{z_{\rho(4)}}{\|X_{\rho(4)}-u_1|^2}\right)^\Delta,
        \end{multline}
while in dimensional regularisation, taking into account the Jacobian just derived, we have instead
	\begin{multline}\label{eq:Witten_diagram_dimreg_trans}
	\cW^{\Delta,D}_L(\zeta,\bar{\zeta}):=\frac{1}{2^{L+1}}\frac{v^\Delta}{(x^2_{12}x^2_{34})^\Delta}
	\int\limits_{(\mathbb{R}^D)^{L+1}}\prod\limits_{i=1}^{L+1}
	\frac{\dd^D X_i}{(u\cdot X_i)^4}\frac{F^{\Delta}(X_1, \dots
		,X_{L+1})}{\norm{X_i-u_1}^{2(D-4)}}\cr
	\times\sum\limits_{\rho\in\mathfrak
		S_4} \frac{\delta(\Gamma_W)}{|\Gamma_W|}\left(\frac{z_{\rho(1)}}{\|X_{\rho(1)}|^2}\right)^\Delta\left(\frac{z_{\rho(2)}}{\|X_{\rho(2)}-u_\zeta|^2}\right)^\Delta
	z^\Delta_{\rho(3)}\left(\frac{z_{\rho(4)}}{\|X_{\rho(4)}-u_1|^2}\right)^\Delta\,.
\end{multline}

%%%%%%%%%%%%%%%%%%%%%%%%%%%%%%%%%%%%%%%%%%%%%%%%%%%%%%%%%%
\subsection{Differential operator relations }
	\label{subsec:diff_operator}
        %%%%%%%%%%%%%%%%%%%%%%%%%%%%%%%%%%%%%%%%%%%%%%%%%%%%%%%%%%

It is possible to obtain the amplitude for the Witten diagrams with
   external dimension $\Delta=2$ from those  with $\Delta=1$ by acting
   with a suitable differential operator on the external points. This
   turns out to be rather useful when working with the dimensional
   regularisation scheme.

   We use the unit vector $u=(0,0,0,1)$ perpendicular to the boundary introduced in section~\ref{subsubsec:dimreg} 
and define the $\tilde X_i=(\vec x_i,w_i)$ associated to the external
   legs which, in this section, we take to lie in the bulk.
   We introduce the operators
\begin{align}
	\label{eq:diff_operator}
	\cH_i&:=\left.u^{\mu}\frac{\partial}{\partial \tilde X_i^\mu}\right\vert_{w_i=0}\,,\qquad
	\cH_{ij}:=\left.u^{\mu}u^{\nu}\frac{\partial}{\partial\tilde X_i^\mu}\frac{\partial}{\partial\tilde  X_j^\nu}\right\vert_{w_i=w_j=0}\,,\cr
	\cH_{ijkl}&:=\left.u^{\mu_1}u^{\mu_2}u^{\mu_3}u^{\mu_4}\frac{\partial}{\partial \tilde X_i^{\mu_1}}\frac{\partial}{\partial \tilde X_j^{\mu_2}}\frac{\partial}{\partial \tilde X_k^{\mu_3}}\frac{\partial}{\partial \tilde X_l^{\mu_4}}\right\vert_{w_i=w_j=w_k= w_l=0}\,.
\end{align}
In order to define the action of these operators on Witten diagrams we
                    move the external legs into the bulk, while
                    keeping the form of the bulk-to-boundary
                    propagator. We consider the generalisation of~\eqref{eq:Witten_diagram_bulk_to_boundary}
\begin{equation}\label{eq:bB_red}
	f^{\Delta}(X_1,\dots,X_4;\tilde{X}_1,\dots,\tilde{X}_4)=\prod\limits_{i=1}^4\left(\frac{u\cdot
            X_i}{\norm{X_i-\tilde{X}_i}^2}\right)^{\Delta},
      \end{equation}
which is not a proper product of bulk-to-bulk propagators, but should rather  be understood as some generating function for bulk-to-boundary propagators obtained by moving the boundary points to a finite value of the radial coordinate. 
It is straightforward to check that the action of the differential operators ~\eqref{eq:diff_operator} on the redefined bulk-to-boundary propagator  ~\eqref{eq:bB_red} gives
\begin{equation}
	\cH_{1234}f^\Delta(X_1,\dots,X_4;\tilde{X}_1,\dots,\tilde{X}_4)=\prod\limits_{i=1}^4\cH_i\left(\frac{u\cdot X_i}{\norm{X_i-\tilde X_i}^2}\right)^{\Delta},
\end{equation}
   so that
   \begin{align}
	\cH_{1234}f^\Delta(X_1,\dots,X_4;\tilde{X}_1,\dots,\tilde{X}_4)&=
	(2\Delta)^4\prod\limits_{i=1}^4\left(\frac{u\cdot X_i}{\norm{X_i-\vec{x}_i}^2}\right)^{\Delta+1},\\
	&=(2\Delta)^4 f^{\Delta+1}(X_1,\dots,X_4;\vec x_1,\dots,\vec x_4)\,.\nonumber
\end{align}
In the preceding section we have shown that the four-point Witten diagrams with external points on the conformal boundary depend only on the cross-ratios~\eqref{eq:crossratio}. If the external points are moved into the bulk, as above, we have to reconsider the transformations leading to this, more precisely,~\eqref{eq:Witten_bulk_to_boundary_tranformed} and~\eqref{eq:transformation_measure}. Repeating the arguments in section~\ref{subsubsec:conformal_mappings} one can show that the integrals with external points in the bulk again depend only on the cross-ratios $v$ and $Y$ now expressed as
\begin{equation}
	\label{eq:crossratio_bulk}
	v=\frac{\norm{\tilde X_{12}}^2\norm{\tilde X_{34}}^2}{\norm{\tilde X_{14}}^2\norm{\tilde X_{23}}^2}\,; \qquad
	1-Y=\frac{\norm{\tilde X_{13}}^2\norm{\tilde X_{24}}^2}{\norm{\tilde X_{14}}^2\norm{\tilde X_{23}}^2}\,.
\end{equation}
Some of the operators in ~\eqref{eq:diff_operator} have simple expressions in terms of the conformal cross-ratios. In particular, 
\begin{align}
	x_{12}^2\cH_{12}=&x_{34}^2\cH_{34}=-2v\frac{\partial}{\partial v}\,,\cr
	x_{13}^2\cH_{13}=&x_{24}^2\cH_{24}=2\left(1-Y\right)\partial_Y\,,\cr
	x_{14}^2\cH_{14}=&x_{23}^2\cH_{23}=2v\partial_v-2(1-Y)\partial_Y\,,\cr
	\text{and}\qquad\qquad\qquad\qquad\quad&\cr
	(x_{12}x_{34})^2\cH_{1234}=&4v\left(v(1+v)\frac{\partial^2}{\partial v^2}+(1-Y)(2-Y)\frac{\partial^2}{\partial Y^2}-2v(1-Y)\frac{\partial^2}{\partial v\partial Y}\right.\cr
	&\qquad\left.+(1+v)\frac{\partial}{\partial v}-(2-Y)\frac{\partial}{\partial Y}\right)\,.\label{eq:H1234_crossratios}
\end{align}
We will use the differential operators $\cH_{14}$ and $\cH_{12}$ to obtain the finite part for $\Delta=2$ at one-loop in~\eqref{e:W2fin} from the simpler auxiliary integral~\eqref{eq:W1_fin_auxs}.

Acting with $\cH_{1234}$ as in~\eqref{eq:diff_operator} on $f^\Delta$ gives
\begin{multline}
    \cH_{1234} f^\Delta(X_1,\dots,X_4;\tilde
   X_1,\dots,\tilde
   X_4)=\frac{1}{(x_{12}^2x_{34}^2)^{\Delta+1}}\cr\times
   \left[4\Delta^2+2\Delta x_{12}^2\cH_{12}+2\Delta
     x_{34}^2\cH_{34}+(x^2_{12}x^2_{34})\cH_{1234}\right]\left(\frac{v\,
         u\cdot
       X_1 \cdots u\cdot X_4}{\norm{X_1}^2\norm{X_2-\uZ}^2
		\norm{X_4-u_1}^2}\right)^{\Delta}\,,
 \end{multline}
   Plugging in equations~\eqref{eq:H1234_crossratios} we obtain
\begin{multline}
	\cH_{1234}f^\Delta=\frac{4}{(x_{12}^2x_{34}^2)^{\Delta+1}}\left[v\left(v(1+v)\partial^2_v+(1-Y)(2-Y)\partial^2_Y-2v(1-Y)\partial_v\partial_Y\right.\right.\cr
\left.\left.+(1+v-2\Delta)\partial_v-(2-Y)\partial_Y\right)+\Delta^2\right]\left(\frac{v\,
         u\cdot
       X_1 \cdots u\cdot X_4}{\norm{X_1}^2\norm{X_2-\uZ}^2
		\norm{X_4-u_1}^2}\right)^{\Delta}\,.
\end{multline}
We will apply this differential operator for evaluating the diverging part for $\Delta=2$ at one-loop in~\eqref{e:W2div} from the $\Delta=1$ result as we will describe in section~\ref{subsubsec:dimreg_oneloop}.
        
%-------------------------------------------------------------------------%

		%%%%%%%%%%%%%%%%%%%%%%%%%%%%%%%%%%%%%%%%%%%%%%%%%%%%%%%%%%%%%%%%%%%%%%%%%%%%%%%%%%%%%%%%%%%%%%%%
	\section{Loop corrections to Witten diagrams}
	\label{sec:calculation_witten_diagram}
	%%%%%%%%%%%%%%%%%%%%%%%%%%%%%%%%%%%%%%%%%%%%%%%%%%%%%%%%%%%%%%%%%%%%%%%%%%%%%%%%%%%%%%%%%%%%%%%%
	
	We are now ready to calculate loop corrections to  Witten diagrams for a $\lambda\phi^4$ theory and make their dependence on conformal cross ratios explicit. Below we will use two different regularisation schemes to establish scheme independence of our results. 

	%%%%%%%%%%%%%%%%%%%%%%%%%%%%%%%%%%%%%%%%%%%%%%%%%%%%%%%%%%
	\subsection{The tree-level cross diagram}
	\label{subsec:cross_diagram}
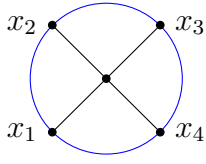
\begin{figure}[ht]
		\centering
		\begin{tikzpicture}[baseline=(z)]
			\begin{feynman}[inline=(z)]
				\tikzfeynmanset{every vertex=dot}
				\vertex (z);	
				\vertex [above left=0.71cm and 0.71cm of z, label=180:$x_2$] (x2);
				\vertex [below left=0.71cm and 0.71cm of z, label=180:$x_1$] (x1);
				\vertex [above right=0.71cm and 0.71cm of z, label=0:$x_3$] (x3);
				\vertex [below right=0.71cm and 0.71cm of z, label=0:$x_4$] (x4);
				\tikzfeynmanset{every vertex={empty dot,minimum size=0mm}}
				\diagram* {
					(x2)--(z),
					(x1)--(z),
					(x3)--(z),
					(x4)--(z),
				};
			\end{feynman}
			\begin{pgfonlayer}{bg}
				\draw[blue] (z) circle (1cm);
			\end{pgfonlayer}
		\end{tikzpicture}
		\caption{Cross diagram}
		\label{fig:cross}
	\end{figure}

	%%%%%%%%%%%%%%%%%%%%%%%%%%%%%%%%%%%%%%%%%%%%%%%%%%%%%%%%%%
	We start with the evaluation of the cross diagram for general integer dimensions $\Delta\geq 1$.\footnote{The cross diagram is referred to as the $D$-function in~\cite{Dolan:2000ut}. We will
		not use this notation, reserving the name of $D(\zeta,\zetab)$ for the
		Bloch-Wigner single-valued dilogarithm function defined in~\eqref{eq:BlochWigner}.} This is the first order perturbation in $\lambda \phi^4$ theory and depicted in figure~\ref{fig:cross}. 

	%%%%%%%%%%%%%%%%%%%%%%%%%%%%%%%%%%%%%%%%%%%%%%%%%%%%%%%%%%
	\subsubsection{General dimensions}
	\label{subsubsec:Cross_generaldim}
	%%%%%%%%%%%%%%%%%%%%%%%%%%%%%%%%%%%%%%%%%%%%%%%%%%%%%%%%%%
	
	The integral corresponding to this Witten diagram as
   defined in~\eqref{eq:Witten_diagram_dimreg_trans} is finite and
   therefore does not have to be regulated. Since this diagram only involves bulk-to-boundary propagators it takes a simple form in any dimension $D$ and for general $\Delta$,
	\begin{align}\label{eq:Cross_allDelta_int}
		&\cW^{\Delta,D}_0(\zeta,\zetab)=\frac12\frac{v^\Delta}{(x^2_{12}x^2_{34})^\Delta}
		\int\limits_{\mathbb{R}^D}
		\frac{\dd^D X}{(u\cdot X)^D}\left(\frac{(u\cdot X)^4}{\|X|^2\|X-u_\zeta|^2\|X-u_1|^2}\right)^\Delta\,.
	\end{align}
	We can evaluate this integral by using the parametric representation which is based on the fact that for $A>0$
	\begin{align}
		\frac{1}{A^n}=\frac{1}{\Gamma(n)}\int_0^\infty\dd\alpha\,\eul^{-\alpha A}\alpha^{n-1}\,.
	\end{align}
	In this representation ~\eqref{eq:Cross_allDelta_int} becomes
	\begin{align}
		\label{eq:Cross_allDelta_param}
		\cW_0^{\Delta,D}&=\frac12\frac{i^{4\Delta-D}\pi^{\frac{D+1}{2}}}{\Gamma\left(\frac{D+1}{2}-2\Delta\right)\Gamma(\Delta)^2}\frac{v^\Delta}{(x^2_{12}x^2_{34})^\Delta}I_\times^\Delta(\zeta,\zetab)\,,
	\end{align}
	with
	\begin{align}
	\label{eq:Cross_allDelta_param_explicit}
		I_\times^\Delta(\zeta,\zetab)&=\int\limits_{(\mathbb{RP}^+)^2}\frac{\prod\limits_{i=1}^3\dd\alpha_i \alpha_i^{\Delta-1}}{(\alpha_1+\alpha_2+\alpha_3)^\Delta(\alpha_1\alpha_2+\alpha_1\alpha_3\zeta\zetab+\alpha_2\alpha_3(1-\zeta)(1-\zetab))^\Delta}\,,
	\end{align}
	where $(\mathbb{RP}^+)^2$ indicates that the integral is taken over the positive real projective space defined as
	\begin{align}
		(\mathbb{RP}^+)^{n-1}:=\{[\alpha_1, \dots ,\alpha_n]\in\mathbb{RP}^{n-1}: \alpha_1, \dots ,\alpha_n\geq 0\}\,.
	\end{align}
	Note that the only dependence on the spacetime dimension is contained in the pre-factor.

We show in the appendix~\ref{sec:cross_mpl} that  for $\Delta\geq1$ the cross integral takes the form
\begin{align}
I_\times^\Delta(\zeta,\zetab)=&\frac{c_1^{\Delta}(\zeta,\zetab)}{(\zeta-\zetab)^{4(\Delta-1)}}
\frac{4i D(\zeta,\zetab)}{\zeta-\zetab} +\frac{c_2^{\Delta}(\zeta,\zetab)}{(\zeta-\zetab)^{4(\Delta-1)}} \log(\zeta\zetab)\nonumber\\*
&+\frac{c_3^\Delta(\zeta,\zetab)}{(\zeta-\zetab)^{4(\Delta-1)}}
\log((1-\zeta)(1-\zetab))
+\frac{c_4^\Delta(\zeta,\zetab)}{(\zeta-\zetab)^{4(\Delta-1)}} .
\end{align}
where $c_r^\Delta(\zeta,\zetab)$ are polynomial in $\zeta$ and
       $\zetab$, and 	with $D(\zeta,\zetab)$ is the Bloch-Wigner dilogarithm defined in equation~\eqref{eq:BlochWigner}.
Despite the apparent singularity for $\zetab=\zeta$ the expression is
       regular on the real slice.
            As expected
            $I_\times^\Delta(\zeta,\zeta^*)$, with  $\zetab=\zeta^*$ being
            complex conjugate of $\zeta$, 
            is a single-valued function on $\mathbb C\setminus\{0,1\}$.

\medskip
In the rest of the paper we will make use of the result for
            $\Delta=1$ which reads
\begin{equation}
		\label{eq:crossDelta1}
		\cW_0^{1,4}(\zeta,\zetab)=\frac{\pi^2}{x_{12}^2x_{34}^2}\zeta\zetab\frac{ 2 i D(\zeta,\zetab)}{\zeta-\zetab},
	\end{equation}
and for $\Delta=2$, given by	{\footnotesize\begin{multline}
		\label{eq:crossDelta2}
		\cW_0^{2,4}(\zeta,\zetab)=\frac{3\pi^2(\zeta\zetab)^2}{4 x_{12}^4x_{34}^4}\cr 
		\times\Big(\frac{4 \zeta ^2 \zetab^2-(\zeta +\zetab)^3+2 \zeta  \zetab (\zeta +\zetab)^2+2 (\zeta +\zetab)^2-8 \zeta  \zetab (\zeta +\zetab)+4\zeta  \zetab}{(\zeta-\zetab)^4}\frac{2i D(\zeta,\zetab)}{\zeta-\zetab}\cr
		+\frac{(\zeta +\zetab)^2-3 \zeta  \zetab (\zeta +\zetab)+2 \zeta  \zetab}{(\zeta-\zetab)^4}\log(\zeta\zetab)\cr
		+\frac{3 \zeta  \zetab (\zeta +\zetab)-2 (\zeta +\zetab)^2+3 (\zeta +\zetab)-4 \zeta  \zetab}{(\zeta-\zetab)^4}\log((1-\zeta)(1-\zetab))+\frac{1}{(\zeta-\zetab)^2}\Big)\,.
	\end{multline}}
$\cW_0^{2,4}(\zeta,\zetab)$ can equivalently be obtained by acting on
           $\cW_0^{1,4}(\zeta,\zetab)$ with the differential operator
           $\cH_{1234}$ in~\eqref{eq:diff_operator}. This is a simple application of the method described in section~\ref{subsec:diff_operator}. 
	
	%%%%%%%%%%%%%%%%%%%%%%%%%%%%%%%%%%%%%%%%%%%%%%%%%%%%%%%%%%
	\subsubsection{Dimensional regularisation}
	\label{subsubsec:Cross_dimreg}
	%%%%%%%%%%%%%%%%%%%%%%%%%%%%%%%%%%%%%%%%%%%%%%%%%%%%%%%%%%
	Even though the Witten cross diagram is finite and does not need to be regularised, we will need the higher terms in the $D-4$ expansion, for the renormalisation of the  one-loop diagrams.  In order to restore AdS-invariance after regularisation, we need to evaluate the cross diagram in $D=4-4\epsilon$ dimensions.\footnote{See Remark 2 at the end of section~\ref{subsubsec:dimreg}.} For $\Delta=1$ the integral is 
	\begin{equation}
	    \label{eq:Cross_dimreg}
		\cW_0^{1,4-4\epsilon}(\zeta,\zetab)=\frac12\frac{\zeta\zetab}{(x_{12}x_{34})^2}\int\limits_{\mathbb{R}^4}\frac{\dd^{4-4\epsilon}X}{\|X|^2\|X-u_1|^{2(1-4\epsilon)}\|X-u_{\zeta}|^2}\,.
	\end{equation}
	Making use of the parametric representation ~\eqref{eq:crossD1_dimreg_param} we can  expand in $\epsilon$. Again the resulting integrand is linearly reducible and we can evaluate the integral by using \texttt{HyperInt}~\cite{Panzer:2014caa}, resulting in
	\begin{equation}\label{eq:w1}
	W_0^{1,4-4\epsilon}(\zeta,\zetab)=\frac{2^4 a^{4+4\epsilon}}{(2\pi)^8}	\cW_0^{1,4-4\epsilon}(\zeta,\zetab)=\frac{2^4 a^{4+4\epsilon}}{(2\pi)^8}\left(\cW_0^{1,4}(\zeta,\zetab)+\epsilon \cW_{0,\epsilon}^{1,4}(\zeta,\zetab)+O(\epsilon^2)\right)\,,
	\end{equation}
	with $W_0^{1,4}(\zeta,\zetab)$ given in ~\eqref{eq:crossDelta1} and 
	\begin{equation}
		\cW_{0,\epsilon}^{1,4}(\zeta,\zetab)=\frac{\zeta\zetab \pi^{2}}{x_{12}^2x_{34}^2}\left(\frac{f_1(\zeta,\zetab)}{\zeta-\zetab}-\frac{2iD(\zeta,\zetab)}{\zeta-\zetab}\log(\zeta\zetab)+
			\frac{2iD(\zeta,\zetab)}{\zeta-\zetab}\log((1-\zeta)(1-\zetab))\right)\,,
	\end{equation}
	where the function $f_1(\zeta,\zetab)$ can be found in equation ~\eqref{eq:Omega1}. The corresponding result for $\Delta=2$ can then be obtained by acting on the parametric representation for $\Delta=1$ with $\cH_{1234}$ before expanding in $\epsilon$. After integration over the Feynman parameters (see ~\eqref{eq:crossD2_dimreg_param}) we find
	\begin{equation}\label{eq:w2}
	W_0^{2,4-4\epsilon}(\zeta,\zetab)	=\frac{2^8 a^{4+4\epsilon}}{(2\pi)^8}\cW_0^{2,4-4\epsilon}(\zeta,\zetab)=\frac{2^8 a^{4+4\epsilon}}{(2\pi)^8}\left(\cW_0^{2,4}(\zeta,\zetab)+\epsilon \cW_{0,\epsilon}^{2,4}(\zeta,\zetab)+O(\epsilon^2)\right)\,,
	\end{equation}
	with $\cW_0^{2,4}(\zeta,\zetab)$ given in~\eqref{eq:crossDelta2} and $\cW_{0,\epsilon}^{2,4}(\zeta,\zetab)$ given by equation~\eqref{eq:crossD2_epsilon}.

	%%%%%%%%%%%%%%%%%%%%%%%%%%%%%%%%%%%%%%%%%%%%%%%%%%%%%%%%%%
	\subsection{The one-loop diagram}
	\label{subsec:bubble_diagram}
	%%%%%%%%%%%%%%%%%%%%%%%%%%%%%%%%%%%%%%%%%%%%%%%%%%%%%%%%%%

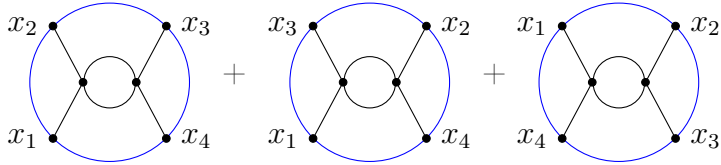
\begin{figure}[ht]
	\centering
	\begin{tikzpicture}[baseline=(z)]
		\begin{feynman}[inline=(z)]
			\vertex (z);	
			\tikzfeynmanset{every vertex=dot}
			\vertex [above left=0.71cm and 0.71cm of z, label=180:$x_2$] (x2);
			\vertex [below left=0.71cm and 0.71cm of z, label=180:$x_1$] (x1);
			\vertex [above right=0.71cm and 0.71cm of z, label=0:$x_3$] (x3);
			\vertex [below right=0.71cm and 0.71cm of z, label=0:$x_4$] (x4);
			\vertex [left=0.3cm of z] (y1);
			\vertex [right=0.3cm of z] (y2);
			\tikzfeynmanset{every vertex={empty dot,minimum size=0mm}}
			\diagram* {
				(x2)--(y1),
				(x1)--(y1),
				(x3)--(y2),
				(x4)--(y2),
			};
		\end{feynman}
		\begin{pgfonlayer}{bg}
			\draw[blue] (z) circle (1.05cm);
			\draw (z) circle (0.34cm);
		\end{pgfonlayer}
	\end{tikzpicture}+
	\begin{tikzpicture}[baseline=(z)]
		\begin{feynman}[inline=(z)]
			\vertex (z);	
			\tikzfeynmanset{every vertex=dot}
			\vertex [above left=0.71cm and 0.71cm of z, label=180:$x_3$] (x2);
			\vertex [below left=0.71cm and 0.71cm of z, label=180:$x_1$] (x1);
			\vertex [above right=0.71cm and 0.71cm of z, label=0:$x_2$] (x3);
			\vertex [below right=0.71cm and 0.71cm of z, label=0:$x_4$] (x4);
			\vertex [left=0.3cm of z] (y1);
			\vertex [right=0.3cm of z] (y2);
			\tikzfeynmanset{every vertex={empty dot,minimum size=0mm}}
			\diagram* {
				(x2)--(y1),
				(x1)--(y1),
				(x3)--(y2),
				(x4)--(y2),
			};
		\end{feynman}
		\begin{pgfonlayer}{bg}
			\draw[blue] (z) circle (1.05cm);
			\draw (z) circle (0.34cm);
		\end{pgfonlayer}
	\end{tikzpicture}+\begin{tikzpicture}[baseline=(z)]
		\begin{feynman}[inline=(z)]
			\vertex (z);	
			\tikzfeynmanset{every vertex=dot}
			\vertex [above left=0.71cm and 0.71cm of z, label=180:$x_1$] (x2);
			\vertex [below left=0.71cm and 0.71cm of z, label=180:$x_4$] (x1);
			\vertex [above right=0.71cm and 0.71cm of z, label=0:$x_2$] (x3);
			\vertex [below right=0.71cm and 0.71cm of z, label=0:$x_3$] (x4);
			\vertex [left=0.3cm of z] (y1);
			\vertex [right=0.3cm of z] (y2);
			\tikzfeynmanset{every vertex={empty dot,minimum size=0mm}}
			\diagram* {
				(x2)--(y1),
				(x1)--(y1),
				(x3)--(y2),
				(x4)--(y2),
			};
		\end{feynman}
		\begin{pgfonlayer}{bg}
			\draw[blue] (z) circle (1.05cm);
			\draw (z) circle (0.34cm);
		\end{pgfonlayer}
	\end{tikzpicture}
	\caption{One-loop Witten diagrams}
	\label{fig:oneloopW}
\end{figure}
	At one-loop level there are numerous diagrams to be evaluated but, as we argued in section~\ref{sec:QFT_in_AdS}, tadpoles and self-energy corrections only contribute to the mass shift at this level so we can reabsorb them into the conformal dimension of the boundary operator. In this section we fix this dimension to $\Delta=1$ and $\Delta=2$ and the only remaining connected one-loop diagrams in $\lambda\phi^4$ theory are the three channels of the one-loop bubble diagram depicted in figure~\ref{fig:oneloopW}. 
	
	%%%%%%%%%%%%%%%%%%%%%%%%%%%%%%%%%%%%%%%%%%%%%%%%%%%%%%%%%%
	\subsubsection{Dimensional regularisation}
	\label{subsubsec:dimreg_oneloop}
	%%%%%%%%%%%%%%%%%%%%%%%%%%%%%%%%%%%%%%%%%%%%%%%%%%%%%%%%%%
	In order to restore conformal invariance after regularisation, we calculate these diagrams in dimensional regularisation with $D=4-2\epsilon$ using the general formula~\eqref{eq:Witten_diagram_dimreg_trans} with $\delta(\Gamma_W)=\delta(X_{\sigma(1)}=X_{\sigma(2)})\delta(X_{\sigma(3)}=X_{\sigma(4)})$ to obtain
	\begin{align}\label{eq:OneLoop_diagram_dim_reg}
		&W^{\Delta,4-2\epsilon}_1(\zeta,\bar{\zeta})=\frac{a^{4+4\epsilon}2^{4\Delta}(\zeta\zetab)^\Delta}{4(x_{12}^2x_{34}^2)^\Delta
			(2\pi)^{8}}
		\int\limits_{(\mathbb{R}^D)^{2}}
		\dd^{4-2\epsilon} X_1 \dd^{4-2\epsilon} X_2\;\cdot\nonumber\\*
		&\cdot\frac{(u\cdot X_1)^{2\Delta-4} (u\cdot X_2)^{2\Delta-4}\tilde\Lambda(\mathbf X_1,\mathbf X_2;\Delta)^2}{\|X_1-u_1|^{-4\epsilon}\|X_2-u_1|^{-4\epsilon}}
		\Bigg(\frac{1}{\|X_1|^{2\Delta}\|X_{2}-u_1|^{2\Delta} 
			\|X_{2}-u_\zeta|^{2\Delta}}\Bigg.\nonumber\\*&+ \frac{1}{\|X_2|^{2\Delta}\|X_{2}-u_1|^{2\Delta} 
			\|X_{1}-u_\zeta|^{2\Delta}}+\frac{1}{\|X_2|^{2\Delta}\|X_{1}-u_1|^{2\Delta} 
			\|X_{2}-u_\zeta|^{2\Delta}}\Bigg)\,,
	\end{align}
	where $\tilde{\Lambda}$ is the propagator~\eqref{eq:LambdaSpecial_iep2} without the normalization factor $a^2/4\pi^2$ which has been pulled out of the integral. 
	Expanding the square with the help of the identity in~\eqref{eq:Gidentity}
	one finds
	\begin{multline}
		\tilde\Lambda(\mathbf X_1,\mathbf
		X_2;\Delta)^2=\frac{(u\cdot X_1)^2 \,
			(u\cdot X_2)^2}{\|X_1-X_2|^4}+\frac{(u\cdot X_1)^2 \,
			(u\cdot \sigma(X_2))^2}{\|X_1-\sigma(X_2)|^4}
		\cr-\frac{(-1)^\Delta}{2}\left( \frac{u\cdot X_1 \,
			u\cdot X_2}{\|X_1-X_2|^2}+ \frac{u\cdot X_1 \,
			u\cdot \sigma(X_2)}{\|X_1-\sigma(X_2)|^2} \right).
	\end{multline}
	Upon substitution into~\eqref{eq:OneLoop_diagram_dim_reg} we arrive at 
	\begin{equation}\label{eq:W1_fin_div_I}
		W^{\Delta,4-2\epsilon}_1(\zeta,\bar{\zeta})=\frac{2^{4\Delta}a^{4+4\epsilon}}{	(2\pi)^{12}}\sum\limits_{i\in\{s,t,u\}}\left(\cW_{1,\mathrm{div}}^{\Delta,4-2\epsilon,i}(\zeta,\bar{\zeta})-\frac{(-1)^\Delta}{2}\cW_{1,\mathrm{fin}}^{\Delta,4,i}(\zeta,\bar{\zeta})\right)\,,
	\end{equation}
	where the integral in $W_{1,\mathrm{div}}^{\Delta,4-2\epsilon,i}(\zeta,\bar{\zeta})$ requires regularisation while  $W_{1,\mathrm{fin}}^{\Delta,4,i}(\zeta,\bar{\zeta})$ does not. For instance, in the $s$-channel
		\begin{equation}
		\label{eq:W1_fin_divs}
		\cW_{1,\mathrm{div}}^{\Delta,4-2\epsilon,s}(\zeta,\bar{\zeta})=\frac12\frac{(\zeta\zetab)^\Delta}{(x_{12}^2x_{34}^2)^\Delta}\int\limits_{\mathbb{R}^{2D}}\frac{\dd^{4-2\epsilon}
                  X_1\dd^{4-2\epsilon} X_2(u\cdot
                  X_1)^{2\Delta-2}(u\cdot X_2)^{2\Delta-2} \|X_1-u_1|^{4\epsilon}}{\|X_1|^{2\Delta}\|X_1-u_\zeta|^{2\Delta}\|X_2-u_1|^{2\Delta-4\epsilon}\|X_1-X_2|^4},
		\end{equation}
	and 
	\begin{equation}
	    \label{eq:W1_fins}
	    \cW_{1,\mathrm{fin}}^{\Delta,4,s}(\zeta,\bar{\zeta})=\frac12\frac{(\zeta\zetab)^\Delta}{(x_{12}^2x_{34}^2)^\Delta}\int_{\mathbb{R}^{8}}\frac{\dd^{4} X_1\dd^{4} X_2(u\cdot X_1)^{2\Delta-3}(u\cdot X_2)^{2\Delta-3}}{\|X_1|^{2\Delta}\|X_1-u_\zeta|^{2\Delta}\|X_2-u_1|^{2\Delta}\|X_1-X_2|^2}\,,
	\end{equation}
	with similar expression for the other channels listed in equation~\eqref{eq:W1_fin_div}.

	\paragraph{For $\Delta$=1} the evaluation of the divergent part is straightforward. In the parametric representation (see~\eqref{eq:W1div_param}) we can integrate using \texttt{HyperInt}~\cite{Panzer:2014caa} giving
	\begin{multline}
		\cW_{1,\mathrm{div}}^{\Delta,4-2\epsilon,s}(\zeta,\zetab)=-\frac{\pi^{4-2\epsilon}\eul^{-2\gamma\epsilon}\zeta\zetab}{2x_{12}^2x_{34}^2}\Big(\frac{1}{\epsilon}\frac{4i D(\zeta,\zetab)}{\zeta-\zetab}+\frac{f_1(\zeta,\zetab)}{\zeta-\zetab}-\frac{2i D(\zeta,\zetab)}{\zeta-\zetab}\log(\zeta\zetab)\cr+\frac{4i D(\zeta,\zetab)}{\zeta-\zetab}\log((1-\zeta)(1-\zetab))\Big)\,.
	\end{multline}
Adding the corresponding contributions form the $t$- and $u$-channel from appendix~\ref{subsubsec:Delta1_exact} we end up with
	\begin{multline}
		W^{1,4-2\epsilon}_{1,\mathrm{div}}(\zeta,\bar{\zeta})=\frac{2^4a^{4+4\epsilon}}{	(2\pi)^{12}}\sum\limits_{i\in\{s,t,u\}}\cW_{1,\mathrm{div}}^{1,4-2\epsilon,i}(\zeta,\bar{\zeta})\cr=
		\frac{2^4a^{4+4\epsilon}}{(2\pi)^{12}}\left(-\frac{3\pi^2}{\epsilon}\cW_0^{1,4-4\epsilon}(\zeta,\bar{\zeta})+\frac{\pi^4v}{2x_{12}^2x_{34}^2}\sum\limits_{i\in\{s,t,u\}}L_{0}^{1,i}(\zeta,\bar{\zeta})\right)\,,
	\end{multline}
where the $L_{0}^{\Delta,i}(\zeta,\bar{\zeta})$ terms are regular for $\epsilon\to 0$. Their expressions are given in appendix~\ref{subsubsec:L0integrals}.

    The finite piece $\cW_{1,\mathrm{fin}}^{1,4,i}$ is harder to solve exactly. In the parametric representation it can be rewritten as (see appendix~\ref{subsubsec:L0primeintegrals} for details)
    \begin{equation}
        \cW_{1,\mathrm{fin}}^{1,4,i}(v,Y)=\frac{2\pi^4v^\Delta}{(x_{12}x_{34})^\Delta}
        \begin{cases}
          L_0'(v,1-Y,1)& i=s\cr
L_0'(1-Y,1,v)& i=t\cr
L_0'(1,v,1-Y)& i=u
        \end{cases}\,,
    \end{equation}
    with
    \begin{equation}
    L'_0(x,y,z)= \int_1^\infty d\lambda\int_0^\infty ds \int_0^1 dr   \frac{
    \log(1+\lambda s)}{4\lambda  \sqrt{(1+s) (1+\lambda s)} (sr(1-r)   x+ r y+(1-r)z)}\,,
    \end{equation}
    and $v=\zeta\zetab$ and $\zeta+\zetab=v+Y$.
    
	The integral is an elliptic    polylogarithm obtained by integrating the dilogarithm in~\eqref{e:Isigmadef} over the elliptic curve~\eqref{eq:elliptic_curve}. Since we want to calculate anomalous
               dimensions, which are related to the coeffcients of the
               terms proportional to $\log(v)$, we are not actually interested
               in the complete result of the integral. In appendix~\ref{subsubsec:L0primeintegrals} we provide an efficient
               way to extract the coefficients of the $\log(v)^2$ and $\log(v)$ terms and do an expansion in $v$ and $Y$.
	
	Altogether,  the total one-loop Witten diagram for $\Delta=1$ is given by
	\begin{multline}
		\label{eq:oneloopD1_dimreg}
		W_1^{1,4-2\epsilon}(v,Y)=\frac{2^4a^{4+4\epsilon}}{(2\pi)^{12}}\Big(-\frac{3\pi^2}{\epsilon}\cW_0^{1,4-4\epsilon}(v,Y)
		+\frac{\pi^4v}{2x_{12}^2x_{34}^2}\sum\limits_{i\in\{s,t,u\}}L_{0}^{1,i}(v,Y)+\cr\frac{\pi^4v}{x_{12}^2x_{34}^2}\sum\limits_{i\in\{s,t,u\}}{L_{0}'}^{i}(v,Y)+\mathcal{O}(\epsilon)\Big).
	\end{multline}

	\paragraph{For $\Delta=2$} we start with the calculation of
               the finite part. There are no elliptic integrals
               to compute and we can find closed form expressions in
               terms of single-valued polylogarithms of
               weight up to three.
               
               To obtain the parametric representation of the finite part \eqref{eq:W1_fins} we introduce the auxiliary integrals $\tilde\cW_{1,\mathrm{fin}}^{2,4,i}$ for each channel, with the s-channel given by
               \begin{align}
               \label{eq:W1_fin_auxs}
                   \tilde \cW_{1,\mathrm{fin}}^{2,4,s}&=\frac{1}{8}\int_{\mathbb{R}^8}\frac{\dd^4 X_1\dd^4 X_2}{\|X_1-\vec x_1|^2\|X_1-\vec x_2|^4\|X_2-\vec x_3|^4\|X_2-\vec x_4|^2\|X_1-X_2|^2}\cr
		&=\frac{1}{8}\frac{x_{14}^2}{x_{12}^4x_{34}^4}(\zeta\zetab)^2\int_{\mathbb{R}^8}\frac{\dd^4 X_1\dd^4 X_2}{\|X_1|^2\|X_1-\uZ|^4\|X_2-u_1|^2\|X_1-X_2|^2}\,,
               \end{align}
               and the other channels displayed in~\eqref{eq:W12finite_aux}. The second line in equation~\eqref{eq:W1_fin_auxs} is obtained by performing the conformal mappings as described in section~\ref{subsubsec:conformal_mappings}.
               Considering the discussion in section~\ref{subsec:diff_operator} it is straightforward to see, that the finite part of the one-loop integral in each channel is given by the action of the differential operator $\cH_{ij}$ on the corresponding auxiliary integral by
	\begin{equation}\label{eq:diffact+1}
		\cW_{1,\mathrm{fin}}^{2,4,s}=\cH_{14}\tilde\cW_{1,\mathrm{fin}}^{2,4,s};\qquad
                \cW_{1,\mathrm{fin}}^{2,4,t}=\cH_{12}\tilde\cW_{1,\mathrm{fin}}^{2,4,t};
                \qquad \cW_{1,\mathrm{fin}}^{2,4,u}=\cH_{12}\tilde\cW_{1,\mathrm{fin}}^{2,4,u}\,.
	\end{equation}
 In equation~\eqref{eq:W2fin_param} we give the result of~\eqref{eq:diffact+1} in the parametric representation.  Integrating over the Feynman parameters we obtain 
	\begin{multline}\label{e:W2fin}
			\cW_{1,\mathrm{fin}}^{2,4,s}(\zeta,\zetab)=\frac{\pi^4}{8}\frac{(\zeta\zetab)^2}{(x_{12}^2x_{34}^2)^2}\left(\frac{(\zeta+\zetab-2)8iD(\zeta,\zetab)}{(\zeta-\zetab)^3}\right.\cr
			\left.+\frac{2(2\zeta\zetab-\zeta-\zetab)}{\zeta\zetab(\zeta-\zetab)^2}\log((1-\zeta)(1-\zetab))
			-\frac{4\log(\zeta\zetab)}{(\zeta-\zetab)^2}\right)\,,
	\end{multline}
    for the $s$ channel. The results for the other channels are given in appendix~\ref{subsubsec:Delta2_exact}.
    
	The divergent integrals in~\eqref{eq:W1_fin_div_I} can be calculated by acting with $\cH_{1234}$ on the corresponding expressions for $\Delta=1$. Some care has to be taken since the action of $\cH_{1234}$ and the $\epsilon$ expansion do not commute: We have to act on the parametric representation of the $\Delta=1$ expressions which gives us the parametric representation of the $\Delta=2$ expressions. These can then be expanded in $\epsilon$. The explicit expressions are given in equation~\eqref{eq:W2div_param}. Integrating over the Feynman parameters and summing over the three channels we end up with
	\begin{multline}\label{e:W2div}
		W^{2,4-2\epsilon}_{1,\mathrm{div}}(\zeta,\bar{\zeta})=\frac{2^8a^{4+4\epsilon}}{(2\pi)^{12}}\Big(-\frac{3\pi^2}{\epsilon}\cW_0^{2,4-4\epsilon}+3\pi^2\cW_0^{2,4}\cr+\frac12\sum\limits_{j\in\{s,t,u\}}\cW_{1,\mathrm{fin}}^{2,4,j}
		+\frac{3\pi^4v^2}{8(x_{12}^2x_{34}^2)^2}\sum\limits_{i\in\{s,t,u\}}L_{0}^{2,i}+\mathcal{O}(\epsilon)\Big).
	\end{multline}
	In sum, the total one-loop Witten diagram for $\Delta=2$ is given by
	\begin{equation}
		\label{eq:oneloopD2_dimreg}
		W_1^{2,4-2\epsilon}(\zeta,\zetab)=\frac{2^8a^{4+4\epsilon}}{(2\pi)^{12}}\left(-\frac{3\pi^2}{\epsilon}\cW_0^{2,4-4\epsilon}+3\pi^2\cW_0^{2,4}
		+\frac{3\pi^4v^2}{8(x_{12}^2x_{34}^2)^2}\sum\limits_{i\in\{s,t,u\}}L_{0}^{2,i}+\mathcal{O}(\epsilon)\right)\,,
	\end{equation}
	with the expressions for $L_{0}^{\Delta,i}$ given in appendix~\ref{subsubsec:L0integrals}.

	\paragraph{Renormalisation:} In order to subtract the UV-divergences in our dimensional re-gularisation we define the bare coupling constant $\lambda$ as usual through $\lambda= \lambda_R\,(a\mu)  \mu^{2\epsilon}+\delta\lambda$. The bare coupling is divergent but gives finite four-point functions by choosing the divergent counter-term $\delta\lambda$ accordingly. The renormalised coupling $\lambda_R$ is finite and dimensionless in any dimension due to the factor $\mu^{2\epsilon}$ where $\mu$ has the dimension of length, which accounts for the scaling correction due to dimensional regularisation.
Summing the tree-level (cross) and the one-loop  (bubble) diagram contributions from~\eqref{eq:w2} and~\eqref{eq:oneloopD2_dimreg} above, we have, for the connected part of the four-point function, up to finite terms,
	\begin{align}
		\lambda_R\mu^{4\epsilon} W_0^{\Delta,4-4\epsilon}-\frac{\lambda_R^2\mu^{4\epsilon}}{2}W_1^{\Delta,4-2\epsilon}&=
		\frac{2^{4\Delta}a^4}{(2\pi)^{8}}\lambda_R\cdot(a\mu)^{4\epsilon}\left(1+\frac{3\lambda_R}{32\pi^2}\frac{1}{\epsilon}\right)\cW_0^{\Delta,4-4\epsilon}\cr
		&\equiv\mu^{2\epsilon}\lambda\cW_0^{\Delta,4-4\epsilon}\,.
	\end{align}
The extra factor $\mu^{2\epsilon}$ in front of $\lambda$ arises
           since we have chosen the measure $d^{4-4\epsilon}X$, rather
           than $d^{4-2\epsilon}X$ for the cross diagram  in~\eqref{eq:Cross_dimreg}.  Focusing on the $1/\epsilon$ pole then fixes the value of the counter-term 
           \begin{equation}
           \delta\lambda= -\frac{3\lambda_R^2\mu^{2\epsilon}}{32\pi^2}\frac{1}{\epsilon}\,.
           \end{equation}
           On the other hand the $\log\mu$ contribution to the finite part in $\delta\lambda$ gives rise to the Callan-Symanzik equation 
	\begin{align}
		\label{eq:Callan_Symanzik}
		0=\mu\frac{\dd}{\dd{\mu}}\lambda=2\epsilon\mu^{2\epsilon}\left(\lambda_R-\frac{3\lambda_R^2}{32\pi^2}\frac{1}{\epsilon}\right)
		+\mu^{2\epsilon}{\mu}\frac{\partial\lambda_R}{\partial{\mu}}\frac{\partial}{\partial\lambda_R}\left(\lambda_R-\frac{3\lambda_R^2}{32\pi^2}\frac{1}{\epsilon}\right),
	\end{align}
	from which we read of the beta function 
	\begin{align}
		{\beta}=\frac{3\lambda_R^2}{16\pi^2}+\Op(\lambda^3)\,.
	\end{align}
	This coincides with the $\beta$ function of $\lambda\phi^4$ theory in flat space.

	%%%%%%%%%%%%%%%%%%%%%%%%%%%%%%%%%%%%%%%%%%%%%%%%%%%%%%%%%%
	\subsubsection{AdS invariant regularisation}
	\label{subsubsec:AdSinvariant_oneloop}
	%%%%%%%%%%%%%%%%%%%%%%%%%%%%%%%%%%%%%%%%%%%%%%%%%%%%%%%%%%
	Let us compare the results obtained so far to the AdS-invariant regularisation method described in section~\ref{subsubsec:covaraint_reg}, which was used in~\cite{Bertan:2018afl,Bertan:2018khc,Heckelbacher:2020nue}. The one-loop Witten diagram associated to the graphs in
	figure~\ref{fig:oneloopW}, with the regularisation given by~\eqref{eq:Witten_diagram_delta_trans}, again consists of the sum over the contributions from the three channels 
	\begin{equation}
		\label{e:W1deltaDef}
		W_1^{\Delta,\delta}(\zeta,\zetab)= \frac{2^{4\Delta}a^4}{(2\pi)^{12}}\sum\limits_{i\in\{s,t,u\}}\cW_1^{\Delta,\delta,i}(\zeta,\zetab)\,,
	\end{equation}
	with the contribution to the s-channel given by
	\begin{align}
		\label{eq:W1_AdSinv}
		\cW_{1}^{\Delta,\delta,s}=&\frac14\frac{(\zeta\zetab)^\Delta}{(x_{12}^2x_{34}^2)^\Delta}\int_{\mathbb{R}^8}\frac{\dd^{4} X_1\dd^{4} X_2 z_1^{2\Delta-4}z_2^{2\Delta-4}}{\|X_1|^{2\Delta}\|X_1-u_\zeta|^{2\Delta}\|X_2-u_1|^{2\Delta}}\left(\frac{K^\delta(\mathbf X_1,\mathbf
			X_2)^{\Delta}}{1-K^\delta(\mathbf X_1,\mathbf X_2)^2}\right)^2,
	\end{align}
	and the integrals for the other channels given in~\eqref{eq:W1_AdSinv_appendix}.
	
\begin{figure}[ht]
		\begin{center}
			% S Channel diagram
			$\begin{tikzpicture}[baseline=(z)]
				\begin{feynman}[inline=(z)]
					\tikzfeynmanset{every vertex=dot}
					\vertex [label=180:$\vec w_1$] (x1);
					\tikzfeynmanset{every vertex={empty dot,minimum size=0mm}}
					\vertex [below right=1.06cm and 1.06cm of x1] (z);
					\tikzfeynmanset{every vertex=dot}
					\vertex [below left=1.06cm and 1.06cm of z, label=180:$\vec w_2$] (x2);
					%\vertex [above right=1.06cm and 1.06cm of z, label=0:$x_3$] (x3);
					%\vertex [below right=1.06cm and 1.06cm of z, label=0:$x_4$] (x4);
					\vertex [left=0.4cm of z, label=180:$X_1$] (y1);
					\vertex [right=1cm of z, label=180:$X_2$] (y3);
					\tikzfeynmanset{every vertex={empty dot,minimum size=0mm}}
					\vertex [right=0.3cm of z] (x);
					\diagram* {
						(x1)--(y1),
						(x2)--(y1),
						%	(x3)--(y3),
						%	(x4)--(y3),
					};
				\end{feynman}
				\begin{pgfonlayer}{bg}
					\draw[blue] (z) circle (1.5cm);
					\draw (x) circle (0.7cm);
				\end{pgfonlayer}
			\end{tikzpicture}$
			\caption{Fish diagram}
			\label{fig:fish_diagram}
		\end{center}
	\end{figure}
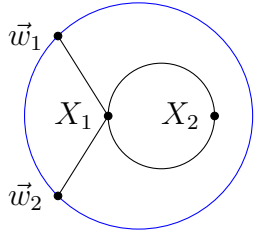

	In order to simplify the calculation we will separate these double integrals into an integral with the two external legs connected to $X_1$ and perform the integration over $X_2$ later as
	\begin{align}\label{eq:spl_1}
		\cW_{1}^{\Delta,\delta,i}=&\frac12\int\limits_{\mathbb R^4} {\dd^4X_2} \hat \cW_{1}^{\Delta,\delta,i}(\vec{w}_1,\vec{w}_2,X_2)\frac{z_2^{2\Delta-4}}{\|\vec{w}_3-X_2|^{2\Delta}}\,,
	\end{align}
	with the intermediate integral
	\begin{equation}
		\hat \cW_{1}^{\Delta,\delta,i}(\vec{w}_1,\vec{w}_2,X_2)=\frac12\frac{v^\Delta}{(x_{12}^2x_{34}^2)^\Delta}\int\limits_{\mathbb{R}^4}\frac{{\dd^4X_1}\;\;z_1^{2\Delta-4}}{\|\vec{w}_1-X_1|^{2\Delta}\|\vec{w}_2-X_1|^{2\Delta}}
		\left(\frac{K^{\delta}(\mathbf
			X_1,\mathbf X_2)^\Delta}{1-K^{\delta}(\mathbf
			X_1,\mathbf X_2)^2}\right)^2,
	\end{equation}
	associated to the fish diagram in figure~\ref{fig:fish_diagram}.
	Comparing to the expressions in equation~\eqref{eq:W1_AdSinv} it is straightforward to identify the three channels as:
	\begin{itemize}
		\item $s$-channel: $\vec{w}_1\to0$, $\vec{w}_2\to  \uZ$ and $\vec{w}_3\to u_1$
		\item $t$-channel: $\vec{w}_1\to \uZ$, $\vec{w}_2\to  u_1$ and $\vec{w}_3\to 0$
		\item $u$-channel: $\vec{w}_1\to0$, $\vec{w}_2\to  u_1$ and $\vec{w}_3\to \uZ$
	\end{itemize}
	
	For $\Delta=1$ and
	$\Delta=2$, this integral can be further
	simplified by rewriting the square of the bulk-to-bulk propagator
	in~\eqref{eq:W1_AdSinv} in terms of euclidean propagators
	as described in~\ref{subsec:mappropagators}.
	The fish diagram  is then given by
	\begin{multline}
		\label{eq:fish}
		\hat\cW_{1}^{\Delta,\delta,i}(\vec{w}_1,\vec{w}_2,X_2)=\frac{v^\Delta}{(x_{12}^2x_{34}^2)^\Delta}\frac14
		\int_{\mathbb{R}^4}\frac{\dd^4X_1}{z_1^4} 
		\prod\limits_{i=1}^2
		\frac{z_1^{\Delta}}{\|\vec{w}_i-X_1|^{2\Delta}}\cr
		\times\left(\frac{K^{\delta}(\mathbf
			X_1,\mathbf X_2)^2}{(1-K^{\delta}(\mathbf
			X_1,\mathbf X_2))^2}-(-1)^{\Delta}\frac{K^{\delta}(\mathbf
			X_1,\mathbf X_2)}{1-K^{\delta}(\mathbf
			X_1,\mathbf X_2)}\right)\,.
	\end{multline}
	
	\paragraph{ For $\Delta=2$:}
	We split the integral into a first piece that diverges when $\delta\to0$
	\begin{align}
		\hat\cW_{1}^{\Delta,\delta,i}(\vec{w}_1,\vec{w}_2,X_2)|_1:=&\frac{v^2}{(x_{12}^2x_{34}^2)^2}\frac14\int_{\mathbb{R}^4}\dd^4X_1
		\prod\limits_{i=1}^2\frac{1}{\|\vec{w}_i-X_1|^4}\frac{K(\mathbf
			X_1,\mathbf X_2)^2}{(1-K(\mathbf
			X_1,\mathbf X_2)+\delta)^2}\cr
		=&\frac{\pi^2v^2}{(x_{12}^2x_{34}^2)^2}\Bigg[\frac18  \prod\limits_{i=1}^2\frac{z_2}{\|\vec{w}_i-X_2|^2}-\prod\limits_{i=1}^2\frac{z_2^2}{\|\vec{w}_i-X_2|^4}\cdot\cr
		&\qquad\qquad\;\;\;\left(\log\left(\frac{z_2^2  \abs{\vec{w}_1-\vec{w}_2}^2}{\|\vec
			w_1-X_2|^2\|\vec w_2-X_2|^2}
		\right)+\log2\delta+2\right)\Bigg]
	\end{align}
	and a second piece that is finite when $\delta\to0$:
	\begin{align}
		\hat\cW_{1}^{\Delta,\delta,i}(\vec{w}_1,\vec{w}_2,X_2)|_2:=&\lim_{\delta\to
			0}\frac{v^2}{(x_{12}^2x_{34}^2)^2}\frac14\int_{\mathbb{R}^4}\dd^4X_1
		\prod\limits_{i=1}^2\frac{1}{\|\vec w_i-X_1|^4}\frac{K(\mathbf
			X_1,\mathbf X_2)}{1-K(\mathbf  X_1,\mathbf X_2)+\delta}\cr
		&
		=\frac{\pi^2v^2}{(x_{12}^2x_{34}^2)^2}\frac18\prod\limits_{i=1}^2\frac{z_2}{\|\vec{w}_{i}-X_2|^2}\,,
	\end{align}
	so that the complete result for the fish diagram  becomes
	\begin{align}
		&\hat\cW_{1}^{\Delta,\delta,i}(\vec{w}_1,\vec{w}_2,X_2)=\hat\cW_{1}^{\Delta,\delta,i}(\vec{w}_1,\vec{w}_2,X_2)|_1-\hat\cW_{1}^{\Delta,\delta,i}(\vec{w}_1,\vec{w}_2,X_2)|_2\cr
		&\qquad=-\frac{\pi^2v^2}{(x_{12}^2x_{34}^2)^2}\prod\limits_{i=1}^2\frac{z_2^2}{\|\vec{w}_i-X_2|^4}\left(\log\left(\frac{z_2^2  \abs{\vec{w}_1-\vec{w}_2}^2}{\|\vec w_1-X_2|^2\|\vec w_2-X_2|^2}
		\right)+\log2\delta+2\right)\,.\cr
	\end{align}
	Finally, using~\eqref{eq:spl_1} we attach the remaining bulk-to-boundary propagator to
	obtain the integral for the one-loop diagram  for $\Delta=2$
	\begin{align}
		\cW_{1}^{\Delta,\delta,i}=-\frac{v^2}{(x_{12}^2x_{34}^2)^2}\frac{\pi^2}{2}\int\limits_{\mathbb{R}^4}\dd^4 X& z^4\prod\limits_{i=1}^3\frac{1}{\|X-\vec w_i|^4}\;\cdot\\&\left(\log\left(\frac{z_2^2  \abs{\vec{w}_1-\vec{w}_2}^2}{\|\vec
			w_1-X|^2\|\vec w_2-X|^2}
		\right)+\log2\delta+2\right)\,,
	\end{align}
	which evaluates to 
	\begin{equation}
		\cW_{1}^{\Delta,\delta,i}=-\pi^2\Big(\log\left(\frac{\delta}{2}\right)+\frac{11}{3}\Big)\cW_0^{2,\delta}+\frac{3\pi^4v^2}{8(x_{12}^2x_{34}^2)^2}L_0^{2,i}\,.
	\end{equation}
	Restoring the prefactors, the complete one-loop diagram is thus
	\begin{equation}
		\label{eq:oneloopD2_delta}
		W_1^{2,\delta}=\frac{2^8a^4\pi^2}{(2\pi)^{12}}\left(-3\log\left(\frac{\delta}{2}\right)\cW_0^{2,\delta}-11\cW_0^{2,\delta}+\frac{3\pi^4v^2}{8(x_{12}^2x_{34}^2)^2}\sum\limits_{i\in\{s,t,u\}}L_0^{2,i}\right)\,,
	\end{equation}
	where the $L_0^{\Delta,i}$ terms are given in~\ref{subsubsec:L0integrals} and $\cW_0^{\Delta,\delta}$ is the cross
  diagram  evaluated in section~\ref{subsubsec:Cross_generaldim}.
	
	\paragraph{For $\Delta=1$:}We split the integral in a first piece that diverges when $\delta\to0$
	\begin{align}
		&\hat\cW_{1}^{1,\delta,i}(\vec{w}_1,\vec{w}_2,X_2)|_1:=\frac{v}{(x_{12}^2x_{34}^2)}\frac14\int_{\mathbb{R}^4}\dd^4X_1
		\prod\limits_{i=1}^2\frac{1}{\|\vec{w}_i-X_1|^2}\frac{K(\mathbf
			X_1,\mathbf X_2)^2}{(1-K(\mathbf
			X_1,\mathbf X_2)+\delta)^2}\cr
		&=-\frac{\pi^2v}{(x_{12}^2x_{34}^2)}\prod\limits_{i=1}^2\frac{z_2}{\|\vec{w}_i-X_2|^2}\left(\log\left(\frac{z_2^2 \abs{\vec{w}_1-\vec{w}_2}^2}{\|\vec	w_1-X_2|^2\|\vec w_2-X_2|^2}
		\right)+\log2\delta\right)\,,
	\end{align}
	and the finite piece when $\delta\to0$:
		\begin{align}
		&\hat\cW_{1}^{1,\delta,i}(\vec{w}_1,\vec{w}_2,X_2)|_2:=\lim_{\delta\to
			0}\frac{v}{(x_{12}^2x_{34}^2)}\frac14\int_{\mathbb{R}^4}\frac{\dd^4X_1}{z_1^2}
		\prod\limits_{i=1}^2\frac{1}{\|\vec w_i-X_1|^2}\frac{K(\mathbf
			X_1,\mathbf X_2)}{1-K(\mathbf  X_1,\mathbf X_2)+\delta}\cr
		&\qquad\qquad\qquad
		=\frac{2\pi^2v^2}{(x_{12}^2x_{34}^2)^2}\prod\limits_{i=1}^2\frac{z_2}{\|\vec{w}_{i}-X_2|^2}\int_0^1\dd u\frac{\mathrm{arctanh(u)}}{4u^2
			+(1-u^2) \frac{\abs{\vec w_1-\vec w_2}^2
				4z_2^2}{\|\vec w_1-X_2|^2 \|\vec w_2-X_2|^2}}\,.
	\end{align}
	Thus the complete integral for the fish diagram is
	\begin{align}
		&\hat\cW_{1}^{\Delta,\delta,i}(\vec{w}_1,\vec{w}_2,X_2)=\hat\cW_{1}^{1,\delta,i}(\vec{w}_1,\vec{w}_2,X_2)|_1-\hat\cW_{1}^{1,\delta,i}(\vec{w}_1,\vec{w}_2,X_2)|_2\cr
		&\qquad=-\frac{\pi^2v}{(x_{12}^2x_{34}^2)}\prod\limits_{i=1}^2\frac{z_2}{\|\vec{w}_i-X_2|^2}\Bigg[\left(\log\left(
		\frac{z_2^2  \abs{\vec{w}_1-\vec{w}_2}^2}{\|\vec
			w_1-X_2|^2\|\vec w_2-X_2|^2}
		\right)+\log2\delta\right)\cr
		&\qquad\qquad\qquad\qquad\qquad\qquad\qquad\qquad-2\int_0^1\dd u\frac{\mathrm{arctanh(u)}}{4u^2
			+(1-u^2) \frac{\abs{\vec w_1-\vec w_2}^2
				4z_2^2}{\|\vec w_1-X_2|^2 \|\vec w_2-X_2|^2}}\Bigg]\,.\cr
	\end{align}
	Finally we attach the remaining bulk-to-boundary propagator to
	obtain the full one-loop diagram for $\Delta=1$
	\begin{align}
		\cW_1^{1,\delta,i}=\frac12\frac{\pi^2v}{(x_{12}^2x_{34}^2)}\int\limits_{\mathbb{R}^4}\dd^4 X_2\prod\limits_{i=1}^3&\frac{1}{\|\vec{w}_i-X_2|^2}\Bigg[2\int_0^1\dd u\frac{\mathrm{arctanh(u)}}{4u^2
			+(1-u^2) \frac{\abs{\vec w_1-\vec w_2}^2
				4z_2^2}{\|\vec w_1-X_2|^2 \|\vec w_2-X_2|^2}}\nonumber\\*&\qquad-\left(\log\left(\frac{z_2^2  \abs{\vec{w}_1-\vec{w}_2}^2}{\|\vec	w_1-X_2|^2\|\vec w_2-X_2|^2}
		\right)+\log2\delta\right)
		\Bigg],
	\end{align}
	with the result
	\begin{equation}
		\cW_{1}^{1,\delta,i}=-\pi^2\log\left(\frac{\delta}{2}\right)\cW_0^{1,\delta}+\frac{\pi^4v}{2x_{12}^2x_{34}^2}L_0^{1,i}+\frac{\pi^4v}{x_{12}^2x_{34}^2}{L_0'}^i\,.
	\end{equation}
	Restoring the prefactors, the complete one-loop diagram is then
		\begin{equation}
			\label{eq:oneloopD1_delta}
		W_1^{1,\delta}=\frac{2^4a^4\pi^2}{(2\pi)^{12}}\left(-3\log\left(\frac{\delta}{2}\right)\cW_0^{1,\delta}+\frac{\pi^2v}{2x_{12}^2x_{34}^2}\sum\limits_{i\in\{s,t,u\}}L_0^{1,i}+\frac{\pi^2v}{x_{12}^2x_{34}^2}\sum\limits_{i\in\{s,t,u\}}{L_0'}^{i}+\mathcal{O}(\delta)\right)\,,
	\end{equation}
	where $L_0^{\Delta,i}$ and ${L_0'}^i$ are given in~\ref{subsubsec:L0integrals} and~\ref{subsubsec:L0primeintegrals} respectively and $\cW_0^{\Delta,\delta}$ is the cross diagram evaluated in section~\ref{subsubsec:Cross_generaldim}.
	
	Note that since the finite terms in both regulariation schemes corresponding to $W_{1,\mathrm{fin}}^{\Delta,4,i}$ and the second term in ~\eqref{eq:fish} are the same, we can conclude immediately that $L_0'$ is the same in both regularisation schemes.

	\paragraph{Renormalisation:}
	\label{sec:renormalisation}
	As expected the UV divergent part is proportional to the cross
                diagram  and can therefore be absorbed in the coupling constant $\lambda$, which makes the coupling constant scale dependent.
	
	To understand how this works in the AdS-invariant regularisation we expand the regularised inverse geodesic distance around the coincidence points
	\begin{equation}
		\frac{K}{1+\delta}=\frac{1}{1+\delta}\frac{1}{\sqrt{1+a^2R^2}}\to1-\frac12 a^2R^2-\delta+\Op(a^4R^4,\delta^2),
	\end{equation}
	where $\delta$ is a dimensionless quantity and $R=\sqrt{(\mathbf X^0-\mathbf Y^0)^2+\cdots+(\mathbf X^3-\mathbf Y^3)^2}$. If we write it as $\delta=\frac12a^2r^2$ we see that this regularisation procedure corresponds to cutting out a ball of radius $r$ around the coinciding points. The quantity $a$ would be the renormalisation scale in usual flat space renormalisation theory, corresponding to the energy at which the physical scattering experiment is performed. In our case, where we are merely interested in boundary to boundary correlation functions, the only physically relevant length scale is the AdS radius and we can therefore identify $a$ with the inverse AdS radius.
	
	To perform the renormalisation we write the connected part of the four-point correlator up to order $\lambda^2$:
	\begin{equation}
		\label{eq:4point_connected}
		\lambda W_0^{\Delta,\delta}-\frac{\lambda^2}{2}W_1^{\Delta,\delta}=
		\frac{2^{4\Delta}a^4}{(2\pi)^{8}}\left[\lambda_R\cdot\left(1+\frac{3\lambda_R}{32\pi^2}\log\left(\frac{\delta}{2}\right)\right)\cW_0^{\Delta,\delta}+\text{finite terms}\right]\,.
	\end{equation}
	To absorb the divergent part, it is straightforward to see that we can choose a counterterm of the form
	\begin{equation}
		\delta\lambda\,\cW_0^{\Delta,\delta}=-\frac{3\lambda_R^2}{32\pi^2}\log\delta\,\cW_0^{\Delta,4}\,.
	\end{equation}
	The renormalised coupling is then related to the bare coupling $\lambda$ through
	\begin{equation}
		\lambda=\lambda_R-\frac{3\lambda_R^2}{2(4\pi)^2}\log\delta+\Op(\lambda_R^3)\,.
	\end{equation}
	This regularises the expression~\eqref{eq:4point_connected} up to order $\lambda_R^2$. The beta function can now be calculated as
	\begin{equation}
		\beta(\lambda)=-\frac{\dd\lambda}{\dd\log r}=\frac{3\lambda^2}{16\pi^2}+\Op(\lambda^3)\,,
	\end{equation}
	which is again consistent with the flat space $\lambda\phi^4$ theory. In this equation we used the fact that $\delta$ is defined as $\delta=\frac12 r^2 a^2$ as described above.
	
	Comparing~\eqref{eq:oneloopD1_delta} and~\eqref{eq:oneloopD2_delta} with~\eqref{eq:oneloopD2_dimreg} and~\eqref{eq:oneloopD1_dimreg} makes it clear that both regularisation schemes are equivalent up to addition 
	of a cross diagram $W_0^{\Delta}$. Since these are the tree-level contributions they can always be absorbed into
	the coupling constant by choosing a non-minimal subtraction scheme. 
	
	In the following we will choose our counter-term such that the finite piece only contains the $L_0^\Delta$ and ${L_0'}^\Delta$ terms. Therefore the renormalised one-loop contributions are given by:
	\begin{align}
		W_1^{1,\mathrm{ren}}&=\frac{2^4a^4\pi^4}{(2\pi)^{12}}\frac{v}{x_{12}^2x_{34}^2}\left(\frac12\sum\limits_{i\in\{s,t,u\}}L_0^{1,i}+\sum\limits_{i\in\{s,t,u\}}{L_0'}^{i}\right)\\
		W_1^{2,\mathrm{ren}}&=\frac{2^8a^4\pi^4}{(2\pi)^{12}}\frac{3v^2}{8(x_{12}^2x_{34}^2)^2}\sum\limits_{i\in\{s,t,u\}}L_0^{2,i}\,.
	\end{align}
	Note that this differs from the scheme used
   in~\cite{Heckelbacher:2020nue,Bertan:2018afl,Bertan:2018khc}
   where contributions from the
   cross diagram have been integrated into the finite piece. For the anomalous dimensions the effect of different renormalisation schemes can always be absorbed into a redefinition of the coupling constant, that is, a change in parametrization, as we will discuss in section~\ref{sec:conformal_block_expansion}.
	
%%%%%%%%%%%%%%%%%%%%%%%%%%%%%%%%%%%%%%%%%%%%%%%%%%%%%%%%%%
\subsection{Two loop diagrams}
\label{subsec:two_loop_calculation}
%%%%%%%%%%%%%%%%%%%%%%%%%%%%%%%%%%%%%%%%%%%%%%%%%%%%%%%%%%
To give an outlook on how to proceed to higher-loop integrals, we
   present the integral expressions
   of the two-loop contributions to
   the four-point function in terms
   of the euclidean propagators
   from
   section~\ref{sec:propagators}
   but  leave the evaluations of
   the integrals for future work.

There are two topologies contributing, which we will refer to as the necklace and the ice cream diagram.

\begin{figure}[ht]
  \centering
 \begin{tikzpicture}[baseline=(z)]
	\begin{feynman}[inline=(z)]	
	\tikzfeynmanset{every vertex=dot}
	\vertex (z);
	\vertex [above left=0.71cm and 0.71cm of z, label=180:$x_2$] (x2);
	\vertex [below left=0.71cm and 0.71cm of z, label=180:$x_1$] (x1);
	\vertex [above right=0.71cm and 0.71cm of z, label=0:$x_3$] (x3);
	\vertex [below right=0.71cm and 0.71cm of z, label=0:$x_4$] (x4);
	\vertex [left=0.4cm of z] (y1);
	\vertex [right=0.4cm of z] (y2);
	\tikzfeynmanset{every vertex={empty dot,minimum size=0mm}}
	\vertex [left=0.2cm of z] (z1);
	\vertex [right=0.2cm of z] (z2);	
	\diagram* {
		(x2)--(y1),
		(x1)--(y1),
		(x3)--(y2),
		(x4)--(y2),
	};
	\end{feynman}
	\begin{pgfonlayer}{bg}
	\draw[blue] (z) circle (1cm);
	\draw (z1) circle (0.2cm);
	\draw (z2) circle (0.2cm);
	\end{pgfonlayer}
      \end{tikzpicture}\qquad\begin{tikzpicture}[baseline=(z)]
	\begin{feynman}[inline=(z)]
	\vertex (z);	
	\tikzfeynmanset{every vertex=dot}
	\vertex [above left=0.71cm and 0.71cm of z, label=180:$x_2$] (x2);
	\vertex [below left=0.71cm and 0.71cm of z, label=180:$x_1$] (x1);
	\vertex [above right=0.71cm and 0.71cm of z, label=0:$x_3$] (x3);
	\vertex [below right=0.71cm and 0.71cm of z, label=0:$x_4$] (x4);
	\vertex [left=0.4cm of z] (y1);
	\vertex [above right=0.17cm and 0.17cm of z] (y2);
	\vertex [below right=0.17cm and 0.17cm of z] (y3);
	\tikzfeynmanset{every vertex={empty dot,minimum size=0mm}}
	\vertex [left=0.2cm of z] (z1);
	\vertex [right=0.2cm of z] (z2);	
	\diagram* {
		(x2)--(y1),
		(x1)--(y1),
		(y1)--(y2),
		(y1)--(y3),
		(x3)--(y2),
		(x4)--(y3),
	};
	\end{feynman}
	\begin{pgfonlayer}{bg}
	\draw[blue] (z) circle (1.05cm);
	\draw (z2) circle (0.2cm);
	\end{pgfonlayer}
	\end{tikzpicture}
  \caption{One channel of the two-loop Necklace (left) and Ice cream (right) diagram. The other channels can be obtained by permutations of the boundary points.}
  \label{fig:twoloopNecklaceW}
\end{figure}
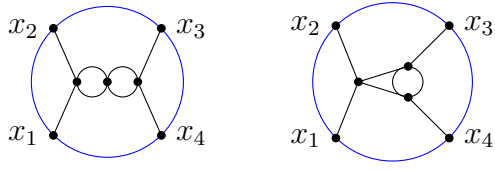
The necklace diagram is depicted in figure~\ref{fig:twoloopNecklaceW} on the left. In dimensional regularisation it leads to the integral
\small{\begin{multline}
	\cW^{\Delta,D}_{2,\circ\circ}(\zeta,\zetab)=\frac{v^\Delta}{8(x^2_{12}x^2_{34})^\Delta}\int\limits_{(\mathbb{R}^D)^3}\prod\limits_{i=1}^{3}
	\frac{\dd^{D} X_i}{(u\cdot X_i)^4\norm{X_i-u_1}^{2(D-4)}}f^{\Delta}_{\circ\circ}(X_1,X_3;\zeta,\bar{\zeta})\nonumber\\*
	\left(\frac{(u\cdot X_1)^2(u\cdot X_2)^2}{\norm{X_1-X_2}^4}+\frac{(-1)^{\Delta}}{2}\frac{(u\cdot X_1)(u\cdot X_2)}{\norm{X_1-X_2}^2}\right)
	\left(\frac{(u\cdot X_2)^2(u\cdot X_3)^2}{\norm{X_2-X_3}^4}+\frac{(-1)^{\Delta}}{2}\frac{(u\cdot X_2)(u\cdot X_3)}{\norm{X_2-X_3}^2}\right),
\end{multline}}
where the bulk-to-boundary part
\begin{multline}
   f^\Delta_{\circ\circ}(X_1,X_3;\zeta,\zetab)= 	\frac{(u\cdot X_1)^{2\Delta} (u\cdot X_3)^{2\Delta}}{\|X_1|^{2\Delta}\|X_{3}-u_1|^{2\Delta} 
			\|X_{3}-u_\zeta|^{2\Delta}}\cr \frac{(u\cdot X_1)^{2\Delta} (u\cdot X_3)^{2\Delta}}{\|X_3|^{2\Delta}\|X_{3}-u_1|^{2\Delta} 
			\|X_{1}-u_\zeta|^{2\Delta}}+\frac{(u\cdot X_1)^{2\Delta} (u\cdot X_3)^{2\Delta}}{\|X_3|^{2\Delta}\|X_{1}-u_1|^{2\Delta} 
			\|X_{3}-u_\zeta|^{2\Delta}},
\end{multline}
is the same as for the one loop diagram in equation~\eqref{eq:OneLoop_diagram_dim_reg}. 

The ice-cream diagram is depicted in figure~\ref{fig:twoloopNecklaceW} on the right. In dimensional regularisation it corresponds to the integral
\begin{multline}
		\cW^{\Delta,D}_{2,\triangleleft\circ}(\zeta,\zetab)=\frac{v^\Delta}{8(x^2_{12}x^2_{34})^\Delta}\int\limits_{(\mathbb{R}^D)^3}\prod\limits_{i=1}^{3}
	\frac{\dd^{D} X_i}{(u\cdot X_i)^4\norm{X_i-u_1}^{2(D-4)}}f^{\Delta}_{\triangleleft\circ}(X_1,X_2,X_3;\zeta,\bar{\zeta})\cr
\times	\frac{(u\cdot X_1)^4(u\cdot X_2)^2(u\cdot X_3)^2}{\norm{X_1-X_2}^4\norm{X_1-X_3}^4}\left(\frac{(u\cdot X_2)^2(u\cdot X_3)^2}{\norm{X_2-X_3}^4}+\frac{(-1)^{\Delta}}{2}\frac{(u\cdot X_2)(u\cdot X_3)}{\norm{X_2-X_3}^2}\right),
\end{multline}
where the bulk-to-boundary part can easily be read-off from the general formula ~\eqref{eq:Witten_bulk_to_boundary_tranformed}.

It is easy to see, that when $D$ approaches 4, these diagrams  diverge like $(D-4)^{-2}$, with coefficients proportional to  the cross diagram, and a sub-leading divergence of order $(D-4)^{-1}$ proportional to the one-loop Witten diagram.  In order to restore the AdS invariance of the renormalised   four-point function,  we will need to evaluate these  divergences in $D=4-4\epsilon/3$ dimensions.

In principle, solving these integrals can be done by following the
   same steps as for the one-loop case, the main difference being that
   the integrals are more complicated and that we will have elliptic
   polylogarithms appearing for the $\Delta=2$ case in the necklace diagram integrals. For
   $\Delta=1$ we meet integrals beyond elliptic integrals whose
   analysis is beyond the scope of the present work.

%%%%%%%%%%%%%%%%%%%%%%%%%%%%%%%%%%%%%%%%%%%%%%%%%%%%%%%%%%
\section{Discontinuities and unitarity of Witten diagrams}
\label{sec:unitarity_methods}
%%%%%%%%%%%%%%%%%%%%%%%%%%%%%%%%%%%%%%%%%%%%%%%%%%%%%%%%%%
In this section we discuss how unitarity can be used to extract the
   prefactors of the $\log(v)^n$ terms in Witten diagrams, by
   calculating the discontinuity in $v$.

\subsection{Discontinuities}
\label{sec:discontinuities}

   On general grounds, to any
   loop order the Witten diagrams have a small $v$ expansion  of the form
\begin{equation}
	\mathcal{W}^\Delta_L(v,Y)=\frac{1}{2^{L+1}}\frac{v^\Delta}{(x^2_{12}x^2_{34})^{\Delta}}\sum\limits_{n=0}^{L+1}\log^n(v) p_L^{(n)}(v,Y;\Delta)+O(v)\,,
\end{equation}
where $p^{(n)}_L(v,Y;\Delta)$  is an analytic function in $v$ and $Y$ for $v$  and
   $Y$ small. The (sequential) discontinuity in $v$ of the Witten diagram  is therefore contained in the $\log^n(v)$ terms. More precisely, 
\begin{align}
	\label{eq:discontinuity_W}
	\mathrm{Disc}_{v}\cW_L^\Delta(v,Y)&=\frac{1}{2^{L+1}}\frac{v^\Delta}{(x^2_{12}x^2_{34})^{\Delta}}\sum\limits_{n=1}^{L+1}\mathrm{Disc}_v\left(\log^n(v) \right)p_L^{(n)}(v,Y;\Delta)\,.
\end{align}
The discontinuity of a function $f(v)$ is defined by
         \begin{equation}\label{e:discDef}
           \mathrm{Disc}_vf(v\pm i0):=\lim_{\varepsilon\to0} \left(f(v+i\varepsilon)-f(v-i\varepsilon)\right).   
         \end{equation}
We use the \emph{principal branch} for the logarithm which is a
   continuous function on the complex plane except for the negative real axis.         
Thus, the discontinuities of $\log(v)$ and $\log^2(v)$ are 
\begin{align}
	\mathrm{Disc}_v\log(v)&=\lim_{\varepsilon\to0}\left(\log(v+i\varepsilon)-\log(v-i\varepsilon)\right)=2\pi i\Theta(-v)\,,\\
	\mathrm{Disc}_v\log^2(v)&=4\pi i\Theta(-v)\log(\abs{v})\,,
\end{align}
 while the sequential double discontinuity is given by
 \begin{align}
	\mathrm{Disc}_v\mathrm{Disc}_v\log(v)&=0\,,\\
\mathrm{Disc}_v	\mathrm{Disc}_v\log^2(v)&=2(2\pi i)^2\Theta(-v)\,.
\end{align}                                 
Here we are only concerned with Witten diagrams up to loop order
         $L=1$, therefore only
         terms which are maximally
         quadratic in $\log(v)$ can
       appear.
In this case the (sequential) discontinuities with respect to $v$,
       applied to the one-loop
       Witten diagrams
       in~\eqref{eq:discontinuity_W},
       lead to 
\begin{align}
  \label{eq:discontinuity_1loop}
  	\mathrm{Disc}_v\cW_0^\Delta(v,Y)&=\frac{1}{2}\frac{v^\Delta}{(x^2_{12}x^2_{34})^{\Delta}}2\pi
       i\Theta(-v) p_0^{(1)}(v,Y;\Delta),\cr
       	\mathrm{Disc}_v\cW_1^\Delta(v,Y)&=\frac{1}{4}\frac{v^\Delta}{(x^2_{12}x^2_{34})^{\Delta}}2\pi i\Theta(-v)\left(2\log(\abs{v})p_1^{(2)}(v,Y;\Delta)+p_1^{(1)}(v,Y;\Delta)\right),\cr
  \mathrm{Disc}_v
  \mathrm{Disc}_v\cW_1^\Delta(v,Y)&=\frac{1}{2}\frac{v^\Delta}{(x^2_{12}x^2_{34})^{\Delta}}(2\pi
 i)^2\Theta(-v) p_1^{(2)}(v,Y;\Delta).
\end{align}
From these expressions we can read-off the coefficients of $\log(v)^2$ and $\log(v)$ which, in turn, provide us with the information about the second order anomalous dimensions of the double-trace operators of the boundary theory.

As we will discuss in section~\ref{sec:conformal_block_expansion} a direct consequence of the conformal symmetry at the boundary is the fact, that the sequential discontinuities of the Witten diagrams can be expanded in terms of conformal blocks of a generalized free field
\begin{align}
    \frac{1}{2\pi i}\mathrm{Disc}_v\cW_0^\Delta=\sum_{n,l\geq0}c_{0,n,l}^\Delta G_{\Delta_{n,l}};\qquad
    \frac{1}{2(2\pi i)^2}\mathrm{Disc}_v\mathrm{Disc}_v\cW_1^\Delta=\sum_{n,l\geq0}c_{1,n,l}^\Delta G_{\Delta_{n,l}}\,.
\end{align}
and furthermore, that the expansion coefficients of the renormalised Witten diagrams are related by the simple relation
\begin{equation}
    \label{eq:unitarity_coeffs}
    c_{1,n,l}^\Delta=-\frac14\left(c_{0,n,l}^\Delta\right)^2\,,
\end{equation}
This relation (and its generalisation to  higher-loop order) follows directly from the way the perturbative bulk interactions generate the anomalous dimensions in~\eqref{e:Deltacorrect}. For example in the $\Delta=2$ case, since $c_{1,n,l}^2=c_{0,n,l}^2=1$, we have the following relation between the discontinuities of the tree-level and one-loop Witten diagram 
\begin{align}
    \frac{1}{2\pi i}\mathrm{Disc}_v\cW_0^\Delta(v,Y)=-\frac14\frac{1}{2(2\pi i)^2}\mathrm{Disc}_v\mathrm{Disc}_v\cW_1^\Delta(v,Y)\,.
\end{align}

In the following we will show how to use the relation between the sequential discontinuities and
multiple unitarity cuts developed in~\cite{Cutkosky:1960sp,Abreu:2014cla,Bourjaily:2020wvq} for  flat space Feynman integrals in momentum space to extract the coefficient of the $\log(v)$. We will demonstrate the success of the method with two examples and compare them to our exact results from section~\ref{sec:calculation_witten_diagram}. Note that we did not have to use this method, since we were able to solve the integrals for the Witten diagrams exactly. However, for higher loops and different conformal weights $\Delta$, where solving the integrals exactly might be more challenging, this method could turn out to be useful.

\subsection{Unitarity cuts}
\label{sec:unitarity-cuts}
We notice that we can interpret the dimensionally regulated  $L$-loop
  Witten diagrams
    in~\eqref{eq:Witten_diagram_dim_reg}
     as three-point momentum Feynman
     integrals in flat space, with
    external ``momenta'' $k_1=\u1-\uZ,
       k_2=\uZ$ and $k_3=-\u1$ where we
   integrate over $L+1$ loop momenta.

  Because of this interpretation, we want to apply the relation between the discontinuity of the Witten
  diagrams with
  respect to the variable $v$ and
  unitarity cuts $
  \mathrm{Disc}_v W_L^\Delta(v,Y) =
  \mathrm{Cut}W_L^\Delta(v,Y)$ along the
  lines of~\cite{Cutkosky:1960sp,Abreu:2014cla}.
For being able to apply the standard methods of calculating the
 Cutkosky discontinuities to the
 Witten diagrams, we need
 to perform a Wick rotation to go
 to Lorentzian AdS, meaning, that
 in this section the conformal flat
 propagator in~\eqref{eq:Gdef} is
 given by
\begin{equation}
			G(X,Y):=\frac{zw}{\|X-Y|^2-i\varepsilon},
                        \qquad \|X-Y|^2=(X_1-Y_1)^2-\sum\limits_{i=2}^4(X_i-Y_i)^2,
\end{equation}
 and 
$\uZ=\frac12(\zeta+\zetab,\zeta-\zetab,0,0)$. 
We have introduced a Feynman  $-i\varepsilon$ prescription following~\cite{Avis:1977yn},  which provides the correct flat limit. 
  
We only consider the case $\Delta=1$ because the $\Delta=2$ case is obtained by acting with $\cH_{1234}$ introduced in section~\ref{subsec:diff_operator}.  

\subsubsection{Unitarity cuts of the cross Witten diagram}
\label{sec:unitairity-cut-cross}
 
\begin{figure}[ht]
	\begin{center}
		% S Channel diagram
		$\begin{tikzpicture}[baseline=(z)]
			\begin{feynman}[inline=(z)]
				\tikzfeynmanset{every vertex=dot}
				\vertex  (x1);
				\vertex [below right=0.75cm and 1.5cm of x1](x2);
				\vertex [below=1.5cm of x1]  (x3);
				\tikzfeynmanset{every vertex={empty dot,minimum size=0mm}}
				\vertex [above left=0.5cm and 2cm of x1] (p1);
				\vertex [right=2cm of x2] (p2);
				\vertex [below left=0.5cm and 2cm of x3] (p3);
				\vertex [above right=0.5cm and 0.6cm of x1](d1);
				\vertex [below=2.5cm of d1](d2);
				\diagram* {
					(p1)--[fermion, edge label={$k_1$}](x1),
					(p2)--[fermion, edge label'={$k_2$}](x2),
					(p3)--[fermion, edge label'={$k_3$}](x3),
					(x2)--[fermion, edge label'={\footnotesize$k_2-l$}](x1),
					(x2)--[fermion, edge label={\footnotesize$l$}](x3),
					(x3)--[fermion, edge label={\footnotesize$k_3+l$}](x1),
					(d1)--[color=red](d2),
				};
			\end{feynman}
		\end{tikzpicture}$
		\caption{Cross diagram as a flat space three point function with $k_1=u_1-\uZ$, $k_2=\uZ$, $k_3=-u_1$ and $l=X$. The red line corresponds to the unitarity cut in the $k_2^2=\zeta\zetab$-channel.}
		\label{fig:CrossToTriangle}
	\end{center}
\end{figure}
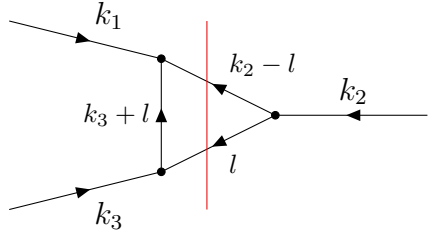

As an example, consider the tree-level cross diagram in AdS from equation~\eqref{eq:Cross_dimreg}. Identifying the bulk point $X$ with the loop momentum $l$, this is equivalent to the flat space diagram depicted in figure~\ref{fig:CrossToTriangle}.

We are interested in the unitarity cut with respect to $k_2^2=\uZ^2=\zeta\zetab=v$. The corresponding cut we have to perform is indicated in figure~\ref{fig:CrossToTriangle}. The cut diagram is now given by~\cite{Abreu:2014cla}
\begin{align}
	\mathrm{Cut}_{\uZ}\cW_0^{1,4-4\epsilon}=\frac12\frac{v}{x^2_{12}x^2_{34}}(2\pi i)^2\int\dd^{4-4\epsilon}X \frac{\delta^+(\|X|^2)\delta^+(\|X-\uZ|^2)}{\left(\|X-u_1|^2-i\varepsilon\right)^{1-4\epsilon}}\,,
\end{align} 
where $\delta^+(\|X|^2)=\delta(\|X|^2)\Theta(X_1)$. We parametrize the
 loop momentum $X$ by ${X=(x_0,r\cos\theta,0,r\sin\theta)}$. The integration measure is then given by
\begin{align}
	\int\limits_{\mathbb{R}^4}\dd^4 X\delta^+(\|X|^2)&=2\pi^{1-2\epsilon}\eul^{-2\gamma\epsilon}\int\limits_0^\infty\dd x_0\int\limits_0^\infty\dd r r^{2-4\epsilon}\int\limits_{-1}^{+1}\dd\cos\theta\delta(x_0^2-r^2)\,.
\end{align}
With this, the cut diagram becomes
\begin{align}
    \label{eq:Cut_Cross}
	\mathrm{Cut}_{\uZ}\cW_0^{1,4-4\epsilon}&=\frac{(2\pi)^3}{4}\frac{(\pi\eul^\gamma)^{-2\epsilon}v}{x^2_{12}x^2_{34}}\int\limits_0^\infty\dd x_0\int\limits_{-1}^{+1}\dd\cos\theta x_0^{1-4\epsilon}(\sin\theta)^{4\epsilon}\frac{\delta\left(\zeta\zetab-x_0\left(\zeta+\zetab-\cos\theta(\zeta-\zetab)\right)\right)}{(1-2x_0)^{1-4\epsilon}}\cr
	&=\frac{(2\pi)^3}{4}\frac{(\pi\eul^\gamma)^{-2\epsilon}v}{x^2_{12}x^2_{34}}\int\limits_{-1}^{+1}\dd x\frac{(1-x^2)^{-2\epsilon}(\zeta\zetab)^{1-4\epsilon}}{(\zeta+\zetab-x(\zeta-\zetab))(\zeta+\zetab-2\zeta\zetab-x(\zeta-\zetab))^{1-4\epsilon}}\,,
\end{align}
which evaluates to 
\begin{equation}
    \label{eq:Cut_Cross_res}
	\mathrm{Cut}_{\uZ}\cW_0^{1,4-4\epsilon}=-\frac{v\pi^3}{x^2_{12}x^2_{34}}\frac{1}{(\zeta-\zetab)}\log\left(\frac{1-\zeta}{1-\zetab}\right)+\Op(\epsilon)\,,
\end{equation}
where the $\Op(\epsilon)$ term is given in the appendix by
   equation~\eqref{eq:Cut_Cross_full}. 
Comparing the $\Op(\epsilon^0)$ expression to~\eqref{eq:discontinuity_W} we see that the coefficient of $\log(v)$ is given by
\begin{equation}
	p_0^{(1)}(v,Y)=\frac{x^2_{12} x^2_{34}}{v}\frac{1}{2\pi i} \mathrm{Cut}_{\uZ}\cW_0^{1,4}=\frac{i\pi^2}{2}\frac{1}{\zeta-\zetab}\log\left(\frac{1-\zeta}{1-\zetab}\right)\,,
\end{equation}
which coincides with the exact calculation in~\eqref{eq:crossDelta1}
                    up to the additional factor of $i$ which is due to
                    the Lorentzian signature. This is a direct
                    verification of the relation between the $v$
                    discontinuities and the unitarity cuts.

The result for                    $\Delta=2$ can easily be obtained by acting with
                    $\cH_{1234}$ on the $\Delta=1$ result, since there
                    are no terms in the Witten diagram  that would produce extra $\log(v)$ terms due to differentiation. 

\subsubsection{Unitarity cuts of the one-loop Witten diagram}
\label{sec:unitairity-cut-oneloop}

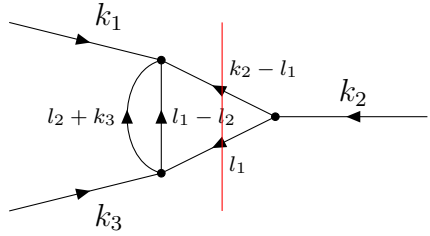
\begin{figure}[ht]
	\begin{center}
		% S Channel diagram
		$\begin{tikzpicture}[baseline=(z)]
			\begin{feynman}[inline=(z)]
				\tikzfeynmanset{every vertex=dot}
				\vertex  (x1);
				\vertex [below right=0.75cm and 1.5cm of x1](x2);
				\vertex [below=1.5cm of x1]  (x3);
				\tikzfeynmanset{every vertex={empty dot,minimum size=0mm}}
				\vertex [above left=0.5cm and 2cm of x1] (p1);
				\vertex [right=2cm of x2] (p2);
				\vertex [below left=0.5cm and 2cm of x3] (p3);
				\vertex [above right=0.5cm and 0.8cm of x1](d1);
				\vertex [below=2.5cm of d1](d2);
				\diagram* {
					(p1)--[fermion, edge label={$k_1$}](x1),
					(p2)--[fermion, edge label'={$k_2$}](x2),
					(p3)--[fermion, edge label'={$k_3$}](x3),
					(x2)--[fermion, edge label'={\scriptsize$k_2-l_1$}](x1),
					(x2)--[fermion, edge label={\scriptsize$l_1$}](x3),
					(x3)--[out=160, in=200, fermion, edge label={\scriptsize$l_2+k_3$}](x1),
					(x3)--[fermion, edge label'={\scriptsize$l_1-l_2$}] (x1),
					(d1)--[color=red](d2),
				};
			\end{feynman}
		\end{tikzpicture}$
		\caption{One-loop s-channel diagram as a two-loop flat space three point function with $k_1=u_1-\uZ$, $k_2=\uZ$, $k_3=-u_1$, $l_2=X_2$ and $l_1=X_1$. The red line corresponds to the unitarity cut in the $k_2^2=\zeta\zetab$ channel.}
		\label{fig:triangle_loop_s}
	\end{center}
\end{figure}

The same method can be applied at one loop, given by the
                    integrals~\eqref{eq:W1_fin_div}. As an example we
                    consider the divergent part of the $s$-channel
                    diagram given by
                    $\cW_{1,\mathrm{div}}^{1,4-2\epsilon,s}$.  The
                    corresponding flat space diagram is now given by a
                    two-loop momentum space integral depicted in
                    figure~\ref{fig:triangle_loop_s}. 

The discontinuity in $v$ can then be calculated by performing the cut as shown in figure~\ref{fig:triangle_loop_s} and we get
\begin{align}
	\mathrm{Cut}_{\uZ}\cW_{1,\mathrm{div}}^{1,4-2\epsilon,s}=&\frac14\frac{(2\pi
           i)^2v}{x^2_{12}x^2_{34}}\int\dd^{4-2\epsilon}X_1\dd^{4-2\epsilon}X_2 \frac{\delta^+(\|X_1|^2)\delta^+(\|X_1-\uZ|^2)\|X_1-u_1|^{4\epsilon}}{(\|X_2-u_1|^2)^{1-2\epsilon}(\|X_1-X_2|^2)^2}\cr
	=&-\frac{\pi^{2-\epsilon}\Gamma(\epsilon)}{\Gamma(1-2\epsilon)}\frac14\frac{(2\pi
           i)^2v}{x^2_{12}x^2_{34}}\int\dd^{4-2\epsilon}X_1\frac{\delta^+(\|X_1|^2)\delta^+(\|X_1-\uZ|^2)}{(\|X_1-u_1|^2)^{1-3\epsilon}}\,,
\end{align}
evaluating the delta-function constraints we have
\begin{align}
	\mathrm{Cut}_{\uZ}\cW_{1,\mathrm{div}}^{1,4-2\epsilon,s}=&-\frac{\pi^{4-2\epsilon}\Gamma(\epsilon)}{\Gamma(1-2\epsilon)\Gamma(1-\epsilon)}\frac14\frac{(2\pi
           i)^2v}{x^2_{12}x^2_{34}}\cr
           &\times\int\limits_{-1}^{+1}\dd x\frac{(1-x^2)^{-\epsilon}(\zeta\zetab)^{1-2\epsilon}}{(\zeta+\zetab-x(\zeta-\zetab))^{1+\epsilon}(\zeta+\zetab-2\zeta\zetab-x(\zeta-\zetab))^{1-3\epsilon}}\cr
	=&-\pi^{4-2\epsilon}\eul^{-4\gamma\epsilon}\frac14\frac{(2\pi
           i)^2v}{x^2_{12}x^2_{34}}\left[\frac{1}{\epsilon}I_{1,\mathrm{div}}^{1}+I_{1,\mathrm{div}}^{1,\epsilon}+\Op(\epsilon)\right]\,.
\end{align}
By comparing the integrand with
           equation~\eqref{eq:Cut_Cross} it is obvious, that
           $I_{1,\mathrm{div}}^{1}$ is given by the $\epsilon^0$ term
           of the cut of the  cross diagram in that equation. The expression for $I_{1,\mathrm{div}}^{1,\epsilon}$ is given in the appendix by equation~\eqref{eq:Cut_oneloop}.

The coefficient of the $\log(v)$ term of the uncut diagram can be extracted from this by comparing $I_{1,\mathrm{div}}^{1,\epsilon}$ to equation~\eqref{eq:discontinuity_1loop}.
\begin{equation}
	p_1^{(1)}(v,Y)=\frac{2}{i\pi} \left.I_{1,\mathrm{div}}^{1,\epsilon}\right\vert_{\log(v=\zeta\zetab)=0}.
\end{equation}
Comparing to the exact result in equation \eqref{eq:W1div_res} we see
                  that the $\log(v)$ coefficients coincide, which is direct
                    verification of the relation between the $v$
                    discontinuities and the unitarity cuts.

This method can be applied to all other integrals to extract the $\log(v)$ coefficients. As mentioned above we will not proceed here since we were able to calculate the exact expressions. We merely want to propose this technique, since it might be useful in future work to go to higher-loop orders, where calculating the exact expressions is much harder.

We note, in passing, that this approach differs from the AdS unitarity methods developed in~\cite{Liu:2018jhs,Meltzer:2019nbs,Meltzer:2020qbr,Meltzer:2021bmb} where the double discontinuity of a Witten diagram is calculated using the split representation of the propagator and the Lorentzian inversion formula~\cite{Caron-Huot:2017vep}. 
 While that method generalizes straightforwardly to general $\Delta$ and gives the result in terms of conformal blocks right away, it is much harder to compute anomalous dimensions beyond tree-level since they would involve cuts in the external bulk to boundary propagators.

In the language of~\cite{Meltzer:2019nbs} we are performing external cuts and therefore calculate the single discontinuity, which lets us extract the information about loop corrections to the anomalous dimensions.

%%%%%%%%%%%%%%%%%%%%%%%%%%%%%%%%%%%%%%%%%%%%%%%%%%%%%%%%%%
\section{Conformal block expansion}
\label{sec:conformal_block_expansion}
%%%%%%%%%%%%%%%%%%%%%%%%%%%%%%%%%%%%%%%%%%%%%%%%%%%%%%%%%%
In order to extract the conformal dimensions of the ``double-trace''
           operators in the conformal field dual to $\phi^4$-theory in
           AdS we now compare the bulk calculation of the latter to
           the conformal block expansion of the former. First, let us
           note that the free part of the four-point correlation function,
           i.e. the disconnected part of
           equation~\eqref{eq:four_point_Wittendiagrams} has the form
           of a generalized free field, meaning that it consists of
           the sum over all permutations of products of two point
           functions, but no classical CFT action exists which would
           generate these two-point correlation functions
\begin{align}
\langle\Op_\Delta(\vec x_1)\Op_\Delta(\vec x_2)\rangle=\lim\limits_{z_1,z_2\to0}(z_1z_2)^{-\Delta}\Lambda(\mathbf{X}_1,\mathbf{X}_2,\Delta)=2^\Delta\mathcal{N}_\Delta\frac{1}{(x_{12}^{2})^{\Delta}}\,.
\end{align}
Summing over the three permutations, as shown in
           equation~\eqref{eq:four_point_Wittendiagrams}, the
           disconnected part of the four-point correlation function becomes
\begin{align}
\label{eq:generalized_free_field}
\langle\Op_\Delta(\vec{x}_1)\Op_\Delta(\vec{x}_2)\Op_\Delta(\vec{x}_3)\Op_\Delta(\vec{x}_4)\rangle_{\mathrm{disc}}&=
\frac{2^{2\Delta}\mathcal{N}_\Delta^2}{(x^2_{12}x^2_{34})^{\Delta}}\left(1+v^\Delta+\left(\frac{v}{1-Y}\right)^\Delta\right)\\
\nonumber &=\frac{2^{2\Delta}\mathcal{N}_\Delta^2}{(x^2_{12}x^2_{34})^{\Delta}}\left(1+v^\Delta\left(2+\sum\limits_{n=1}^\infty\frac{\Gamma(\Delta+n)}{\Gamma(\Delta)\Gamma(n+1)}Y^n\right)\right)\,.
\end{align}
In the last step we expand in $\vec x_1\to\vec x_2$ and $\vec x_3\to
            \vec x_4$, which translates into a small $v$ and $Y$ expansion
\begin{equation}v=\frac{x_{12}^2x_{34}^2}{x_{14}^2x_{23}^2}, \qquad Y=1- \frac{x_{13}^2x_{24}^2}{x_{14}^2x_{23}^2}.
\end{equation}
From the perspective of the CFT this corresponds to the double operator product expansion (OPE)
\begin{align}
&\Op_\Delta(\vec x_1)\Op_\Delta(\vec x_2)=\sum\limits_{\tilde\Op}a_{\Delta_{\tilde\Op}}\mathcal{D}_{\tilde\Op}(x_{12},\partial_2)\tilde{\Op}(\vec x_2)\,,\cr
&\Op_\Delta(\vec x_3)\Op_\Delta(\vec x_4)
=\sum\limits_{\tilde\Op}a_{\Delta_{\tilde\Op}}\mathcal{D}_{\tilde\Op}(x_{34},\partial_4)\tilde{\Op}(\vec x_4)\,,
\end{align}
where $\mathcal{D}_{\tilde\Op}(x_{ij},\partial_i)$ is a differential operator given by a power series in $\partial_i$ of the form
\begin{equation}
    \mathcal{D}_{\tilde\Op}(x_{ij},\partial_j)=(x^2_{ij})^{-\Delta+\frac12\Delta_{\tilde\Op}}
    \left(1+a \, x_{ij}\cdot\partial_j+b\, x^2_{ij}\partial_j^2+\cdots\right),
\end{equation}
where the expansion coefficients $a,b,\dots$ are completely fixed by conformal symmetry.\\
The four-point function then becomes
\begin{align}
&\langle\Op_\Delta(\vec x_1)\Op_\Delta(\vec x_2)\Op_\Delta(\vec x_3)\Op_\Delta(\vec x_4)\rangle=
\sum\limits_{\tilde{\Op},\tilde{\tilde{\Op}}}a_{\Delta_{\tilde\Op}}a_{\Delta_{\tilde{\tilde{\Op}}}}\mathcal{D}(x_{12},\partial_2)\mathcal{D}(x_{34},\partial_4)\langle\tilde\Op(\vec x_2)\tilde{\tilde{\Op}}(\vec x_4)\rangle\nonumber\\*
&\qquad\qquad\qquad\qquad\qquad\qquad\qquad=\frac{2^{2\Delta}\mathcal{N}_\Delta^2}{(x_{12}^2x^2_{34})^\Delta}\left(1+\sum\limits_{\tilde\Op}A_{\Delta_{\tilde\Op}}G_{\Delta_{\tilde\Op},l}(v,Y)\right)  ,  
\end{align}
where we used that $\langle\tilde\Op(\vec x_2)\tilde{\tilde{\Op}}(\vec x_4)\rangle$ vanishes for $\tilde\Op\neq\tilde{\tilde{\Op}}$. Here $G_{\Delta_{\tilde\Op},l}(v,Y)$ are conformal blocks, see e.g.~\cite{Dolan:2000ut}, that contain the information about the entire multiplet of a primary operator $\tilde\Op$ and its descendants appearing in the OPE. They are eigenfunctions of the quadratic Casimir of the conformal group and depend on the conformal dimension $\Delta_{\tilde\Op}$ and the spin $l$ of $\tilde\Op$. In three dimensions the conformal blocks can be obtained from the formula for general dimensions, which has been calculated in~\cite{Li:2019cwm}. We list the relevant formula from this calculation in appendix~\ref{sec:OPE}. In the following we will refer to $A_{\Delta_\Op}\equiv a^2_{\Delta_\Op}$ as the OPE coefficients. 
The normalization of the expansion is fixed by our bulk theory.

For a generalized free field the conformal block expansion can be determined exactly: The spectrum of primary ``double-trace'' operators is given by  $:\Op_\Delta\square^n\partial^l\Op_\Delta:$, with conformal dimension $\Delta_{(n,l)}=2\Delta+2n+l$, where $n,l/2\in\mathbb{N}$. The OPE coefficients $A_{n,l}$ for these operators are known as well~\cite{Fitzpatrick:2011dm} and given in appendix~\ref{sec:OPE}. We can therefore immediately write down the conformal block expansion for the generalized free field
\begin{equation}\label{e:ConfBlockExpand}
    \langle\Op_\Delta(\vec x_1)\Op_\Delta(\vec x_2)\Op_\Delta(\vec x_3)\Op_\Delta(\vec x_4)\rangle
=\frac{2^{2\Delta}\mathcal{N}_\Delta^2}{(x_{12}^2x^2_{34})^\Delta}\left(1+\sum\limits_{n,l}A_{n,l} G_{\Delta_{(n,l)},l}(v,Y)\right)\,.
\end{equation}

By adding the interaction term $\lambda\phi^4$ in the bulk we deform the four-point function, such that the deformation is parametrized by an expansion in the renormalized bulk coupling constant $\lambda_R$. From the calculation in section~\ref{sec:calculation_witten_diagram} we obtained the following four-point function up to $\Op(\lambda_R^2)$:
\begin{multline}
\label{eq:fourpoint_deformed}
    \langle\Op_\Delta(\vec{x}_1)\Op_\Delta(\vec{x}_2)\Op_\Delta(\vec{x}_3)\Op_\Delta(\vec{x}_4)\rangle
    =\frac{2^{2\Delta}\mathcal{N}_\Delta^2}{(x^2_{12}x^2_{34})^{\Delta}}\left[1+v^\Delta\left(2+\sum\limits_{n=1}^\infty\frac{\Gamma(\Delta+n)}{\Gamma(\Delta)\Gamma(n+1)}Y^n\right.\right.\cr
    \left.\left.-\frac{\lambda_R}{(4\pi)^2}\frac{2^{2\Delta}\sqrt{\pi}}{2\Gamma(\frac52-2\Delta)\Gamma(\Delta)^2}I_\times^\Delta(v,Y)
    +\frac{\lambda_R^2}{(4\pi)^4}\sum\limits_{i\in\{s,t,u\}}
    \left\{\begin{array}{cl}
        L_0^{1,i}+2{L'_0}^i &\text{for }\Delta=1  \\
        3L_0^{2,i} &\text{for }\Delta=2 
    \end{array}\right.\right)\right]\,,
\end{multline}

From the CFT side the deformation generated by the bulk interaction term generates anomalous dimensions for the double-trace operators 
\begin{equation}\label{e:Deltacorrect}
    \Delta_{(n,l)}\to\Delta_{(n,l)}+\sum\limits_{p=0}^\infty\gamma_{n,l}^{(p)}(\Delta)\,,
\end{equation}
where $\gamma_{n,l}^{(p)}(\Delta)$ is of order $\lambda_R^p$ in the renormalized bulk coupling constant $\lambda_R$.  
In order to match the conformal block expansion to the deformed
                     four-point correlation function in equation~\eqref{eq:fourpoint_deformed}, we expand both, the OPE coefficients and conformal blocks in powers of the anomalous dimensions up to $\Op(\lambda_R^2)$
\begin{align}
	\mathcal{A}_{n,l}(\Delta)=&A_{n,l}(\Delta)+(\gamma^{(1)}_{n,l}(\Delta)+\gamma^{(2)}_{n,l}(\Delta))A^{(1)}_{n,l}+\frac12(\gamma^{(1)}_{n,l}(\Delta))^2A^{(2)}_{n,l}+ \cdots \\
\nonumber	\mathcal{G}_{\Delta_{(n,l)},l}=&G_{\Delta(n,l),l}+(\gamma^{(1)}_{n,l}(\Delta)+\gamma^{(2)}_{n,l}(\Delta))\underbrace{\left.\frac{\partial G_{\Delta,l}}{\partial\Delta}\right\vert_{\Delta(n,l)}}_{G'_{\Delta(n,l),l}}
	+\frac12(\gamma^{(1)}_{n,l}(\Delta))^2\underbrace{\left.\frac{\partial^2G_{\Delta,l}}{\partial\Delta^2}\right\vert_{\Delta(n,l)}}_{G''_{\Delta(n,l),l}}+ \cdots\,, 
\end{align}
so that 
\begin{align}
    \mathcal{A}_{n,l}\mathcal{G}_{\Delta(n,l),l}=&A_{n,l}G_{\Delta(n,l),l}+\gamma^{(1)}_{n,l}(\Delta)\left(A_{n,l}G'_{\Delta(n,l),l}+C^{(1)}_{n,l}G_{\Delta(n,l),l}\right)\cr
	&+\frac{1}{2}(\gamma^{(1)}_{n,l}(\Delta))^2\left(A_{n,l}G''_{\Delta(n,l),l}+A^{(2)}_{n,l}G_{\Delta(n,l),l}+2A^{(1)}_{\Delta(n,l),l}G'_{\Delta(n,l),l}\right)\cr
	&+\gamma^{(2)}_{n,l}(\Delta)\left(A_{n,l}G'_{\Delta(n,l),l}+A^{(1)}_{n,l}G_{\Delta(n,l),l}\right)+\Op(\lambda^3)\,.
	\label{eq:CBExp}
\end{align}
The conformal blocks are of the form  $G_{\Delta,l}(v,Y)=v^{\Delta/2}f(v,Y)$ so that the derivatives contain terms like
\begin{equation}
    G'_{\Delta,l}(v,Y)=v^{\Delta/2}\log(v)f(v,Y)+ \cdots ;\quad G''_{\Delta,l}(v,Y)=v^{\Delta/2}\log^2(v)f(v,Y)+ \cdots 
\end{equation}
Comparing this to equation~\eqref{eq:CBExp} we realize that the terms
   proportional to $\log(v)$ in~\eqref{eq:CBExp} give us access to the
   anomalous dimensions at a given order in $\lambda_R$, while the
   $\log^2(v)$ term provides a consistency check  that the boundary
   Witten diagrams correspond to a consistent CFT. Consistency between
   the first and second order calculation in $\lambda_R$ require that the $\log^2(v)$ term has to be proportional $(\gamma_{n,l}^{(1)})^2$. This is the basis for equation \eqref{eq:unitarity_coeffs} as well. The contributions without log's then provide information about the OPE coefficients. Thus we can expand the exact expressions for the Witten diagrams we calculated in section~\ref{sec:calculation_witten_diagram} in $v,Y$ and compare them to the conformal block expansion to extract the anomalous dimensions and OPE coefficients.

By extracting the coefficient of $\log(v)=\log(\zeta\zetab)$ in the
   analytic expressions for Witten diagrams up to one-loop
   order, and comparing with the expansion of the four-point
   correlation function, we can extract the $L$-loop contributions to the anomalous dimensions $\gamma^{(L)}_{n,l}(\Delta)$. 
These contributions to the anomalous dimensions depend on the renormalised coupling  
\begin{equation}
    \gamma:=\frac{\lambda_R}{16\pi^2},
\end{equation}
such that at loop order $L$ the ratio $\gamma_{n,l}^{(L)}(\Delta)/\gamma^L$ is independent of the renormalised coupling.  We will comment more about the renormalisation scheme dependence below.

\paragraph{Anomalous dimensions for $\mathbf{\Delta=1}$} 
The anomalous dimensions for $\Delta=1$ are given by 
\begin{align}
&\gamma_{n,l}^{(1)}(1)=\gamma\, (1+\delta_{n,0})\delta_{l,0};\label{eq:gamma11}\\
&\gamma^{(2)}_{n,l>0}(1)=\gamma^2\begin{cases}
	\frac{-2}{l(l+1)}+\frac{4}{2l+1}\left(H^{(2)}_l-\zeta(2)\right)& \text{for }n=0 \\
T^1_{n,l}& \text{for }n>0
\end{cases}\\
&\gamma^{(2)}_{n,0}(1)=\gamma^2\begin{cases}
	-4+\frac{4}{2l+1}\left(H^{(2)}_l-\zeta(2)\right)& \text{for }n=0 \\
	\frac{(6n^2-3n-2)}{n(2n+1)}H^{(1)}_{2n}-1& \text{for }n>0
\end{cases}\,,
\end{align}
where the generalized harmonic numbers are given by {$H^{(k)}_i=\sum_{n=1}^i n^{-k}$} and the rational piece  $T_{n,l}^\Delta$ is given by
{\footnotesize\begin{equation}
\label{eq:Tcoeff}
T^\Delta_{n,l}=-\frac{2(l^2+(2\Delta+2n-1)(\Delta+n+l-1))}{l(l+1)(2\Delta+2n+l-1)(2\Delta+2n+l-2)}
-\frac{2(-1)^\Delta(H^{(1)}_l-H^{(1)}_{2\Delta+2n+l-2})}{(2\Delta+2n+2l-1)(\Delta+n-1)}\,.
\end{equation}}

The tree level results agree with~\cite{Heemskerk:2009pn}. The OPE coefficients at order $\lambda$ for $l=0$ are given by the known formula~\cite{Heemskerk:2009pn,Fitzpatrick:2011dm}
\begin{align}
\label{eq:firstorderOPE}
A^{(1)}_{n,0}(\Delta)=\frac12\frac{\partial A_{n,0}(\Delta)}{\partial n},
\end{align}
For the second order OPE coefficients and the first order OPE coefficients at $l>0$  one needs to expand the finite piece of the $L_0'$ integral, which we leave to a further study.

\paragraph{Anomalous dimensions for $\mathbf{\Delta=2}$} 
Similarly  we have the following results for the anomalous dimensions
\begin{align}
\gamma^{(1)}_{n,l}(2)&=\gamma\,\delta_{l,0} \quad\text{for}\ n\geq0;\\
\gamma^{(2)}_{n,l}(2)&= \gamma^2
                         \begin{cases}
                           T^2_{n,l}& \text{for}\ l>0\cr
\frac{2\left(6 n^2+15 n+11\right) H^{(1)}_{2 n+2}-\left(26
      n^2+65 n+41\right)}{2 (n+1) (2 n+3)}& \text{for}~ l=0
                         \end{cases}
\end{align}
where $T^2_{n,l}$ is given by equation~\eqref{eq:Tcoeff}.\footnote{The OPE coefficients at first order and $l=0$ obey equation~\eqref{eq:firstorderOPE} as well. The OPE coefficients $A_{n,l}^{(1)}(\Delta)$ up to spin 200 can be downloaded \href{https://github.com/pierrevanhove/AdS4/blob/main/AdS4.ipynb}{here}.}
We thus obtained closed expressions for the anomalous dimensions of all double trace operators appearing in the OPE expansion of the single trace operator $\mathcal{O}_\Delta$ for $\Delta=1,2$. To our knowledge, these have not been obtained before.

\paragraph{Renormalisation scheme dependence}
Note, that the first order anomalous dimension, which is generated by the cross Witten diagram, has only a non-zero constant contribution for $l=0$. Changing the renormalisation scheme, i.e. adding a cross term to the finite piece of the one loop contribution therefore only shifts the $\gamma^{(2)}_{n,0}(\Delta)$ part of the second order anomalous dimensions by a constant, which can always be absorbed by redefining the coupling constant. The anomalous dimensions for $l>0$ are completely scheme independent.

In the $\Delta=1$ case we find an anomalous piece in the $n=0$ trajectory given by
\begin{align}
\frac{4\gamma^2}{2l+1}\left(H^{(2)}_l-\zeta(2)\right)=-\frac{4\gamma^2\psi^{(1)}(l+1)}{2l+1}\,,
\end{align}
where $\psi^{(1)}(l+1)$ is the digamma function, which is absent in the $\Delta=2$ case. This is consistent with the result obtained in~\cite{Bertan:2018afl}.

In both cases the anomalous dimensions of the scalar operators $:\Op\square^n\Op:$ are positive and have different behaviour compared to the operators with non-vanishing spin. The behaviour for the latter can be summarized into equation~\eqref{eq:Tcoeff}, applicable to both cases. It is consistent with previous results for the $n=0$ trajectory in~\cite{Bertan:2018afl,Bertan:2018khc} and for the subleading trajectories obtained in~\cite{Heckelbacher:2020nue}.

\paragraph{Regge trajectories}
We can use equation~\eqref{eq:Tcoeff} to compare our result to previous results for large $l$ obtained by bootstrap methods~\cite{Basso:2006nk,Alday:2015eya,Alday:2017xua}. Expanding around $l\to\infty$ we obtain
\begin{equation}
\gamma^{(2)}_{n,l}(\Delta)=\gamma^2\sum\limits_{k=0}^{\infty}\frac{q^\Delta_{k}(n)}{l^{2\Delta+k}},
\end{equation}
where the $q^\Delta_{k}(n)$ are polynomials in $n$ of order $2\Delta+k-2$, which can easily be extracted from the exact expressions.

It is also straightforward to express the anomalous dimensions in terms of the conformal spin 
\begin{equation}
J^2=(l+\Delta+n)(l+\Delta+n-1).
\end{equation}
Expanding the anomalous dimensions in large $J$ we obtain
\begin{equation}
\gamma^{(2)}_{n,J}(\Delta)=\gamma^2\sum\limits_{k=0}^\infty\frac{Q^\Delta_k(n)}{J^{2\Delta+2k}},
\end{equation}
where the $Q^\Delta_k(n)$ are polynomials in $n$ of order $2\Delta+2k-2$. For $\Delta=1$ these polynomials only contain even powers of $n$. These behaviours are in agreement with the results from~\cite{Basso:2006nk,Alday:2015eya,Alday:2015ewa}.

Another interesting limit to explore would be the behaviour at
  $n\to\infty$. Taking the limit $n\to\infty$ in equation~\eqref{eq:Tcoeff} we obtain
\begin{equation}
\label{eq:large_n_gamma}
\lim\limits_{n\to\infty}\gamma^{(2)}_{n,l>0}(\Delta)=-\gamma^2\frac{1}{l(1+l)}.
\end{equation}
For $\Delta=1$ the limit is approached from below, while for $\Delta=2$ it is reached from above, as can be understood from the $(-1)^\Delta$ factor in \eqref{eq:Tcoeff}. Curiously the limit does not appear to depend on the value of $\Delta$. It would be interesting to test this observation for other values of $\Delta$.

%%%%%%%%%%%%%%%%%%%%%%%%%%%%%%%%%%%%%%%%%%%%%%%%%%%%%%%%%%
\section{Outlook}
\label{sec:conclusions}
%%%%%%%%%%%%%%%%%%%%%%%%%%%%%%%%%%%%%%%%%%%%%%%%%%%%%%%%%%

One of the main goals of this work is to build a bridge between  Witten diagrams and  flat space multi-loop Feynman integral techniques.
To this end,  we have presented a formulation of the Witten diagrams as 
combinations of dimensionally regularised\footnote{To restore conformal invariance of the renormalised four-point functions we had to use a loop depend regularisation $D=4-{4\epsilon\over L+1}$. This situation is somewhat similar to the one with the critical vector model \cite{Vasiliev:1975mq}, where the interaction is logarithmic (conformal) in any dimension and, hence, the usual replacement $d\rightarrow d-2\epsilon$ does not regularize the model. One can employ the analytic regularization by shifting the dimension of one of the fields by $\epsilon$. As a result, contributions to the physical quantities, e.g. anomalous dimensions, are proportional to the number of regulated lines in a diagram. } flat space Feynman integrals of the type
\begin{multline}
I(\underline n,\underline m, \underline \eta,D)=\int \prod_{i=1}^L \frac{ \dd^D X_i}{(u\cdot X_i)^{n_i} }
 \prod_{1\leq i<j\leq L}\frac{1}{\|X_k-X_j|^{2n_{kj}+\eta_{k,j}}}\\*
 \times\frac{1}{(\|X_{a_1}|^2)^{m_1}(\|X_{a_2}-\u1|^2)^{m_2+2(D-4)} (\|X_{a_3}-\uZ|^2)^{m_3}}
\end{multline} 
where $n_{kj}\in\mathbb Z$ are integers, $n_i$, $m_1$, $m_2$ and $m_3$ are powers depending on the conformal dimensions $\Delta$ and $\eta_{kj}$ are analytic parameters. The value of the Witten diagram  is the mutli-linear contribution  $\prod_{k,j}\partial_{\eta_{k,j}} I(\underline n,\underline m, \underline \eta,D)|_{\eta_{k,j}=0}$ in the analytic parameters $\eta_{k,j}$. 

With this reformulation, one can analyze the Witten diagrams using the standard methods for evaluating Feynman integrals~\cite{Smirnov:2004ym} and apply the flat space unitarity methods after performing the Wick rotation to Lorentz signature as described in~\cite{Cutkosky:1960sp,Abreu:2014cla,Bourjaily:2020wvq}. We hope that this approach will be useful for  extracting the higher-loop corrections to the anomalous dimensions.

As an application, we found analytic and closed expressions to almost all integrals involved and, furthermore, found closed expressions for the anomalous dimensions for all values of $n$ and $l$ of the ``double-trace'' operators $:\Op\square^n\partial^l\Op:$ up to second order in the coupling constant. To our knowledge, these have not been obtained before. In the process we formulated a version of dimensional regularisation in AdS which keeps the finite piece of the result AdS invariant. We checked this by comparing to an AdS invariant regularisation method and testing some CFT consistency conditions.

We also showed how unitarity can be used to extract the coefficients of the $\log(v)$ in the conformal block expansion. This should be useful to extract the higher-loop corrections to the anomalous dimensions. 

The techniques presented in this work give a systematic way of analyzing the loop corrections to the anomalous dimensions of double-trace operators and their Regge trajectories. We hope that they will be useful in improving the understanding of string theory in AdS-space.

There are several interesting directions to proceed. 
The most obvious next application of our method is to continue with the calculation of higher loop corrections. Let us emphasize that, since the integrals involved will be significantly harder to solve, the method of choice would be the unitarity cuts, as proposed in section~\ref{sec:unitarity_methods}.

Another straightforward application is the generalization to different values of the conformal dimension, especially $\Delta\geq 3$. In section~\ref{sec:bulkpropagator}, we explained that those cases could be treated using our method if we consider flat space propagators with  additional analytic parameters.  It would be interesting to compare the anomalous dimensions obtained like this with the results of~\cite{Carmi:2020ekr}, in the same way as it would be interesting to check if the bootstrap methods of~\cite{Carmi:2020ekr} can be used to reproduce our results for subleading Regge trajectories described in section~\ref{sec:conformal_block_expansion}.

A further potentially fruitful way to proceed is to use different new techniques to calculate Witten diagrams, like the differential representation~\cite{Armstrong:2020woi,Herderschee:2021jbi,Herderschee:2022ntr} and unitarity methods based on the split representation of the propagator~\cite{Liu:2018jhs,Meltzer:2019nbs,Meltzer:2020qbr} in combination with our method.

Finally, we would like to mention some recent developments in the calculation of cosmological correlators in de Sitter~\cite{Hogervorst:2021uvp,DiPietro:2021sjt,Sleight:2019hfp,Sleight:2021plv}. It could be interesting to explore, if our method can be applied in that framework as well.

%%%%%%%%%%%%%%%%%%%%%%%%%%%%%%%%%%%%%%%%%%%%%%%%%%%%%%%%%%
\section*{Acknowledgments}

We thank Samuel Abreu, Claude Duhr, \"Omer G\"urdo\v gan, Shota Komatsu, Roman Lee, Oliver Schnetz and Vladimir Smirnov for discussions. T.H. thanks Mart\'in Enr\'iquez Rojo for helpful feedback on the manuscript. The research of P.V. has received funding from the ANR
grant ``Amplitudes'' ANR-17- CE31-0001-01, and the ANR grant ``SMAGP''
ANR-20-CE40-0026-01 and is partially supported by the Laboratory of Mirror Symmetry NRU HSE, RF Government grant, ag. No 14.641.31.0001. P.V. is grateful to the I.H.E.S. for the use of their computer resources. This project has received funding from the European Research Council (ERC) under the European Union’s Horizon 2020 research and innovation programme (grant agreement No 101002551). The work of T.H. and I.S.  
 was funded by the Excellence Cluster Origins of the DFG under Germany’s Excellence Strategy EXC-2094 390783311.
T.H. was partially supported by the Hans-B\"ockler-Stiftung of the German Trade Union Confederation.

%%%%%%%%%%%%%%%%%%%%%%%%%%%%%%%%%%%%%%%%%%%%%%%%%%%%%%%%%%

%%%%%%%%%%%%%%%%%%%%%%%%%%%%%%%%%%%%%%%%%%%%%%%%%%%%%%%%%%
\begin{appendix}\label{app}
	\renewcommand{\thesection}{\Alph{section}}
	\renewcommand{\theequation}{\Alph{section}.\arabic{equation}}
	\setcounter{equation}{0}\setcounter{section}{0}
	%%%%%%%%%%%%%%%%%%%%%%%%%%%%%%%%%%%%%%%%%%%%%%%%%%%%%%%%%%
	\newpage
	
	%%%%%%%%%%%%%%%%%%%%%%%%%%%%%%%%%%%%%%%%%%%%%%%%%%%%%%%%%%
\section{Multiple polylogarithms}	
\label{sec:MPLs}
%%%%%%%%%%%%%%%%%%%%%%%%%%%%%%%%%%%%%%%%%%%%%%%%%%%%%%%%%%
In the evaluation of the Witten diagrams, we encountered multiple polylogarithms as the results of linearly reducible Witten diagrams in the parametric representation. Following the convention used by Panzer in \texttt{HyperInt}~\cite{Panzer:2014caa}, they are defined by the nested sum
\begin{equation}
    \mathrm{Li}_{s_1, \dots,  s_k}(x_1, \dots ,x_k):=\sum\limits_{0<p_1 <\cdots <p_k}^\infty\frac{x_1^{p_1}}{p_1^{s_1}} \cdots \frac{x_k^{p_k}}{p_k^{s_k}}\quad\text{for }\abs{x_1\cdots x_i}<1,\quad\forall i\in\{1,..,k\}\,.
\end{equation}
The sum $s_1+s_2+ \dots +s_k$ is referred to as the weight of the multiple polylogarithm.

Some useful definitions and identities are
\begin{align}
    \Mpl{1}{x}&=-\log(1-x)\,,\\
    \Mpl{1,1}{y,x}&=\Mpl{2}{\frac{x(y-1)}{1-x}}-\Mpl{2}{\frac{x}{x-1}}-\Mpl{2}{xy}\,,
\end{align}
and the Bloch-Wigner dilogarithm given by:
	\begin{align}
		\label{eq:BlochWigner}
		D(\zeta,\zetab)&=\frac{1}{2i}\left(\Mpl{2}{\zeta}-\Mpl{2}{\zetab}-\frac12\log(\zeta\zetab)\left(\Mpl{1}{\zeta}-\Mpl{1}{\zetab}\right)\right).
	\end{align}
For a detailed discussion of these functions and their properties we refer the interested reader to~\cite{Zagier:2007knq,goncharov2001multiple,Ablinger:2011te,Remiddi:1999ew}.
	
	%%%%%%%%%%%%%%%%%%%%%%%%%%%%%%%%%%%%%%%%%%%%%%%%%%%%%%%%%%
	\subsection{Some recurring expressions}
	\label{subsec:recurring_expressions}
	%%%%%%%%%%%%%%%%%%%%%%%%%%%%%%%%%%%%%%%%%%%%%%%%%%%%%%%%%%
We collect recurring expressions that enter the evaluation of the
Witten diagrams:
	\begin{align}
		\label{eq:Omega1}
		f_1(\zeta,\zetab)&=\log(\zeta\zetab)\left(\Mpl{1,1}{\bar\zeta,{\zeta\over\bar\zeta}}-\Mpl{1,1}{\zeta,{\bar\zeta\over\zeta}}
		+\Mpl{1}{\zeta}\Mpl{1}{{\bar\zeta\over\zeta}}-\Mpl{1}{\bar\zeta}\Mpl{1}{{\zeta\over\bar\zeta}}\right)\cr
		&+\Mpl{3}{\zeta}-\Mpl{3}{\bar\zeta}+
		\Mpl{2,1}{1,\zeta}-\Mpl{2,1}{1,\bar\zeta}\cr
		&+2\,\Mpl{2,1}{\zeta,{\frac {\bar\zeta}{\zeta}}}-2\,
		\Mpl{2,1}{\bar\zeta,{\frac {\zeta}{\bar\zeta}}}+
		\Mpl{1,2}{\zeta,{\frac {\bar\zeta}{\zeta}}}
		-\Mpl{1,2}{\bar\zeta,{\frac {\zeta}{\bar\zeta}}}\cr
		&-2\,\Mpl{1}{{\frac {\bar\zeta}{\zeta}}}\Mpl{2}{\zeta}-
		\Mpl{2}{{\frac {\bar\zeta}{\zeta}}}\Mpl{1}{\zeta}+2\,
		\Mpl{1}{{\frac {\zeta}{\bar\zeta}}}\Mpl{2}{\bar\zeta}
		+\Mpl{1}{\bar\zeta}\Mpl{2}{{\frac {\zeta}{\bar\zeta}}},\\
		\label{eq:Omega2}
		f_2(\zeta,\zetab)&=-\frac12f_1(\zeta,\zetab)+\frac12\left(\Mpl{2}{\zeta}\Mpl{1}{\zetab}-\Mpl{2}{\zetab}\Mpl{1}{\zeta}\right)\cr
		&+\Mpl{1,2}{1,\zeta}-\Mpl{1,2}{1,\zetab}+\frac12\left(\Mpl{2,1}{1,\zeta}-\Mpl{2,1}{1,\zetab}\right)\cr
		&+\frac12\log(\zeta\zetab)\left(\Mpl{2}{\zeta}-\Mpl{2}{\zetab}-\Mpl{1,1}{1,\zeta}+\Mpl{1,1}{1,\zetab}\right)\cr
		&-\frac14\log^2(\zeta\zetab)\left(\Mpl{1}{\zeta}-\Mpl{1}{\zetab}\right),\\
		f_3(\zeta,\zetab)&=4i{\zeta+\bar\zeta-2\over\zeta-\bar\zeta}
		D(\zeta,\bar \zeta) +
		\log(\zeta\bar\zeta)\log\left((1-\zeta)(1-\bar\zeta)\over\zeta\bar\zeta\right),\\
		f_4(\zeta,\zetab)&= -4i{\zeta+\bar\zeta\over\zeta-\bar\zeta}D(\zeta,\bar\zeta)-\log((1-\zeta)(1-\bar\zeta))\log\left((1-\zeta)(1-\bar\zeta)\over\zeta\bar\zeta\right)\\
		f_5(\zeta,\zetab)&={4i
			(\zeta+\bar  \zeta
			-2\zeta\bar\zeta)\over  \zeta -\bar
			\zeta }
		D(\zeta,\bar\zeta)-\log(\zeta\bar\zeta)\log((1-\zeta)(1-\bar\zeta)),
	\end{align}

All these expression are single-valued multiple-polylogarithms in
$\mathbb C\backslash\{0,1\}$ where
$\zetab=\zeta^*$ is the complex conjugate of $\zeta$. The
single-valuedness of the expressions are easily checked using the {\tt
  HyperlogProcedures} by Schnetz~\cite{Schnetz}.

	%%%%%%%%%%%%%%%%%%%%%%%%%%%%%%%%%%%%%%%%%%%%%%%%%%%%%%%%%%
	\section{Evaluation of the Witten cross diagram}
	\label{sec:exapnsion_cross}
	In this appendix we collect exact evaluations of the Witten
       cross diagram. In section~\ref{sec:cross_mpl} we given
       an analytic evaluation of the cross diagram for all $\Delta$,
       in section~\ref{subsec:Cross_exact} we give the results
       for the evaluation of the cross diagram  in dimensional
       regularisation for $\Delta=1$ and $\Delta=2$ and in
       section~\ref{subsec:Cross_expand} we give the $v$ and $Y$
       expansion of the cross diagram  for all values of $\Delta$.
\subsection{The analytic evaluation of  cross diagram for all \texorpdfstring{$\Delta$}{Lg}}\label{sec:cross_mpl}

Using the creative telescoping algorithm implemented
             in~\cite{Koutchan:2013} we
             deduce that the
             integral
	\begin{equation}
		I_\times^\Delta(\zeta,\zetab)=\int\limits_{\alpha_i\geq0}\frac{\prod\limits_{i=1}^3\dd\alpha_i \alpha_i^{\Delta-1}}{(\alpha_1+\alpha_2+\alpha_3)^\Delta(\alpha_1\alpha_2+\alpha_1\alpha_3\zeta\zetab+\alpha_2\alpha_3(1-\zeta)(1-\zetab))^\Delta}\,,
	\end{equation}
satisfies the recursion  relation for $\Delta\geq1$
             \begin{equation}
 \sum_{n=0}^4 c(n) I_\times^{\Delta+n}(\zeta,\bar\zeta)=0  ,
\end{equation}
with
\begin{align}
  c(0)&=\Delta ^2 (4 \Delta +7) (4 \Delta +11),\cr
        c(1)&=-\left(\left(64 \Delta ^4+352 \Delta ^3+620 \Delta ^2+410 \Delta +99\right) (2 \zeta 
              \bar\zeta-\zeta -\bar\zeta+2)\right),\cr
                c(2)=&4 \left(16 \Delta ^2+56 \Delta +33\right) (2
                       \Delta +3)^2 \zeta ^2 \bar\zeta^2
                       -4
   \left(16 \Delta ^2+56 \Delta +33\right) (2 \Delta +3)^2 (\zeta
                       +\bar\zeta)\cr
                       &-4
   \left(16 \Delta ^2+56 \Delta +33\right) (2 \Delta +3)^2 \zeta  \bar\zeta (\zeta
   +\bar\zeta)+4 \left(16 \Delta ^2+56 \Delta +33\right) (2 \Delta
                         +3)^2\cr
                         &+\left(96
   \Delta ^4+624 \Delta ^3+1414 \Delta ^2+1322 \Delta +423\right)
                           (\zeta +\bar\zeta)^2\cr
                           &+8 \left(48 \Delta ^4+312 \Delta ^3+715 \Delta ^2+689 \Delta +234\right) \zeta 
   \bar\zeta,\cr
                                 c(3)&=-\left(16 \Delta ^2+48 \Delta +27\right) \Big(4 \left(8 \Delta ^2+36 \Delta
     +39\right) \zeta ^2 \bar\zeta^2\cr
     &+2 \zeta  \bar\zeta \left(\left(4 \Delta
   ^2+18 \Delta +19\right) (\zeta +\bar\zeta)^2-4 \left(6 \Delta ^2+27 \Delta
       +29\right) (\zeta +\bar\zeta)+16 \Delta ^2+72 \Delta +78\right)\cr
       &-\left(4 \Delta
   ^2+18 \Delta +19\right) (\zeta +\bar\zeta-2) (\zeta +\bar\zeta)^2\Big),\cr
           \nonumber                c(4)&=(\Delta +3)^2 \left(16 \Delta ^2+40 \Delta +21\right) (\zeta -\bar\zeta)^4,
\end{align}
which are symmetric polynomials in $v=\zeta\zetab$ and $v+Y=\zeta+\zetab$.
The recursion implies that for $n\geq 0$ integer
\begin{equation}
  I_\times(\Delta+4+n)= \sum_{r=0}^3  \sum_{a_0+\cdots+a_4=n+1\atop a_1+2a_2+3a_3+4a_4=3n+r}  \prod_{i=0}^4c(r)^{a_i}   { I_\times^{\Delta+r} (\zeta,\bar\zeta)\over (\zeta-\bar
    \zeta)^{4(n+1)}} .
\end{equation}

\paragraph{The case of $\Delta$ integer.} When $\Delta$ is a positive integer
we have that for $\Delta\geq5$
\begin{equation}
  I_\times^{\Delta}(\zeta,\zetab)= \sum_{r=0}^3 {\sum_{0\leq a,b\leq \Delta+1}
n_r^{a,b}(\Delta) (\zeta\bar\zeta)^a (\zeta+\bar\zeta)^b\over (\zeta-\bar
    \zeta)^{4(\Delta-4)}}   I_\times^{1+r} (\zeta,\bar\zeta) .
\end{equation}
The evaluation of the integrals $I_\times^r(\zeta,\zetab)$ with $1\leq
        r \leq 4$ is easily done with
{\tt HyperInt}~\cite{Panzer:2014caa}, with the results
\begin{equation}
  I_\times^1(\zeta,\bar\zeta)= {4i D(\zeta,\bar\zeta)\over\zeta-\bar\zeta}, 
\end{equation}
and
\begin{multline}
  I_\times^2(\zeta,\bar\zeta)= {4i\left(-(\zeta+\bar \zeta)^3+2 (\zeta+\bar \zeta)^2 \zeta \bar \zeta+2 (\zeta+\bar \zeta)^2-8 (\zeta+\bar \zeta) \zeta \bar \zeta+4 \zeta^2 \bar \zeta^2+4 \zeta \bar \zeta\right)\over(\zeta-\bar\zeta)^4}
  { D(\zeta,\bar\zeta)\over\zeta-\bar\zeta}\cr
  +{4\left(\left(\zeta +\bar\zeta \right)^{2}-3 \left(\zeta +\bar\zeta
      \right) \zeta \bar\zeta +2 \zeta
      \bar\zeta\right)\over(\zeta-\bar\zeta)^4}\log(\zeta\bar\zeta)\cr
  +{4\left(-2 \left(\zeta +\bar\zeta \right)^{2}+3 \left(\zeta +\bar\zeta
      \right) \zeta \bar\zeta +3 \zeta +3 \bar\zeta -4 \zeta \bar\zeta\right)\over (\zeta-\bar\zeta)^4}\log((1-\zeta)(1-\bar\zeta))+ {2\over(\zeta-\bar\zeta)^2}
\end{multline}
and
\begin{equation}
  I_\times^3(\zeta,\zetab) ={c_1^3(\zeta,\zetab)\over (\zeta-\zetab)^8}   {4i D(\zeta,\zetab)\over \zeta-\zetab}+{c_2^3(\zeta,\zetab)\over (\zeta-\zetab)^8} \log(\zeta\zetab)+{c_3^3(\zeta,\zetab)\over (\zeta-\zetab)^8}
  \log((1-\zeta)(1-\zetab))+   {c_4^3(\zeta,\zetab)\over (\zeta-\zetab)^8},
\end{equation}
with
\begin{align}
  c_1^3(\zeta,\zetab)&=\left(\zeta +\zetab \right)^{6}-6 \left(\zeta
                       +\zetab \right)^{5} \zeta \zetab +6 \left(\zeta
                       +\zetab \right)^{4} \zeta^{2} \zetab^{2}-6
                       \left(\zeta +\zetab \right)^{5}+66 \left(\zeta
                       +\zetab \right)^{4} \zeta \zetab \cr
&-132 \left(\zeta +\zetab \right)^{3} \zeta^{2} \zetab^{2}+72
                            \left(\zeta +\zetab \right)^{2} \zeta^{3} \zetab^{3}+6 \left(\zeta +\zetab \right)^{4}-132 \left(\zeta +\zetab \right)^{3} \zeta \zetab +324 \left(\zeta +\zetab \right)^{2} \zeta^{2} \zetab^{2}\cr
&-216 \left(\zeta +\zetab \right) \zeta^{3} \zetab^{3}+36 \zeta^{4} \zetab^{4}+72 \left(\zeta +\zetab \right)^{2} \zeta \zetab -216 \left(\zeta +\zetab \right) \zeta^{2} \zetab^{2}+104 \zeta^{3} \zetab^{3}+36 \zeta^{2} \zetab^{2}
\cr
  c_2^3(\zeta,\zetab)&=-3 \left(\zeta +\zetab \right)^{5}+22
                       \left(\zeta +\zetab \right)^{4} \zeta \zetab
                       -25 \left(\zeta +\zetab \right)^{3} \zeta^{2}
                       \zetab^{2}+6 \left(\zeta +\zetab \right)^{4}-96
                       \left(\zeta +\zetab \right)^{3} \zeta \zetab\cr
& +204 \left(\zeta +\zetab \right)^{2} \zeta^{2} \zetab^{2}-110 \left(\zeta +\zetab \right) \zeta^{3} \zetab^{3}+72 \left(\zeta +\zetab \right)^{2} \zeta \zetab -198 \left(\zeta +\zetab \right) \zeta^{2} \zetab^{2}+92 \zeta^{3} \zetab^{3}+36 \zeta^{2} \zetab^{2}
\cr
  c_3^3(\zeta,\zetab)&=-6 \left(\zeta +\zetab \right)^{5}+28
                       \left(\zeta +\zetab \right)^{4} \zeta \zetab
                       -25 \left(\zeta +\zetab \right)^{3} \zeta^{2}
                       \zetab^{2}+28 \left(\zeta +\zetab
                       \right)^{4}-192 \left(\zeta +\zetab \right)^{3}
                       \zeta \zetab \cr
&+276 \left(\zeta +\zetab \right)^{2} \zeta^{2} \zetab^{2}-110
        \left(\zeta +\zetab \right)
        \zeta^{3} \zetab^{3}-25
        \left(\zeta +\zetab
        \right)^{3}+276 \left(\zeta
        +\zetab \right)^{2} \zeta
        \zetab -396 \left(\zeta
        +\zetab \right) \zeta^{2}
        \zetab^{2}\cr
&+128 \zeta^{3} \zetab^{3}-110 \left(\zeta +\zetab \right) \zeta \zetab +128 \zeta^{2} \zetab^{2}
\cr
  c_4^3(\zeta,\zetab)&={-13 \left(\zeta +\zetab \right)^{3}+26 \left(\zeta +\zetab \right)^{2} \zeta \zetab +26 \left(\zeta +\zetab \right)^{2}-88 \left(\zeta +\zetab \right) \zeta \zetab +36 \zeta^{2} \zetab^{2}+36 \zeta \zetab\over2}\cr
\end{align}
and
\begin{equation}
  I_\times^4(\zeta,\zetab) = {c_1^4(\zeta,\zetab)\over (\zeta-\zetab)^{12}}  {4i D(\zeta,\zetab)\over \zeta-\zetab}+  {c_2^3(\zeta,\zetab)\over (\zeta-\zetab)^{12}} \log(\zeta\zetab)+ {c_3^3(\zeta,\zetab)\over (\zeta-\zetab)^{12}} 
  \log((1-\zeta)(1-\zetab))+    {c_4^3(\zeta,\zetab)\over (\zeta-\zetab)^{12}} 
\end{equation}
with
\begin{align}
  c_1^4(\zeta,\zetab)&=400 \zeta^{3} \zetab^{3}-5076 \left(\zeta
                       +\zetab \right)^{5} \zeta^{2} \zetab^{2}+9312
                       \left(\zeta +\zetab \right)^{4} \zeta^{3}
                       \zetab^{3}-6900 \left(\zeta +\zetab \right)^{3}
                       \zeta^{4} \zetab^{4}\cr
&+1800 \left(\zeta +\zetab \right)^{2} \zeta^{5} \zetab^{5}-19304
               \left(\zeta +\zetab
               \right)^{3} \zeta^{3}
               \zetab^{3}+15528
               \left(\zeta +\zetab
               \right)^{2} \zeta^{4}
               \zetab^{4}-4800
               \left(\zeta +\zetab
               \right) \zeta^{5}
               \zetab^{5}\cr
&-11136 \left(\zeta +\zetab \right) \zeta^{4} \zetab^{4}+12
                               \left(\zeta +\zetab \right)^{8} \zeta \zetab -30 \left(\zeta +\zetab \right)^{7} \zeta^{2} \zetab^{2}+20 \left(\zeta +\zetab \right)^{6} \zeta^{3} \zetab^{3}\cr
&-234 \left(\zeta +\zetab \right)^{7} \zeta \zetab +948 \left(\zeta
                        +\zetab \right)^{6} \zeta^{2} \zetab^{2}-1320 \left(\zeta +\zetab \right)^{5} \zeta^{3} \zetab^{3}+600 \left(\zeta +\zetab \right)^{4} \zeta^{4} \zetab^{4}\cr
&+948 \left(\zeta +\zetab \right)^{6} \zeta \zetab -1320 \left(\zeta
               +\zetab \right)^{5} \zeta \zetab +9312 \left(\zeta +\zetab \right)^{4} \zeta^{2} \zetab^{2}+600 \left(\zeta +\zetab \right)^{4} \zeta \zetab \cr
&-6900 \left(\zeta +\zetab \right)^{3} \zeta^{2} \zetab^{2}+15528
                          \left(\zeta +\zetab \right)^{2} \zeta^{3} \zetab^{3}+1800 \left(\zeta +\zetab \right)^{2} \zeta^{2} \zetab^{2}\cr
&-4800 \left(\zeta +\zetab \right) \zeta^{3} \zetab^{3}-\left(\zeta +\zetab \right)^{9}+12 \left(\zeta +\zetab \right)^{8}-30 \left(\zeta +\zetab \right)^{7}+20 \left(\zeta +\zetab \right)^{6}+2352 \zeta^{5} \zetab^{5}\cr&+400 \zeta^{6} \zetab^{6}+2352 \zeta^{4} \zetab^{4}
\cr
  c_2^4(\zeta,\zetab)&={1\over3}\Big(11 \left(\zeta +\zetab
                       \right)^{8}-150 \left(\zeta +\zetab \right)^{7}
                       \zeta \zetab +411 \left(\zeta +\zetab
                       \right)^{6} \zeta^{2} \zetab^{2}-294
                       \left(\zeta +\zetab \right)^{5} \zeta^{3}
                       \zetab^{3}-60 \left(\zeta +\zetab
                       \right)^{7}\cr
&+1444 \left(\zeta +\zetab \right)^{6} \zeta \zetab -6390 \left(\zeta
      +\zetab \right)^{5} \zeta^{2}
      \zetab^{2}+9306 \left(\zeta
      +\zetab \right)^{4} \zeta^{3}
      \zetab^{3}-4368 \left(\zeta
      +\zetab \right)^{3} \zeta^{4}
      \zetab^{4}\cr
&+60 \left(\zeta +\zetab \right)^{6}-3060 \left(\zeta +\zetab
                      \right)^{5}
                      \zeta \zetab
                      +18786
                      \left(\zeta
                      +\zetab
                      \right)^{4}
                      \zeta^{2}
                      \zetab^{2}-34920
                      \left(\zeta
                      +\zetab
                      \right)^{3}
                      \zeta^{3}
                      \zetab^{3}\cr
&+24264 \left(\zeta +\zetab \right)^{2} \zeta^{4} \zetab^{4}-5544
    \left(\zeta +\zetab \right) \zeta^{5} \zetab^{5}+1800 \left(\zeta +\zetab \right)^{4} \zeta \zetab -18000 \left(\zeta +\zetab \right)^{3} \zeta^{2} \zetab^{2}\cr &+37984 \left(\zeta +\zetab \right)^{2} \zeta^{3} \zetab^{3}-25680 \left(\zeta +\zetab \right) \zeta^{4} \zetab^{4}+4944 \zeta^{5} \zetab^{5}+5400 \left(\zeta +\zetab \right)^{2} \zeta^{2} \zetab^{2}\cr&-13800 \left(\zeta +\zetab \right) \zeta^{3} \zetab^{3}+6656 \zeta^{4} \zetab^{4}+1200 \zeta^{3} \zetab^{3}\Big)
\cr
  c_3^4(\zeta,\zetab)&={1\over3}\Big(6144 \zeta^{3} \zetab^{3}-9450 \left(\zeta +\zetab \right)^{5} \zeta^{2} \zetab^{2}+11106 \left(\zeta +\zetab \right)^{4} \zeta^{3} \zetab^{3}-4368 \left(\zeta +\zetab \right)^{3} \zeta^{4} \zetab^{4}\cr&-52920 \left(\zeta +\zetab \right)^{3} \zeta^{3} \zetab^{3}+29664 \left(\zeta +\zetab \right)^{2} \zeta^{4} \zetab^{4}-5544 \left(\zeta +\zetab \right) \zeta^{5} \zetab^{5}-39480 \left(\zeta +\zetab \right) \zeta^{4} \zetab^{4}\cr&-210 \left(\zeta +\zetab \right)^{7} \zeta \zetab +471 \left(\zeta +\zetab \right)^{6} \zeta^{2} \zetab^{2}-294 \left(\zeta +\zetab \right)^{5} \zeta^{3} \zetab^{3}+2888 \left(\zeta +\zetab \right)^{6} \zeta \zetab \cr&-9450 \left(\zeta +\zetab \right)^{5} \zeta \zetab +37572 \left(\zeta +\zetab \right)^{4} \zeta^{2} \zetab^{2}+11106 \left(\zeta +\zetab \right)^{4} \zeta \zetab -52920 \left(\zeta +\zetab \right)^{3} \zeta^{2} \zetab^{2}\cr&+75968 \left(\zeta +\zetab \right)^{2} \zeta^{3} \zetab^{3}-4368 \left(\zeta +\zetab \right)^{3} \zeta \zetab +29664 \left(\zeta +\zetab \right)^{2} \zeta^{2} \zetab^{2}-39480 \left(\zeta +\zetab \right) \zeta^{3} \zetab^{3}\cr&-5544 \left(\zeta +\zetab \right) \zeta^{2} \zetab^{2}+22 \left(\zeta +\zetab \right)^{8}-210 \left(\zeta +\zetab \right)^{7}+471 \left(\zeta +\zetab \right)^{6}-294 \left(\zeta +\zetab \right)^{5}+6144 \zeta^{5} \zetab^{5}\cr&+13312 \zeta^{4} \zetab^{4}
\Big)\cr
  c_4^4(\zeta,\zetab)&={1\over18}\Big(193 \left(\zeta +\zetab \right)^{6}-1044 \left(\zeta +\zetab \right)^{5} \zeta \zetab +1044 \left(\zeta +\zetab \right)^{4} \zeta^{2} \zetab^{2}-1044 \left(\zeta +\zetab \right)^{5}\cr&+9384 \left(\zeta +\zetab \right)^{4} \zeta \zetab -17352 \left(\zeta +\zetab \right)^{3} \zeta^{2} \zetab^{2}+8784 \left(\zeta +\zetab \right)^{2} \zeta^{3} \zetab^{3}+1044 \left(\zeta +\zetab \right)^{4}\cr&-17352 \left(\zeta +\zetab \right)^{3} \zeta \zetab +39648 \left(\zeta +\zetab \right)^{2} \zeta^{2} \zetab^{2}-24768 \left(\zeta +\zetab \right) \zeta^{3} \zetab^{3}+3600 \zeta^{4} \zetab^{4}\cr&+8784 \left(\zeta +\zetab \right)^{2} \zeta \zetab -24768 \left(\zeta +\zetab \right) \zeta^{2} \zetab^{2}+11552 \zeta^{3} \zetab^{3}+3600 \zeta^{2} \zetab^{2}
\Big)
\end{align}

	%%%%%%%%%%%%%%%%%%%%%%%%%%%%%%%%%%%%%%%%%%%%%%%%%%%%%%%%%%
	\subsection{Cross in dimensional regularisation}
	\label{subsec:Cross_exact}
	%%%%%%%%%%%%%%%%%%%%%%%%%%%%%%%%%%%%%%%%%%%%%%%%%%%%%%%%%%
	The cross term for $\Delta=1$ in $D=4-4\epsilon$ dimensions is given by:
	\begin{align}
		\label{eq:crossD1_dimreg_param}
		\cW_0^{1,4-4\epsilon}&=\frac12\frac{\zeta\zetab}{x_{12}^2 x_{34}^2}\int_{\mathbb R^{4-4\epsilon}} {\dd^{4-4\epsilon}X (u\cdot X)^4\over \|X|^4\|X-u_1|^{4(1-4\epsilon)}\|X-u_\zeta|^4}\cr
		&=\frac12\frac{\pi^{2-2\epsilon}\zeta\zetab}{x_{12}^2 x_{34}^2}\frac{\Gamma(1-2\epsilon)}{\Gamma(1-4\epsilon)}\int\limits_{(\mathbb{RP}^+)^2}\frac{\dd\alpha_1\dd\alpha_2\dd\alpha_3\alpha_2^{-4\epsilon}}
		{(\alpha_1+\alpha_2+\alpha_3)(\alpha_1\alpha_2+\alpha_1\alpha_3\zeta\zetab+(1-\zeta)(1-\zetab)\alpha_2\alpha_3)^{1-2\epsilon}}\cr
	\end{align}
	Acting on~\eqref{eq:crossD1_dimreg_param} with $\cH_{1234}$ we obtain the parametric representation of the $\Delta=2$ case:
	\begin{align}
		\label{eq:crossD2_dimreg_param}
		\cW_0^{2,4-4\epsilon}&=\frac12\frac{(\zeta\zetab)^2}{x_{12}^4 x_{34}^4}\int_{\mathbb R^{4-4\epsilon}} {\dd^{4-4\epsilon}X (u\cdot X)^4\over \|X|^4\|X-u_1|^{4(1-4\epsilon)}\|X-u_\zeta|^4}\cr
		&=\frac{2\pi^{2-2\epsilon}}{16}\frac{(\zeta\zetab)^2}{x_{12}^4 x_{34}^4}\frac{\Gamma(1-2\epsilon)}{\Gamma(1-4\epsilon)}\times\cr
		&\int\limits_{(\mathbb{RP}^+)^2}\frac{\dd\alpha_1\dd\alpha_2\dd\alpha_3\alpha_2^{-4\epsilon}(C_1\alpha_1\alpha_2^2\alpha_3+C_2\alpha_1^2\alpha_2\alpha_3+C_3\alpha_2^2\alpha_3^2+C_4\alpha_1\alpha_2\alpha_3^2+C_5\alpha_1^2\alpha_3^2)}
		{(\alpha_1+\alpha_2+\alpha_3)(\alpha_1\alpha_2+\alpha_1\alpha_3\zeta\zetab+(1-\zeta)(1-\zetab)\alpha_2\alpha_3)^{3-2\epsilon}}\cr
	\end{align}
The coefficients in the parametric integral~\eqref{eq:crossD2_dimreg_param} are given by:
	\begin{align}
		\label{eq:Ccoefficients}
		C_1&= (1-6 \epsilon ) (\zeta +\zetab-\zeta  \zetab)+8 \epsilon -2\cr
		C_2&=-(1-6\epsilon)\zeta\zetab-1+2\epsilon\cr
		C_3&=(4 \zeta  \zetab \epsilon ^2-4 \epsilon ^2 (\zeta +\zetab)+8 \epsilon ^2-4 \epsilon +1)(1-\zeta)(1-\zetab)\cr
		C_4&=8 \zeta ^2 \zetab^2 \epsilon ^2-8 \zeta  \zetab \epsilon ^2 (\zeta +\zetab)+\zeta  \zetab \left(8 \epsilon ^2+4 \epsilon -2\right)+(1-2 \epsilon ) (\zeta +\zetab)+2 \epsilon -1\cr
		C_5&=4 \zeta ^2 \zetab^2 \epsilon ^2+\zeta  \zetab \left(4 \epsilon ^2-4 \epsilon +1\right)
	\end{align}
The $\Op(\epsilon)$ term of the result of equation~\eqref{eq:crossD2_dimreg_param} is given by
	{\scriptsize\begin{align}
			\label{eq:crossD2_epsilon}
			&\cW_{0,\epsilon}^{2,4}=\frac{3(\zeta\zetab)^2\left(-\left(\zeta+\zetab\right)^{3}+2 \left(\zeta+\zetab\right)^{2} \zeta \zetab+2 \left(\zeta+\zetab\right)^{2}-8 \zeta \zetab \left(\zeta+\zetab\right)+4 \zeta^{2} \zetab^{2}+4 \zeta \zetab\right)}{2(\zeta-\zetab)^5}f_2\cr
			&-\frac{4i(\zeta\zetab)^2\left(-3 \left(\zeta+\zetab\right)^{3}+5 \left(\zeta+\zetab\right)^{2} \zeta \zetab+5 \left(\zeta+\zetab\right)^{2}-12 \zeta \zetab \left(\zeta+\zetab\right)+4 \zeta^{2} \zetab^{2}+4 \zeta \zetab\right)D(\zeta,\zetab)}{(\zeta-\zetab)^5}\cr
			&+\frac{3(\zeta\zetab)^2\left(-2 \left(\zeta+\zetab\right)^{2}+3 \zeta \zetab \left(\zeta+\zetab\right)+3 \zeta+3 \zetab-4 \zeta \zetab\right)}{2(\zeta-\zetab)^4}\left(\Mpl{1}{\zeta}\Mpl{1}{\zetab}+\Mpl{1,1}{1,\zeta}+\Mpl{1,1}{1,\zetab}\right)\cr
			&-\frac{3\zeta\zetab\left(-\left(\zeta+\zetab\right)^{2} \zeta \zetab+3 \left(\zeta+\zetab\right) \zeta^{2} \zetab^{2}-2 \zeta^{2} \zetab^{2}\right)}{2(\zeta-\zetab)^4}\log(\zeta\zetab)\log((1-\zeta)(1-\zetab))\cr
			&-\frac{\zeta\zetab\left(\left(\zeta+\zetab\right)^{3} \zeta \zetab+\left(\zeta+\zetab\right)^{3}-18 \left(\zeta+\zetab\right)^{2} \zeta \zetab+8 \left(\zeta+\zetab\right) \zeta^{2} \zetab^{2}+8 \zeta \zetab \left(\zeta+\zetab\right)+24 \zeta^{2} \zetab^{2}\right)}{4(\zeta-\zetab)^4}\log((1-\zeta)(1-\zetab))\cr
			&+\frac{3(\zeta\zetab)^2\left(-\left(\zeta+\zetab\right)^{2}+3 \zeta \zetab \left(\zeta+\zetab\right)-2 \zeta \zetab\right)}{4(\zeta-\zetab)^4}\log^2(\zeta\zetab)+\cr
			&+\frac{(\zeta\zetab)^2\left(-\left(\zeta+\zetab\right)^{4}+\left(\zeta+\zetab\right)^{3} \zeta \zetab+10 \left(\zeta+\zetab\right)^{3}-18 \left(\zeta+\zetab\right)^{2} \zeta \zetab+8 \left(\zeta+\zetab\right) \zeta^{2} \zetab^{2}-8 \left(\zeta+\zetab\right)^{2}-4 \zeta \zetab \left(\zeta+\zetab\right)+16 \zeta^{2} \zetab^{2}+8 \zeta \zetab\right)\log(\zeta\zetab)}{4(\zeta-\zetab)^4(1-\zeta)(1-\zetab)}\cr
	\end{align}}

\subsection{The expansion of the cross diagram}\label{subsec:Cross_expand}

Here we rederive the cross term in general $\Delta\geq1$ as an expansion in $v$ and $Y$.

We start from equation \eqref{eq:Cross_allDelta_param_explicit}, replace $v=\zeta\zetab$ and $Y=1-(1-\zeta)(1-\zetab)$ and make the coordinate transformation $\alpha_i\to\alpha_i^{-1}$. Setting $\alpha_1=1$ due to the projectivity of the integral and expanding in $Y$ we arrive at
\begin{align}
	I_\times^\Delta&=\sum\limits_{m=0}^{\infty}\frac{Y^m}{m!}\frac{\Gamma(\Delta+m)}{\Gamma(\Delta)}\int\limits_0^\infty\frac{\dd\alpha_2\dd\alpha_3(\alpha_2\alpha_3)^{\Delta-1}}{(\alpha_2+\alpha_3+\alpha_2\alpha_3)^\Delta(1+\alpha_3+\alpha_2 v)^{\Delta+m}}\cr
	&=\sum\limits_{m=0}^{\infty}\frac{Y^m}{m!}\frac{\Gamma(\Delta+m)^2}{\Gamma(2\Delta+m)}\int\limits_0^\infty\frac{\dd\alpha_3\alpha_3^{\Delta-1}}{(1+\alpha_3)^{2\Delta+m}}
	{}_2F_1\left({\Delta,\Delta+m\atop 2\Delta+m},1-\frac{\alpha_3 v}{(1+\alpha_3)^2}\right)
\end{align}
For $a,b\in\mathbb{N}$ the hypergeometric function can be expanded as
\begin{multline}
	{}_2F_1\left({a,b \atop a+b},1-z\right)=- {\Gamma(a+b)\over
		\Gamma(a)\Gamma(b)}\sum_{n\geq0}\left(\log(z)+H^{(1)}_{a+n-1}+H^{(1)}_{b+n-1}-2H^{(1)}_{n}\right)\cr
	\times{\Gamma(a+n)\Gamma(b+n)\over\Gamma(a)\Gamma(b)}  {z^n  \over n!^2},
\end{multline}
where $H^{(1)}_{n}=\sum_{r=1}^n 1/r$ is the harmonic number.
Using that
\begin{equation}
	\int_0^\infty {\alpha_3^{n+\Delta-1}\over (1+\alpha_3)^{2n+m+2\Delta}}
	\dd \alpha_3={\Gamma(\Delta+n)\Gamma(\Delta+m+n)\over\Gamma(2\Delta+2n+m)},
\end{equation}
and
\begin{multline}
	\int_0^\infty {\alpha_3^{n+\Delta-1}\over (1+\alpha_3)^{2n+m+2\Delta}}\log\left(\alpha_3\over(1+\alpha_3)^2\right)
	\dd \alpha_3={\Gamma(\Delta+n)\Gamma(\Delta+m+n)\over\Gamma(2\Delta+2n+m)}
	\cr
	\times \left(H^{(1)}_{\Delta+n-1}+H^{(1)}_{\Delta+m+n-1}-2H^{(1)}_{2\Delta+m+2n-1}\right),
\end{multline}
the expansion of $I_\times^\Delta$ reads
\begin{multline}
	I_\times^\Delta=-\sum_{n,m\geq0}
	{\Gamma(\Delta+n)^2\Gamma(\Delta+m+n)^2\over\Gamma(\Delta)^2\Gamma(2\Delta+m+2n)} {v^n Y^m\over n!^2m!} \cr
	\times\left(\log(v)+2H^{(1)}_{\Delta+n-1}+2H^{(1)}_{\Delta+m+n-1}-2H^{(1)}_{n}-2H^{(1)}_{2\Delta+m+2n-1}
	\right).
\end{multline} 
This expression matches the one given in~\cite{Dolan:2000ut}.

	%%%%%%%%%%%%%%%%%%%%%%%%%%%%%%%%%%%%%%%%%%%%%%%%%%%%%%%%%%
	\section{Evaluation of the one-loop Witten bubble diagram}
	\label{sec:special functions}
	%%%%%%%%%%%%%%%%%%%%%%%%%%%%%%%%%%%%%%%%%%%%%%%%%%%%%%%%%%
	In this appendix we give the expressions for the evaluation of
        the one-loop Witten bubble diagram  in dimensional
        regularisation for $\Delta=1$ and $\Delta=2$.

	%%%%%%%%%%%%%%%%%%%%%%%%%%%%%%%%%%%%%%%%%%%%%%%%%%%%%%%%%%
	\subsection{The one-loop  diagram}
	\label{subsec:Bubble_exact}
	%%%%%%%%%%%%%%%%%%%%%%%%%%%%%%%%%%%%%%%%%%%%%%%%%%%%%%%%%%
	The general integrals to be solved in dimensional regularisation are given by:
	\begin{align}
		\label{eq:W1_fin_div}
   	\cW_{1,\mathrm{div}}^{\Delta,4-2\epsilon,s}=&\frac12\frac{(\zeta\zetab)^\Delta}{(x_{12}^2x_{34}^2)^\Delta}\int_{\mathbb{R}^{2D}}\frac{\dd^{4-2\epsilon} X_1\dd^{4-2\epsilon} X_2(u\cdot X_1)^{2\Delta-2}(u\cdot X_2)^{2\Delta-2}}{\|X_1|^{2\Delta}\|X_1-u_\zeta|^{2\Delta}\|X_2-u_1|^{2\Delta-4\epsilon}\|X_1-u_1|^{-4\epsilon}\|X_1-X_2|^4}\cr
		\cW_{1,\mathrm{div}}^{\Delta,4-2\epsilon,t}=&\frac12\frac{(\zeta\zetab)^\Delta}{(x_{12}^2x_{34}^2)^\Delta}\int_{\mathbb{R}^{2D}}\frac{\dd^{4-2\epsilon} X_1\dd^{4-2\epsilon} X_2(u\cdot X_1)^{2\Delta-2}(u\cdot X_2)^{2\Delta-2}}{\|X_1|^{2\Delta}\|X_2-u_\zeta|^{2\Delta}\|X_2-u_1|^{2\Delta-4\epsilon}\|X_1-u_1|^{-4\epsilon}\|X_1-X_2|^4}\cr
		\cW_{1,\mathrm{div}}^{\Delta,4-2\epsilon,u}=&\frac12\frac{(\zeta\zetab)^\Delta}{(x_{12}^2x_{34}^2)^\Delta}\int_{\mathbb{R}^{2D}}\frac{\dd^{4-2\epsilon} X_1\dd^{4-2\epsilon} X_2(u\cdot X_1)^{2\Delta-2}(u\cdot X_2)^{2\Delta-2}}{\|X_1|^{2\Delta}\|X_2-u_\zeta|^{2\Delta}\|X_1-u_1|^{2\Delta-4\epsilon}\|X_2-u_1|^{-4\epsilon}\|X_1-X_2|^4}\cr
		\cW_{1,\mathrm{fin}}^{\Delta,4,s}=&\frac12\frac{(\zeta\zetab)^\Delta}{(x_{12}^2x_{34}^2)^\Delta}\int_{\mathbb{R}^{8}}\frac{\dd^{4} X_1\dd^{4} X_2(u\cdot X_1)^{2\Delta-3}(u\cdot X_2)^{2\Delta-3}}{\|X_1|^{2\Delta}\|X_1-u_\zeta|^{2\Delta}\|X_2-u_1|^{2\Delta}\|X_1-X_2|^2}\cr
		\cW_{1,\mathrm{fin}}^{\Delta,4,t}=&\frac12\frac{(\zeta\zetab)^\Delta}{(x_{12}^2x_{34}^2)^\Delta}\int_{\mathbb{R}^{8}}\frac{\dd^{4} X_1\dd^{4} X_2(u\cdot X_1)^{2\Delta-3}(u\cdot X_2)^{2\Delta-3}}{\|X_1|^{2\Delta}\|X_2-u_\zeta|^{2\Delta}\|X_2-u_1|^{2\Delta}\|X_1-X_2|^2}\cr
		\cW_{1,\mathrm{fin}}^{\Delta,4,u}=&\frac12\frac{(\zeta\zetab)^\Delta}{(x_{12}^2x_{34}^2)^\Delta}\int_{\mathbb{R}^{8}}\frac{\dd^{4} X_1\dd^{4} X_2(u\cdot X_1)^{2\Delta-3}(u\cdot X_2)^{2\Delta-3}}{\|X_1|^{2\Delta}\|X_2-u_\zeta|^{2\Delta}\|X_1-u_1|^{2\Delta}\|X_1-X_2|^2}
	\end{align}
	
	The auxiliary integrals used to obtain the parametric representation of the finite integrals for $\Delta=2$ are given by
		\begin{align}
		\label{eq:W12finite_aux}
		\tilde \cW_{1,\mathrm{fin}}^{2,4,s}&=\frac{1}{8}\int_{\mathbb{R}^8}\frac{\dd^4 X_1\dd^4 X_2}{\|X_1-\vec x_1|^2\|X_1-\vec x_2|^4\|X_2-\vec x_3|^4\|X_2-\vec x_4|^2\|X_1-X_2|^2}\cr
		&=\frac{1}{8}\frac{x_{14}^2}{x_{12}^4x_{34}^4}(\zeta\zetab)^2\int_{\mathbb{R}^8}\frac{\dd^4 X_1\dd^4 X_2}{\|X_1|^2\|X_1-\uZ|^4\|X_2-u_1|^2\|X_1-X_2|^2}\cr
		\tilde \cW_{1,\mathrm{fin}}^{2,4,t}&=\frac{1}{8}\int_{\mathbb{R}^8}\frac{\dd^4 X_1\dd^4 X_2}{\|X_1-\vec x_1|^2\|X_1-\vec x_3|^4\|X_2-\vec x_2|^2\|X_2-\vec x_4|^4\|X_1-X_2|^2}\cr
		&=\frac{1}{8}\frac{\zeta\zetab}{x_{12}^2x_{34}^4}\int_{\mathbb{R}^8}\frac{\dd^4 X_1\dd^4 X_2}{\|X_1|^2\|X_2-u_1|^4\|X_2-\uZ|^2\|X_1-X_2|^2}\cr
		\tilde \cW_{1,\mathrm{fin}}^{2,4,u}&=\frac{1}{8}\int_{\mathbb{R}^8}\frac{\dd^4 X_1\dd^4 X_2}{\|X_1-\vec x_1|^2\|X_1-\vec x_4|^4\|X_2-\vec x_2|^2\|X_2-\vec x_3|^4\|X_1-X_2|^2}\cr
		&=\frac{1}{8}\frac{\zeta\zetab}{x_{12}^2x_{34}^4}\int_{\mathbb{R}^8}\frac{\dd^4 X_1\dd^4 X_2}{\|X_1|^2\|X_1-u_1|^4\|X_2-\uZ|^2\|X_1-X_2|^2}\,.
	\end{align}
	
	The integrals to be solved in the AdS-invariant regularisation are given by:
		\begin{align}
		\label{eq:W1_AdSinv_appendix} 
		\cW_{1}^{\Delta,\delta,s}=&\frac14\frac{(\zeta\zetab)^\Delta}{(x_{12}^2x_{34}^2)^\Delta}\int_{\mathbb{R}^8}\frac{\dd^{4} X_1\dd^{4} X_2 z_1^{2\Delta-4}z_2^{2\Delta-4}}{\|X_1|^{2\Delta}\|X_1-u_\zeta|^{2\Delta}\|X_2-u_1|^{2\Delta}}\left(\frac{K^\delta(\mathbf X_1,\mathbf
			X_2)^{\Delta}}{1-K^\delta(\mathbf X_1,\mathbf X_2)^2}\right)^2\cr
		\cW_{1}^{\Delta,\delta,t}=&\frac14\frac{(\zeta\zetab)^\Delta}{(x_{12}^2x_{34}^2)^\Delta}\int_{\mathbb{R}^8}\frac{\dd^{4} X_1\dd^{4} X_2z_1^{2\Delta-4}z_2^{2\Delta-4}}{\|X_2|^{2\Delta}\|X_1-u_\zeta|^{2\Delta}\|X_1-u_1|^{2\Delta}}\left(\frac{K^\delta(\mathbf X_1,\mathbf
			X_2)^{\Delta}}{1-K^\delta(\mathbf X_1,\mathbf X_2)^2}\right)^2\cr
		\cW_{1}^{\Delta,\delta,u}=&\frac14\frac{(\zeta\zetab)^\Delta}{(x_{12}^2x_{34}^2)^\Delta}\int_{\mathbb{R}^8}\frac{\dd^{4} X_1\dd^{4} X_2z_1^{2\Delta-4}z_2^{2\Delta-4}}{\|X_1|^{2\Delta}\|X_2-u_\zeta|^{2\Delta}\|X_1-u_1|^{2\Delta}}\left(\frac{K^\delta(\mathbf X_1,\mathbf
			X_2)^{\Delta}}{1-K^\delta(\mathbf X_1,\mathbf X_2)^2}\right)^2\,.
	\end{align}
	%%%%%%%%%%%%%%%%%%%%%%%%%%%%%%%%%%%%%%%%%%%%%%%%%%%%%%%%%%
	\subsubsection{\texorpdfstring{$\Delta=1$}{Lg}}
	\label{subsubsec:Delta1_exact}
	%%%%%%%%%%%%%%%%%%%%%%%%%%%%%%%%%%%%%%%%%%%%%%%%%%%%%%%%%%
	\paragraph{The finite integrals} are the $L_0^\prime$ integrals which are discussed in detail in appendix~\ref{subsubsec:L0primeintegrals}
	
	\paragraph{The divergent integrals} in the parametric representation are given by
	{\footnotesize\begin{align}
		\label{eq:W1div_param}
		\cW_{1,\mathrm{div}}^{1,4-2\epsilon,s}&=\frac{\pi^{4-2\epsilon}\zeta\zetab}{\Gamma(-2\epsilon)}
		\int\limits_{(\mathbb{RP}^+)^4}\prod\limits_{i=1}^5\dd\alpha_i\cr
		&\cdot\frac{\alpha_3^{-1-2\epsilon}\alpha_1^{-2\epsilon}\alpha_5((\alpha_2+\alpha_3+\alpha_4)\alpha_5+(\alpha_2+\alpha_3+\alpha_4+\alpha_5)\alpha_1)^{-1-\epsilon}}{\left(\alpha_4(\alpha_3\alpha_5+\alpha_1(\alpha_3+\alpha_5))(1-\zeta)(1-\zetab)+\alpha_2\alpha_4(\alpha_1+\alpha_5)\zeta\zetab+\alpha_2(\alpha_3\alpha_5+\alpha_1(\alpha_3+\alpha_5))\right)^{1-2\epsilon}}\cr
		\cW_{1,\mathrm{div}}^{1,4-2\epsilon,t}&=\frac{\pi^{4-2\epsilon}\zeta\zetab}{\Gamma(-2\epsilon)}
		\int\limits_{(\mathbb{RP}^+)^4}\prod\limits_{i=1}^5\dd\alpha_i\cr &\cdot\frac{\alpha_2^{-1-2\epsilon}\alpha_3^{-2\epsilon}\alpha_5((\alpha_1+\alpha_2)(\alpha_3+\alpha_4)+(\alpha_1+\alpha_2+\alpha_3+\alpha_4)\alpha_5)^{-1-\epsilon}}{\left(\alpha_4((\alpha_1+\alpha_2)\alpha_3+(\alpha_2+\alpha_3)\alpha_5)(1-\zeta)(1-\zetab)+\alpha_1\alpha_2(\alpha_3+\alpha_4+\alpha_5)+\alpha_1\alpha_5(\alpha_3+\alpha_4\zeta\zetab)\right)^{1-2\epsilon}}\cr
		\cW_{1,\mathrm{div}}^{1,4-2\epsilon,u}&=\frac{\pi^{4-2\epsilon}\zeta\zetab}{\Gamma(-2\epsilon)}
		\int\limits_{(\mathbb{RP}^+)^4}\prod\limits_{i=1}^5\dd\alpha_i\cr &\cdot\frac{\alpha_1^{-1-2\epsilon}\alpha_4^{-2\epsilon}\alpha_5((\alpha_1+\alpha_2)(\alpha_3+\alpha_4)+(\alpha_1+\alpha_2+\alpha_3+\alpha_4)\alpha_5)^{-1-\epsilon}}{\left(\alpha_3\alpha_4\alpha_5+\alpha_1\alpha_3(\alpha_4+\alpha_5)+\alpha_2(\alpha_4\alpha_5+\alpha_1(\alpha_3+\alpha_4+\alpha_5))(1-\zeta)(1-\zetab)+\alpha_2\alpha_3(\alpha_4+\alpha_5\zeta\zetab)\right)^{1-2\epsilon}}\cr
\end{align}}

The solution to the integrals \eqref{eq:W1div_param} up to $\Op(\epsilon^0)$ is given by
{
	\begin{align}
	    \label{eq:W1div_res}
		\cW_{1,\mathrm{div}}^{\Delta,4-2\epsilon,s}=&-\frac{\pi^{4-2\epsilon}\eul^{-2\gamma\epsilon}(\zeta\zetab)^{1-{\epsilon\over2}}((1-\zeta)(1-\zetab))^{\epsilon}
                           }{2x_{12}^2x_{34}^2}\left(\frac{1}{\epsilon}\frac{4i
                           D(\zeta,\zetab)}{\zeta-\zetab}+\frac{f_1(\zeta,\zetab)}{\zeta-\zetab}+\mathcal  O(\epsilon)\right),\\
		\cW_{1,\mathrm{div}}^{\Delta,4-2\epsilon,t}=&-\frac{\pi^{4-2\epsilon}\eul^{-2\gamma\epsilon}(\zeta\zetab)^{1-\epsilon}((1-\zeta)(1-\zetab))^{3\epsilon\over2}}{2x_{12}^2x_{34}^2}\left(\frac{1}{\epsilon}\frac{4i D(\zeta,\zetab)}{\zeta-\zetab}+\frac{f_1(\zeta,\zetab)}{\zeta-\zetab}+\mathcal  O(\epsilon)\right),\\
		\cW_{1,\mathrm{div}}^{\Delta,4-2\epsilon,u}=&-\frac{\pi^{4-2\epsilon}\eul^{-2\gamma\epsilon}(\zeta\zetab)^{1-\epsilon}((1-\zeta)(1-\zetab))^{\epsilon}}{2x_{12}^2x_{34}^2}\left(\frac{1}{\epsilon}\frac{4i D(\zeta,\zetab)}{\zeta-\zetab}+\frac{f_1(\zeta,\zetab)}{\zeta-\zetab}+\mathcal  O(\epsilon)\right).
	\end{align}}

	%%%%%%%%%%%%%%%%%%%%%%%%%%%%%%%%%%%%%%%%%%%%%%%%%%%%%%%%%%
	\subsubsection{\texorpdfstring{$\Delta=2$}{Lg}}
	\label{subsubsec:Delta2_exact}
	%%%%%%%%%%%%%%%%%%%%%%%%%%%%%%%%%%%%%%%%%%%%%%%%%%%%%%%%%%
	\paragraph{The finite integrals} are given by:
	{\footnotesize\begin{align}
			\label{eq:W2fin_param}
			W_{1,\mathrm{fin}}^{2,4,s}=&\frac{\pi^4}{2}\frac{(\zeta\zetab)^2}{(x_{12}x_{34})^4}\int\limits_{(\mathbb{RP}^+)^3}\prod\limits_{i=1}^4\dd\alpha_i
			\frac{\alpha_1 \alpha_2 \alpha_3 \alpha_4(\alpha_4 (\alpha_1+\alpha_2+\alpha_3)+\alpha_3 (\alpha_1+\alpha_2))^{-1}}{(\alpha_1 \alpha_2 (\alpha_3+\alpha_4)\zeta\zetab+\alpha_1 \alpha_3 \alpha_4 (1-\zeta)(1-\zetab)+\alpha_2 \alpha_3 \alpha_4)^2},\cr
			W_{1,\mathrm{fin}}^{2,4,t}=&\frac{\pi ^4 }{2}\frac{(\zeta\zetab)^2}{(x_{12}x_{34})^4}\int\limits_{(\mathbb{RP}^+)^3}\prod\limits_{i=1}^4\dd\alpha_i\frac{\alpha_1 \alpha_2 \alpha_3 \alpha_4(\alpha_4 (\alpha_1+\alpha_2+\alpha_3)+\alpha_2 (\alpha_1+\alpha_3))^{-1}}{(\alpha_1 \alpha_3 (\alpha_2+\alpha_4)(1-\zeta)(1-\zetab)+\alpha_1 \alpha_2 \alpha_4 \zeta\zetab+\alpha_2 \alpha_3 \alpha_4)^2},\cr
			W_{1,\mathrm{fin}}^{2,4,u}=&\frac{\pi ^4 }{2}\frac{(\zeta\zetab)^2}{(x_{12}x_{34})^4}\int\limits_{(\mathbb{RP}^+)^3}\prod\limits_{i=1}^4\dd\alpha_i\frac{\alpha_1 \alpha_2 \alpha_3 \alpha_4(\alpha_1 (\alpha_2+\alpha_3+\alpha_4)+\alpha_4 (\alpha_2+\alpha_3))^{-1}}{(\alpha_1 \alpha_2 (\alpha_3+\alpha_4 \zeta\zetab)+\alpha_1 \alpha_3 \alpha_4 (1-\zeta)(1-\zetab)+\alpha_2 \alpha_3 \alpha_4)^2}.
	\end{align}}
	
	The solution to the integrals \eqref{eq:W2fin_param} is given by
	{\footnotesize\begin{align}
			\cW_{1,\mathrm{fin}}^{2,4,s}&=\frac{\pi^4}{8}\frac{(\zeta\zetab)^2}{(x_{12}x_{34})^4}\left(\frac{(\zeta+\zetab-2)8iD(\zeta,\zetab)}{(\zeta-\zetab)^3}+\frac{(4\zeta-2)\zetab-2\zeta}{\zeta\zetab(\zeta-\zetab)^2}\log((1-\zeta)(1-\zetab))
			-\frac{4\log(\zeta\zetab)}{(\zeta-\zetab)^2}\right)\cr
			\cW_{1,\mathrm{fin}}^{2,4,t}&=\frac{\pi^4}{8}\frac{(\zeta\zetab)^2}{(x_{12}x_{34})^4}\left(-\frac{(\zeta+\zetab)8iD(\zeta,\zetab)}{(\zeta-\zetab)^3}+\frac{(4\zeta-2)\zetab-2\zeta}{(1-\zeta)(1-\zetab)(\zeta-\zetab)^2}\log(\zeta\zetab)
			-\frac{4\log((1-\zeta)(1-\zetab))}{(\zeta-\zetab)^2}\right)\cr
			\cW_{1,\mathrm{fin}}^{2,4,u}&=\frac{\pi^4}{8}\frac{(\zeta\zetab)^2}{(x_{12}x_{34})^4}\left(-\frac{((4\zeta-2)\zetab-2\zeta)4iD(\zeta,\zetab)}{(\zeta-\zetab)^3}+\frac{2(\zeta+\zetab)}{(\zeta-\zetab)^2}\log(\zeta\zetab)-\frac{2(\zeta+\zetab-2)\log((1-\zeta)(1-\zetab))}{(\zeta-\zetab)^2}\right)
	\end{align}}

	\paragraph{The divergent integrals} are given by:
	{\footnotesize\begin{align}
			\label{eq:W2div_param}
			\cW_{1,\mathrm{div}}^{2,4-2\epsilon,s}&=\frac{4\pi^{4-2\epsilon}(\zeta\zetab)^2}{16\Gamma(-2\epsilon)}
			\int\limits_{(\mathbb{RP}^+)^4}\prod\limits_{i=1}^5\dd\alpha_i F_s(\zeta,\zetab,\epsilon;\alpha_1,\alpha_2,\alpha_3,\alpha_4,\alpha_5)\cr
			&\times\frac{\alpha_3^{-1-2\epsilon}\alpha_1^{-2\epsilon}\alpha_5((\alpha_2+\alpha_3+\alpha_4)\alpha_5+(\alpha_2+\alpha_3+\alpha_4+\alpha_5)\alpha_1)^{-1-\epsilon}}{\left(\alpha_4(\alpha_3\alpha_5+\alpha_1(\alpha_3+\alpha_5))(1-\zeta)(1-\zetab)+\alpha_2\alpha_4(\alpha_1+\alpha_5)\zeta\zetab+\alpha_2(\alpha_3\alpha_5+\alpha_1(\alpha_3+\alpha_5))\right)^{3-2\epsilon}}\cr
			\cW_{1,\mathrm{div}}^{2,4-2\epsilon,t}&=\frac{4\pi^{4-2\epsilon}(\zeta\zetab)^2}{16\Gamma(-2\epsilon)}
			\int\limits_{(\mathbb{RP}^+)^4}\prod\limits_{i=1}^5\dd\alpha_i F_t(\zeta,\zetab,\epsilon;\alpha_1,\alpha_2,\alpha_3,\alpha_4,\alpha_5)\cr
			&\times\frac{\alpha_2^{-1-2\epsilon}\alpha_3^{-2\epsilon}\alpha_5((\alpha_1+\alpha_2)(\alpha_3+\alpha_4)+(\alpha_1+\alpha_2+\alpha_3+\alpha_4)\alpha_5)^{-1-\epsilon}}{\left(\alpha_4((\alpha_1+\alpha_2)\alpha_3+(\alpha_2+\alpha_3)\alpha_5)(1-\zeta)(1-\zetab)+\alpha_1\alpha_2(\alpha_3+\alpha_4+\alpha_5)+\alpha_1\alpha_5(\alpha_3+\alpha_4\zeta\zetab)\right)^{3-2\epsilon}}\cr
			\cW_{1,\mathrm{div}}^{2,4-2\epsilon,u}&=\frac{4\pi^{4-2\epsilon}(\zeta\zetab)^2}{16\Gamma(-2\epsilon)}
			\int\limits_{(\mathbb{RP}^+)^4}\prod\limits_{i=1}^5\dd\alpha_i F_u(\zeta,\zetab,\epsilon;\alpha_1,\alpha_2,\alpha_3,\alpha_4,\alpha_5)\cr
			&\times\frac{\alpha_1^{-1-2\epsilon}\alpha_4^{-2\epsilon}\alpha_5((\alpha_1+\alpha_2)(\alpha_3+\alpha_4)+(\alpha_1+\alpha_2+\alpha_3+\alpha_4)\alpha_5)^{-1-\epsilon}}{\left(\alpha_3\alpha_4\alpha_5+\alpha_1\alpha_3(\alpha_4+\alpha_5)+\alpha_2(\alpha_4\alpha_5+\alpha_1(\alpha_3+\alpha_4+\alpha_5))(1-\zeta)(1-\zetab)+\alpha_2\alpha_3(\alpha_4+\alpha_5\zeta\zetab)\right)^{3-2\epsilon}}\cr
	\end{align}}

The expansion of the prefactors starts at $\Op(\epsilon)$ so only integrals that diverge at least with $\epsilon^{-1}$ contribute to the final result. When only keeping those terms, the functions $F_s,F_t$ and $F_u$ are given by:
\begin{align}
	F_s&=C_1(\alpha_1^2\alpha_2\alpha_4\alpha_5^2+2\alpha_1\alpha_2\alpha_3\alpha_4\alpha_5^2+\alpha_2\alpha_3^2\alpha_4\alpha_5^2)\cr
	&+C_2(\alpha_1\alpha_2^2\alpha_4\alpha_5^2+\alpha_2^2\alpha_3\alpha_4\alpha_5^2+\alpha_1^2\alpha_2^2\alpha_4\alpha_5)+C_3(\alpha_1^2\alpha_4^2\alpha_5^2+2\alpha_1\alpha_3\alpha_4^2\alpha_5^2+\alpha_3^2\alpha_4^2\alpha_5^2)\cr
	&+C_4(\alpha_1^2\alpha_2\alpha_4^2\alpha_5+\alpha_1\alpha_2\alpha_4^2\alpha_5^2+\alpha_2\alpha_3\alpha_4^2\alpha_5^2)+C_5(2\alpha_1\alpha_2^2\alpha_4^2\alpha_5+\alpha_2^2\alpha_4^2\alpha_5^2+\alpha_1^2\alpha_2^2\alpha_4^2)\cr
	&+\alpha_1^2\alpha_2^2\alpha_5^2+2\alpha_1\alpha_2^2\alpha_3\alpha_5^2+\alpha_2^2\alpha_3^2\alpha_5^2\\
	F_t&=C_1(\alpha_1^2\alpha_3^2\alpha_4\alpha_5+\alpha_1\alpha_2^2\alpha_4\alpha_5^2+2\alpha_1\alpha_2\alpha_3\alpha_4\alpha_5^2+\alpha_1\alpha_3^2\alpha_4\alpha_5^2)+C_2(\alpha_1^2\alpha_2\alpha_4\alpha_5^2+\alpha_1^2\alpha_3\alpha_4\alpha_5^2)\cr
	&+C_3(\alpha_2^2\alpha_4^2\alpha_5^2+2\alpha_2\alpha_3\alpha_4^2\alpha_5^2+\alpha_3^2\alpha_4^2\alpha_5^2+\alpha_1^2\alpha_3^2\alpha_4^2+2\alpha_1\alpha_3^2\alpha_4^2\alpha_5)\cr
	&+C_4(\alpha_1^2\alpha_3\alpha_4^2\alpha_5+\alpha_1\alpha_2\alpha_4^2\alpha_5^2+\alpha_1\alpha_3\alpha_4^2\alpha_5^2)+C_5\alpha_1^2\alpha_4^2\alpha_5^2\cr
	&+\alpha_1^2\alpha_2^2\alpha_5^2+2\alpha_1^2\alpha_2\alpha_3\alpha_5^2+\alpha_1^2\alpha_3^2\alpha_5^2\\
	F_u&=C_1(\alpha_1^2\alpha_2\alpha_3\alpha_5+2\alpha_1\alpha_2\alpha_3\alpha_4\alpha_5^2+\alpha_2^2\alpha_3\alpha_4^2\alpha_5+\alpha_2\alpha_3\alpha_4^2\alpha_5^2)\cr
	&+C_2(\alpha_1\alpha_2\alpha_3^2\alpha_5^2+\alpha_2^2\alpha_3^2\alpha_4\alpha_5+\alpha_2\alpha_3^2\alpha_4\alpha_5^2)+C_3(\alpha_1^2\alpha_2^2\alpha_5^2+2\alpha_1\alpha_2^2\alpha_4\alpha_5^2+\alpha_2^2\alpha_4^2\alpha_5^2)\cr
	&+C_4(\alpha_1\alpha_2^2\alpha_3\alpha_5^2+\alpha_2^2\alpha_3\alpha_4\alpha_5^2)+C_5\alpha_2^2\alpha_3^2\alpha_5^2+\alpha_1^2\alpha_3^2\alpha_5^2+2\alpha_1\alpha_3^2\alpha_4\alpha_5^2+\alpha_2^2\alpha_3^2\alpha_4^2\cr
	&+2\alpha_2\alpha_3^2\alpha_4^2\alpha_5+\alpha_3^2\alpha_4^2\alpha_5^2
\end{align}
with the coeffifients $C_i$ given in~\eqref{eq:Ccoefficients}.

%%%%%%%%%%%%%%%%%%%%%%%%%%%%%%%%%%%%%%%%%%%%%%%%%%%%%%%%%%
\subsubsection{\texorpdfstring{$L_0^\Delta$}{Lg} integrals}
\label{subsubsec:L0integrals}
%%%%%%%%%%%%%%%%%%%%%%%%%%%%%%%%%%%%%%%%%%%%%%%%%%%%%%%%%%
The $L_0^\Delta$ pieces appearing in the finite part of the one-loop bubble integrals of $\Delta=1$ and $\Delta=2$ are given by:
\begin{align}
	\label{eq:L0integral}
	L^{\Delta}_0(\mathbf{x},\mathbf{y},\mathbf{z})&=\int\limits_0^{\infty}\dd\sigma\int\limits_0^1\dd\varrho\frac{(\sigma\varrho(1-\varrho))^{\Delta-1}\log(1+\sigma)}
	{(1+\sigma)^{\Delta}\left(\sigma\varrho(1-\varrho)\mathbf{x}+\varrho\mathbf{y}+(1-\varrho)\mathbf{z}\right)^{\Delta}}
\end{align}
Where the three channels are given by 
\begin{itemize}
	\item $s$-channel: $\mathbf x\to v$, $\mathbf y\to 1-Y$, $\mathbf z\to 1$
	\item $t$-channel: $\mathbf x\to 1-Y$, $\mathbf y\to v$, $\mathbf z\to 1$
	\item $u$-channel: $\mathbf x\to 1$, $\mathbf y\to 1-Y$, $\mathbf z\to v$
\end{itemize}
They are linearly reducible and given by single valued polylogarithms of maximal weight three
\paragraph{For $\Delta=1$} we have
\begin{align}
	L_0^{1,s}(\zeta,\bar\zeta)&={f_1(\zeta,\bar\zeta)-2i
		\log(\zeta\bar\zeta) D(\zeta,\bar\zeta)\over
		\zeta-\bar\zeta}  \\
	L_0^{1,t}(\zeta,\bar\zeta)&={f_1(\zeta,\bar\zeta)-2i
		\log((1-\zeta)(1-\bar\zeta)) D(\zeta,\bar\zeta)\over
		\zeta-\bar\zeta}\\
	L_0^{1,u}(\zeta,\bar\zeta)&={f_1(\zeta,\bar\zeta)\over
		\zeta-\bar\zeta} \,.
\end{align}

\paragraph{ For $\Delta=2$} we have
\begin{multline}
	L_0^{2,s}(\zeta,\bar\zeta)\cdot(\zeta-\bar\zeta)^5= \left(\left( \zeta+\bar\zeta \right) ^{2}-3\, \left( \zeta+\bar\zeta
	\right) \zeta\,\bar\zeta+2\,\zeta\,\bar\zeta\right) f_{3}(\zeta,\bar\zeta)\cr
	+ \left(- \left( \zeta+\bar\zeta \right) ^{3}+2\, \left( \zeta+\bar\zeta
	\right) ^{2}\zeta\,\bar\zeta+2\, \left( \zeta+\bar\zeta \right) ^
	{2}-8\, \left( \zeta+\bar\zeta \right) \zeta\,\bar\zeta+4\,{\zeta}
	^{2}{\bar\zeta}^{2}+4\,\zeta\,\bar\zeta
	\right) f_{1}(\zeta,\bar\zeta)\cr
	-2\,i \left( 2\,{\zeta}
	^{3}\bar\zeta+8\,{\zeta}^{2}{\bar\zeta}^{2}+2\,\zeta\,{\bar\zeta
	}^{3}-{\zeta}^{3}-11\,{\zeta}^{2}\bar\zeta-11\,\zeta\,{\bar\zeta}^
	{2}-{\bar\zeta}^{3}+2\,{\zeta}^{2}+8\,\zeta\,\bar\zeta+2\,{\bar\zeta}^{2}
	\right) \ln  \left( \zeta\,\bar\zeta \right) D(\zeta,\bar\zeta) \cr
	-4\,i
	\left( {\zeta}^{3}\bar\zeta+6\,{\zeta}^{2}{\bar\zeta}^{2}+\zeta\,
	{\bar\zeta}^{3}-{\zeta}^{3}-7\,{\zeta}^{2}\bar\zeta-7\,\zeta\,{\bar\zeta}^{2}-{\bar\zeta}^{3}+2\,{\zeta}^{2}+4\,\zeta\,\bar\zeta
	+2\,{\bar\zeta}^{2} \right) D(\zeta,\bar\zeta)\cr
	-2\, \left( \zeta-\bar\zeta \right) \zeta\,\bar\zeta\, \left( \zeta+\bar\zeta-2
	\right) \ln  \left( \zeta\,\bar\zeta \right) \cr
	+ \left( 2\,\zeta\,\bar\zeta-\zeta-\bar\zeta \right)  \left( \zeta-\bar\zeta
	\right)  \left( \zeta+\bar\zeta-2 \right) \ln  \left(  \left( -1+
	\zeta \right)  \left( -1+\bar\zeta \right)  \right) +2\, \left( 
	\zeta-\bar\zeta \right) ^{3}
\end{multline}
\begin{multline}
	L_0^{2,t}(\zeta,\bar\zeta)\cdot(\zeta-\bar\zeta)^5=\left(  \left( 3\,\zeta-2 \right) {\bar\zeta}^{2}+ \left( 3\,{\zeta
	}^{2}-8\,\zeta+3 \right) \bar\zeta-2\,{\zeta}^{2}+3\,\zeta \right)
	f_4 (\zeta,\bar\zeta)\cr
	+ \left(  \left( 2\,\zeta-1 \right) {\bar\zeta}^{3}+ \left( 8
	\,{\zeta}^{2}-11\,\zeta+2 \right) {\bar\zeta}^{2}+ \left( 2\,{\zeta}
	^{3}-11\,{\zeta}^{2}+8\,\zeta \right) \bar\zeta-{\zeta}^{3}+2\,{
		\zeta}^{2} \right) f_{1}(\zeta,\bar\zeta)\cr
	+  2\,i\big(  \left( -2\zeta+1
	\right) {\bar\zeta}^{3}-\left( 8\,{\zeta}^{2}-11\,\zeta+2
	\right) {\bar\zeta}^{2}-\left( 2\,{\zeta}^{3}-11\,{\zeta}^{2}+
	8\,\zeta \right) \bar\zeta\cr
	+\left( -2+\zeta \right) {\zeta}^{2}
	\big) \ln  \left(  \left( 1-\zeta \right)  \left( 1-\bar\zeta
	\right)  \right)  D(\zeta,\bar\zeta)\cr
	-4\,i\left(\zeta\,{\bar\zeta}^{3}+\left( 6\,{\zeta}
	^{2}-8\,\zeta+1 \right) {\bar\zeta}^{2}+\zeta\, \left( {\zeta}^{
		2}-8\,\zeta+6 \right) \bar\zeta+{\zeta}^{2} \right) D(\zeta,\bar\zeta)\cr
	- \left( 2\,\zeta\,\bar\zeta-\zeta-\bar\zeta \right)  \left( {
		\zeta}^{2}-{\bar\zeta}^{2} \right) \ln  \left( \zeta\,\bar\zeta
	\right) +2\, \left( 1-\bar\zeta \right)  \left( 1-\zeta \right) 
	\left( \zeta-\bar\zeta \right)  \left( \zeta+\bar\zeta \right) 
	\ln  \left(  \left( 1-\zeta \right)  \left( 1-\bar\zeta \right) 
	\right) \cr
	+2\, \left( \zeta-\bar\zeta \right) ^{3}
\end{multline}
\begin{multline}
	L_0^{2,u}(\zeta,\bar\zeta)\cdot(\zeta-\bar\zeta)^5= \left( {\zeta}^{2}+4\,\zeta\,\bar\zeta+{\bar\zeta}^{2}-3\,\zeta-3
	\,\bar\zeta \right) f_{5}(\zeta,\bar\zeta)\cr
	+ \left( 2\,{\zeta}^{3}\bar\zeta+8\,{
		\zeta}^{2}{\bar\zeta}^{2}+2\,\zeta\,{\bar\zeta}^{3}-{\zeta}^{3}-11
	\,{\zeta}^{2}\bar\zeta-11\,\zeta\,{\bar\zeta}^{2}-{\bar\zeta}^{3
	}+2\,{\zeta}^{2}+8\,\zeta\,\bar\zeta+2\,{\bar\zeta}^{2} \right)
	f_{1}(\zeta,\bar\zeta)\cr
	-4\,i \left( 2\,{\zeta}^{3}\bar\zeta+4\,{\zeta}^{2}{\bar\zeta}^{2}+2\,\zeta\,{\bar\zeta}^{3}-{\zeta}^{3}-7\,{\zeta}^{2}\bar\zeta-7\,\zeta\,{\bar\zeta}^{2}-{\bar\zeta}^{3}+{\zeta}^{2}+6
	\,\zeta\,\bar\zeta+{\bar\zeta}^{2} \right) D(\zeta,\bar\zeta)\cr
	- \left( 2
	\,\zeta\,\bar\zeta-\zeta-\bar\zeta \right)  \left( \zeta-\bar\zeta
	\right)  \left( \zeta+\bar\zeta \right) \ln  \left(
	\zeta\,\bar\zeta \right) \cr
	+ \left( 2\,\zeta\,\bar\zeta-\zeta-\bar\zeta
	\right)  \left( \zeta-\bar\zeta \right)  \left( \zeta+\bar\zeta-2
	\right) \ln  \left(  \left( 1-\zeta \right)  \left( 1-\bar\zeta
	\right)  \right) +2\, \left( \zeta-\bar\zeta \right) ^{3}
\end{multline}
%%%%%%%%%%%%%%%%%%%%%%%%%%%%%%%%%%%%%%%%%%%%%%%%%%%%%%%%%%
\subsubsection{\texorpdfstring{$L_0'$}{Lg} integrals}
\label{subsubsec:L0primeintegrals}
%%%%%%%%%%%%%%%%%%%%%%%%%%%%%%%%%%%%%%%%%%%%%%%%%%%%%%%%%%
	The finite integrals for $\Delta=1$ are much harder to evaluate since they involve elliptic integrals in the parametric representation. Therefore we were not able to find closed expressions. But as the main goal of this work is to extract anomalous dimensions of the double-trace operators in the dual CFT, we are mainly interested in the coefficients of the $\log(v)^n$ terms. After identifying these terms the rest of the integral is finite and we can expand the integrand in powers of $v$ and $Y$ and integrate over the coefficients.
	
	Let us first note that the integrals involved in the finite piece are all of the form
	\begin{equation}
		\label{eq:Iv1v2}
		I(v_1,v_2):=\int_{\mathbb R^8}
		{\dd^{4}X\dd^{4}Y\over \|X|^2 \|Y-v_1|^2
			\|Y-v_2|^2 \|X-Y|^2 \, u\cdot X \, u\cdot Y}.
	\end{equation}
	Comparing with~\eqref{eq:W1_fin_div} we recognise the finite pieces of the different channels as:
	\begin{align}
			\cW_{1,\mathrm{fin}}^{1,4,s}=&\frac12\frac{(\zeta\zetab)^\Delta}{(x_{12}^2x_{34}^2)^\Delta}I(u_1,u_1-\uZ),\cr
			\cW_{1,\mathrm{fin}}^{1,4,t}=&\frac12\frac{(\zeta\zetab)^\Delta}{(x_{12}^2x_{34}^2)^\Delta}I(u_1,\uZ),\cr
			\cW_{1,\mathrm{fin}}^{1,4,u}=&\frac12\frac{(\zeta\zetab)^\Delta}{(x_{12}^2x_{34}^2)^\Delta}I(\uZ,\uZ-u_1).
	\end{align}
	A parametric representation is given by
	\begin{equation}
		I(v_1,v_2):=\pi^{4}\int\limits_{(\mathbb{RP}^+)^4}
		{d\alpha_0 \cdots d\alpha_5\over
			({\alpha_1+\alpha_2+\alpha_3\over4}\alpha_4^2+{\alpha_0+\alpha_1\over4}\alpha_5^2+{\alpha_1\over2} \alpha_4\alpha_5+\hat F)^{2}},
	\end{equation}
	with
	\begin{equation}
		\hat F=-(v_1-v_2)^2(\alpha_0+\alpha_1)\alpha_2\alpha_3 -v_1^2\alpha_0\alpha_1\alpha_2-v_2^2\alpha_0\alpha_1\alpha_3\,.
	\end{equation}
	Changing variables to
	\begin{equation}
		{\alpha_1+\alpha_2+\alpha_3\over4}\alpha_4^2+{\alpha_0+\alpha_1\over4}\alpha_4^2+{\alpha_1\over2} \alpha_4\alpha_4=
		{\alpha_1+\alpha_2+\alpha_3\over4} (\beta_4^2+\beta_5^2),
	\end{equation}
	with
	\begin{equation}
		\beta_4=\alpha_4+{\alpha_1\alpha_5\over \alpha_1+\alpha_2+\alpha_3}; \qquad \beta_5={\alpha_5\sqrt{
				\alpha_0(\alpha_1+\alpha_2+\alpha_3)+\alpha_1(\alpha_2+\alpha_3)}\over \alpha_1+\alpha_2+\alpha_3}\,.
	\end{equation}
	Setting
	\begin{equation}
		\beta_4={t \beta_5 \alpha_1\over\sqrt{  \alpha_0(\alpha_1+\alpha_2+\alpha_3)+\alpha_1(\alpha_2+\alpha_3)}},
	\end{equation}
	and
	performing the integration over $\beta_5$, and changing variables to
	$\alpha_i\to 1/\alpha_i$ we get
	\begin{multline}
		I(v_1,v_2;0):=-2\pi^{4}\int_1^\infty dt\int_0^\infty   {
			d\alpha_0 d\alpha_1d\alpha_2d\alpha_3
			\over
			(v_1-v_2)^2(\alpha_0+\alpha_1)+ v_1^2\alpha_3+v_2^2\alpha_2}\cr
		\times  {1\over  
			\alpha_1((\alpha_0+\alpha_1)(\alpha_2+\alpha_3)+\alpha_2\alpha_3)+\alpha_0\alpha_2\alpha_3 t^2 }.
	\end{multline}
	Setting $x:=(v_1-v_2)^2, y:=v_1^2$ and $z:=v_2^2$, this defines the $L_0'(x,y,z):=I(v_1,v2;0)/(4\pi^4)$ integral 
		\begin{equation}
	L_0'(x,y,z)=\int_1^\infty d\lambda\int_0^\infty ds \int_0^1 dr{\log(1+\lambda s)\over4\lambda  \sqrt{(1+s) (1+\lambda s)} (sr(1-r)   x+ r y+(1-r)z)}.
	\end{equation}
	
	For the $s$-channel we have
\begin{equation}
  (x,y,z)=(v,1-Y,1),
  \end{equation}
For the $t$-channel we have 
\begin{equation}
  (x,y,z)=(1-Y,1,v)     ,
\end{equation}
For the $u$-channel we have 
\begin{equation}
  (x,y,z)=(1,v,1-Y) .
\end{equation}

We first evaluate  the integral over $\lambda$  to get
\begin{align}
  I(s)&= \int_1^\infty  {\log(1+s\lambda)\over \lambda\sqrt{1+\lambda s}}d\lambda\\
  \nonumber &=2
   \text{Li}_2\left(\frac{1}{\sqrt{s+1}}\right)-2
                      \text{Li}_2\left(-\frac{1}{\sqrt{s+1}}\right)
  -\log (s+1) \log
   \left(\frac{\sqrt{s+1}-1}{\sqrt{s+1}+1}\right)\,.
\end{align}

For computing this integral we evaluated

\begin{align}
    \int_1^\infty  {(1+s\lambda)^{-\frac12+\epsilon}\over
  \lambda}d\lambda&= -\frac{2 s^{-\frac{1}{2}+\epsilon}
                    {}_{2}F_{1}
                    \left(\frac{1}{2}-\epsilon ,\frac{1}{2}-\epsilon
                    ;\frac{3}{2}-\epsilon ;-\frac{1}{s}\right)}{-1+2
                    \epsilon}\cr
                    &=-\log
                      \left(\frac{\sqrt{s+1}-1}{\sqrt{s+1}+1}\right)
                      +\epsilon  \Big(2
   \text{Li}_2\left(\frac{1}{\sqrt{s+1}}\right)-2
                      \text{Li}_2\left(-\frac{1}{\sqrt{s+1}}\right)\cr
  &-\log (s+1) \log
   \left(\frac{\sqrt{s+1}-1}{\sqrt{s+1}+1}\right)\Big)+O\left(\epsilon ^2\right)
\end{align}
changing variables by setting $s=1/\sigma^2-1$ we have
\begin{equation}\label{e:Isigmadef}
  I(\sigma)=2\Li{2}(\sigma)-2\Li{2}(-\sigma)  +2\log(\sigma)\left(\log(1-\sigma)-\log(1+\sigma)\right)
\end{equation}

\begin{equation}
\label{eq:L0prime_sigma}
L_0'(x,y,z)=\frac14\int_0^1 \int_0^1   {I(\sigma)\over
         \left(\sigma^{2}  -1 \right)x r^{2}+\left(\left(-x +y -z \right) \sigma^{2}+x \right) r +z \,\sigma^{2}}     dr d\sigma
      \end{equation}

      The vanishing locus of the denominator of the integral
      \begin{equation}
      \label{eq:elliptic_curve}
                  \left(\sigma^{2}  -1 \right)x r^{2}+\left(\left(-x +y -z \right) \sigma^{2}+x \right) r +z \,\sigma^{2}=0
      \end{equation}
   defines an elliptic curve. Therefore the result of the integral is
   an elliptic polylogarithm. We are not interested in the exact
   expression but in the degeneration limit of the elliptic curve for
   small $v$ and $Y$. Therefore, we only evaluate the integrals in the
   asymptotic $0\leq v\ll 1$ region.

\paragraph{$s$-channel}
We can perform the integration over $\sigma$ right away. The positive root of equation~\eqref{eq:elliptic_curve} in $\sigma$ is given by 
\begin{equation}
  \sigma(r):=\frac{\sqrt{\left(1-r\right) r v}}{\sqrt{-r^{2} v +Y r +r v -1}} \,.
\end{equation}
Note that the limit $v\to0$ coincides with $\sigma(r)\to0$, which means that the integration in $\sigma$ should provide us with the $\log(v)^2$ and $\log(v)$ divergences of the integral.\\
Indeed, performing the integration over $\sigma$ leads to 
     \begin{multline}
       L_0'(v,1-Y,1)=\frac14 \int_0^1 \log\left(1-\sigma(r)\over
         1+\sigma(r)\right) {dr\over 2 \left(-1-r^{2} v +r \left(v +Y
           \right)\right) \sigma(r)} \log(v)^2\cr
       +\int_0^1 \left(\Li{2}(\sigma(r))-\Li{2}(-\sigma(r))+i \pi
         \Li{1}(-\sigma(r))-i\pi\Li{1}(\sigma(r))\right){\log(v) dr\over
      2   \left(-1-r^{2} v +r \left(v +Y
           \right)\right) \sigma(r)}\cr
+   \int_0^1 \log\left(-i \sqrt{r(1-r)}\over \sqrt{1-Yr}\right) {\log(v) dr\over
      2   \left(-1-r^{2} v +r \left(v +Y
           \right)\right) \sigma(r)}+O(v^0)
     \end{multline}
One can perform the small $v$ series expansion under the integrals and
integrate in $r$ term by term.

  \paragraph{$t$-channel} 
  Here the $\log(v)$ divergence can be extracted from the $r$ integral. We notice that equation~\eqref{eq:L0prime_sigma} can be written as
  \begin{align}
      L_0'(1-Y,1,y)&=\frac14\int_0^1 \int_0^1{I(\sigma)\over
         (r-r_+(\sigma))(r-r_-(\sigma))}\frac{1}{(\sigma^2-1)(1-Y)}dr d\sigma\cr
         &=\frac14\int_0^1\frac{I(\sigma)\log\left(\frac{r_+(1-r_-)}{r_-(1-r_+)}\right)}{r_--r_+}d\sigma
  \end{align}
  with
  \begin{align}
      r_{\pm}&=\frac12\frac{\sigma^2(v-Y)-(1-Y)\pm\sqrt{(\sigma^2(v-Y)-(1-Y))^2-4\sigma^2v(\sigma^2-1)(1-Y)}}{\sqrt{(\sigma^2-1)(1-Y)}}
  \end{align}
  The logarithmic term in the numerator diverges with $\log(v)$ in the limit $v\to0$. The $\log(v)$ term to the integral is therefore given by
  {\footnotesize\begin{multline}
    L_0'(1-Y,1,v) = -\int_0^1 \frac{\left(\Li{2}(\sigma)-\Li{2}(-\sigma)  +\log(\sigma)\left(\log(1-\sigma)-\log(1+\sigma)\right)\right) d\sigma}{\sqrt{\left(Y^{2}+2 Y v +v^{2}-4 v \right) \sigma^{4}-2 \left(Y -1\right) \left(v +Y \right) \sigma^{2}+\left(Y -1\right)^{2}}}    \log(v)+O(v^0)
  \end{multline}}

  The integrand can be expanded for small $v$ and $Y$ and integrated
  term-by-term using the small $\sigma$ expansion
  \begin{equation}
        \Li{2}(\sigma)-\Li{2}(-\sigma)
      +\log(\sigma)\left(\log(1-\sigma)-\log(1+\sigma)\right)=
      2\sum_{n\geq0} \sigma^{2n} \left({1\over
          (2n+1)^2}-{\log(\sigma)\over 2n+1}\right) 
  \end{equation}
so that
  \begin{multline}\label{e:intL}
    \int_0^1     \left(\Li{2}(\sigma)-\Li{2}(-\sigma)
      +\log(\sigma)\left(\log(1-\sigma)-\log(1+\sigma)\right)\right)
    \sigma^{2m} d\sigma\cr ={\pi^2\over 6 (1+2m)}+{1\over 2(1+2m)}
                           \sum_{n=1}^{m} {1\over n^2}.
  \end{multline}

  \paragraph{$u$-channel}
Repeating the same steps as for the $t$ channel we arrive at the integral
   {\footnotesize\begin{multline}
    L_0'(1,v,1-Y) =-\int_0^1 \frac{\left(\Li{2}(\sigma)-\Li{2}(-\sigma)
        +\log(\sigma)\left(\log(1-\sigma)-\log(1+\sigma)\right)\right)
      d\sigma}{\sqrt{1+\left(v^{2}+\left(2 Y -4\right) v +Y^{2}\right)
        \sigma^{4}+\left(-2 Y +2 v \right) \sigma^{2}}} \log(v)+O(v^0)
  \end{multline}}
  The integrand can be expanded for small $v$ and $Y$ and integrated
  term by term using~\eqref{e:intL}.
  
  	%%%%%%%%%%%%%%%%%%%%%%%%%%%%%%%%%%%%%%%%%%%%%%%%%%%%%%%%%%
	\subsection{Expressions from unitarity cuts}
	\label{subsec:unitarity_appendix}
	%%%%%%%%%%%%%%%%%%%%%%%%%%%%%%%%%%%%%%%%%%%%%%%%%%%%%%%%%%
	The unitarity cut of the cross diagram in $D=4-4\epsilon$ dimensions up to order $\epsilon$ is given by
	\begin{equation}\label{eq:Cut_Cross_full_def}
	\mathrm{Cut}_{\uZ}\cW_0^{1,4-4\epsilon}=\frac{(2\pi)^3}{4}\frac{(\pi\eul^\gamma)^{-2\epsilon}v}{(x_{12}x_{34})^2}\int\limits_{-1}^{+1}\dd x\frac{(1-x^2)^{-2\epsilon}(\zeta\zetab)^{1-4\epsilon}}{(\zeta+\zetab-x(\zeta-\zetab))(\zeta+\zetab-2\zeta\zetab-x(\zeta-\zetab))^{1-4\epsilon}}
	\end{equation}
	which evaluates
		\begin{align}\label{eq:Cut_Cross_full}
	\mathrm{Cut}_{\uZ}\cW_0^{1,4-4\epsilon}&=-\frac{v\pi^3}{(x_{12}x_{34})^2}\frac{1}{(\zeta-\zetab)}\left[\log\left(\frac{1-\zeta}{1-\zetab}\right)\right.\cr
	&-2\epsilon\left(\Mpl{1,1}{\bar\zeta,{\zeta\over\bar\zeta}}-\Mpl{1,1}{\zeta,{\bar\zeta\over\zeta}}
	+\Mpl{1}{\zeta}\Mpl{1}{{\bar\zeta\over\zeta}}-\Mpl{1}{\bar\zeta}\Mpl{1}{{\zeta\over\bar\zeta}}\right.\cr
	&\left.-(\Mpl{2}{\zeta}-\Mpl{2}{\zetab})+\log\left(\frac{1-\zeta}{1-\zetab}\right)\log((1-\zeta)(1-\zetab))\right.\cr
	&+\left.\log(\zeta\zetab)\log\left(\frac{1-\zeta}{1-\zetab}\right)\right)+\mathcal O(\epsilon^2)
	\end{align}
	
	The $\Op(\epsilon^0)$ term of the cut one-loop $s$-channel integral is given by
	\begin{multline}
	\label{eq:Cut_oneloop}
	    I_{1,\mathrm{div}}^{1,\epsilon}=\frac{1}{2(\zeta-\zetab)}\Big(\Mpl{1,1}{\bar\zeta,{\zeta\over\bar\zeta}}-\Mpl{1,1}{\zeta,{\bar\zeta\over\zeta}}
	+\Mpl{1}{\zeta}\Mpl{1}{{\bar\zeta\over\zeta}}-\Mpl{1}{\bar\zeta}\Mpl{1}{{\zeta\over\bar\zeta}}\cr
	-(\Mpl{2}{\zeta}-\Mpl{2}{\zetab})+\log\left(\frac{1-\zeta}{1-\zetab}\right)\log((1-\zeta)(1-\zetab))
	-\log(\zeta\zetab)\log\left(\frac{1-\zeta}{1-\zetab}\right)\Big)+\mathcal O(\epsilon).
	\end{multline}

	%%%%%%%%%%%%%%%%%%%%%%%%%%%%%%%%%%%%%%%%%%%%%%%%%%%%%%%%%%
	\section{Conformal blocks and OPE coefficients}
	\label{sec:OPE}
	%%%%%%%%%%%%%%%%%%%%%%%%%%%%%%%%%%%%%%%%%%%%%%%%%%%%%%%%%%
	The OPE coefficients for a generalized free field in $d=3$ dimensions with external conformal dimension $\Delta$ are given by~\cite{Fitzpatrick:2011dm}
	\begin{equation}
	    \label{eq:OPEcoefficients}
	    A_{n,l}(\Delta)=\frac{2^{1+l}(\Delta-1/2)_n^2(\Delta)_{n+l}^2}{l!n!(l+3/2)_n(2\Delta+n-2)_n(2\Delta+n+l-3/2)_n(2\Delta+2n+l-1)_l}\,,
	\end{equation}
where  $(a)_n:=\Gamma(a+n)/\Gamma(a)$ is the Pochhammer symbol.
	The conformal blocks for a multiplet of dimension $\Delta$ and spin $l$ in $d=3$ dimensions have been calculated in~\cite{Li:2019cwm}. In the $v,Y$ expansion we are interested in, they are given by
	\begin{equation}
	    G_{\Delta,l}(v,Y)=\sum\limits_{k=0}^\infty v^{\frac{\Delta-l}{2}+k}\sum\limits_{m=0}^{2k}A_{k,m}f_{k,m}(Y),
	\end{equation}
	with
	\begin{equation}
	    f_{k,m}(Y)=Y^{l-m}
	    {}_2F_1\left({\frac12(\Delta+l)+k-m,\frac12(\Delta+l)+k-m\atop \Delta+l+2k-2m};Y\right)
	    \end{equation}
	    and
	    \begin{multline}
	    A_{k,m}(\Delta)=\sum\limits_{m_1,m_2=0}^{\lfloor\frac{m}{2}\rfloor}{(-1)^{m+m_1+1}4^{m_1+m_2}\over 2^l}
	    \frac{(-l)_m(-\lfloor m/2\rfloor))_{m_1+m_2}(k-\lfloor m/2\rfloor)+1/2)_{m_1}}{m! m_1! m_2!(k-m+m_1)!}\cr
	    \times\frac{(\Delta-1)_{2k-m}(3/2-\Delta)_{m-k-m_1-m_2}(l-\Delta+2)_{2(\lfloor m/2\rfloor-m_2)-n}}{(\Delta+l-m-1)_{2k-m}(\Delta+l)_{2(k+m1-\lfloor m/2\rfloor)-m}}\cr
	    \times\frac{(m+m_2-m_1-k-l-1/2)(\lfloor (m+1)/2\rfloor -l)_{m_2}}{(1/2-l)_{m+m_2-k}(3/2+l-m_2)_{k-m+m_1+m_2}}\cr
	    \times\left(\left(\frac12(\Delta+l)\right)_{k-m+m_1}\left(\frac12(\Delta-l-1)\right)_{m_2}\right)^4.
	\end{multline}
	where we use a slightly different normalization compared to~\cite{Li:2019cwm}.
	
\end{appendix}
%%%%%%%%%%%%%%%%%%%%%%%%%%%%%%%%%%%%%%%%%%%%%%%%%%%%%%%%%%

\newpage
\bibliographystyle{JHEP}
\providecommand{\href}[2]{#2}\begingroup\raggedright\endgroup

\end{document}